\documentclass[12pt]{elsarticle}

\usepackage{lipsum}
\makeatletter
\def\ps@pprintTitle{%
 \let\@oddhead\@empty
 \let\@evenhead\@empty
 \def\@oddfoot{}%
 \let\@evenfoot\@oddfoot}
\makeatother
\newcommand{\RomanNumeralCaps}[1]
    {\MakeUppercase{\romannumeral #1}}

\usepackage{subcaption}
\usepackage [english]{babel}
\usepackage [autostyle, english = american]{csquotes}
\MakeOuterQuote{"}
\usepackage[utf8]{inputenc}
\usepackage[export]{adjustbox}
\usepackage{wrapfig}
\usepackage{graphicx}
\usepackage{amssymb}
\usepackage{amsmath, nccmath}
\usepackage{commath}
\usepackage{amsthm}
\usepackage{geometry}
\usepackage{lineno}

\usepackage{lipsum}
\usepackage{array}
\makeatletter
\def\ps@pprintTitle{%
 \let\@oddhead\@empty
 \let\@evenhead\@empty
 \def\@oddfoot{}%
 \let\@evenfoot\@oddfoot}
\makeatother

\begin{document}

\begin{frontmatter}


\title{Extrapolation from hypergeometric functions, continued functions and Borel-Leroy transformation; Resummation of perturbative renormalization functions from field theories}


\author{Venkat Abhignan}
\address{Department of Physics, National Institute of Technology, Tiruchirapalli - 620015, India}

\begin{abstract}
Physically relevant field-theoretic quantities are usually derived from perturbation techniques. These quantities are solved in the form of an asymptotic series in powers of small perturbation parameters related to the physical system, and calculating higher powers typically results in a higher amount of computational complexity. Such divergent expansions were handled using hypergeometric functions, continued functions, and Borel-Leroy transforms. Hypergeometric functions are expanded as series, and a rough estimate of next-order information is predicted using information from known orders. Continued functions are used for the resummation of these series. The effective nature of extrapolation using such functions is illustrated by taking different examples in field theories. In the vicinity of second-order phase transitions, critical exponents are the most interesting numerical quantities corresponding to a wide range of physical systems. Using the techniques mentioned in this work, precise estimates are obtained for these critical exponents in $\phi^4$ and $\phi^3$ field models. 
\end{abstract}


\end{frontmatter}

\section{Introduction}

One of the most successful outputs of quantum field theory is to study $\phi^4$ scalar models with quartic self-interaction. This theory can produce interesting results to quantify the nature of continuous phase transitions on a wide range of physical systems \cite{kardar2007statistical,kleinert,PELISSETTO2002549}. Perturbative renormalization group and epsilon expansion is an effective approach that can give interesting output in this model \cite{Wilson:1973jj}. Recently with the help of a novel method called graphical functions, more orders of information have been obtained in the renormalization approach for this well known $\phi^4$ theory \cite{seven-O(n),borinsky2021graphical,seven-O(n)-2}, and also interestingly in notoriously hard to handle $\phi^3$ theory \cite{phi^3}. While the interesting quantities like critical exponents are derived close to the non-trivial IR fixed point (at distances $\sim O(4-$dimensions)) in $\phi^4$ theory, $\phi^3$ theory with cubic self-interaction is asymptotically free in 6 dimensions \cite{MACFARLANE197491}, and the quantities are derived in $O(6-$dimensions). Field-theoretic perturbative methods typically produce divergent solutions with numerically growing alternating coefficients for asymptotic series (at the limit of perturbation parameter $\rightarrow 0$), and meaningful estimates are extracted only from resummation procedures \cite{CALICETI20071,physics3040053}.  

The role of the resummation method becomes paramount when one needs to deduce quantities of high precision from divergent perturbative expansions of $O(4-$dimensions) and $O(6-$dimensions) for physically relevant three-dimensional systems where perturbation parameter $\geq1$. A large section of literature related to results of perturbative renormalization in various models is dedicated to carefully performing different resummation methods and comparing these results with experimental or numerical simulations or other recent theoretical approaches (Non-perturbative renormalization group or Conformal Bootstrap calculations) \cite{phi^3,six-O(n),ADZHEMYAN2019,KOMPANIETS2020,kompaniets2020critical,adzhemyan2021sixloop}. Without this, the perturbative results are meaningless and are less significant by themselves. There exist different resummation methods for effective summation of asymptotic series such as widely used Pad\'e approximants \cite{baker_graves-morris_1996}, Borel summation \cite{kleinert,34,36}  and Pad\'e-Borel-Leroy transformation \cite{phi^3,six-O(n),ADZHEMYAN2019,KOMPANIETS2020,kompaniets2020critical,adzhemyan2021sixloop,RIM2021}. More recently derived methods are based on conformal mapping, Borel technique with hypergeometric functions \cite{hyp1,hyp2,hyp3,hyp4,hyp5,hyp6} and continued functions \cite{abhignan2020continued,abhignan2021}. When dealing with an appropriate resummation technique for desired divergent quantities, more information regarding them would be ideal, but very few properties and information are available. So it is, therefore, useful to try and compare different approaches, which helps in removing inconsistencies and strengthening existing opinions regarding such $\phi^4$ and $\phi^3$ field theories. 

Having the lower-order or weak coupling information, strong coupling information, and asymptotic large-order information would be ideal when handling a divergent field-theoretic quantity. However, all this information is not known in most cases for different field theories. Typically used Pad\'e based approximants cannot account for strong coupling information and are not uniquely defined \cite{baker_graves-morris_1996}. Shalaby introduced a new parametrization to include all this information in resummation methods using hypergeometric functions \cite{shalaby2020,shalaby2020critical,SHALABY202010,Shalaby2022,Shalaby2022arxiv}. Initially, hypergeometric approximants were implemented by Mera et al. \cite{hyp1}. Later efficient hypergeometric analytic continuation was obtained compared to typical Pad\'e and Pad\'e-Borel methods in various physically relevant systems \cite{hyp2,hyp3,hyp4,hyp5,hyp6}. Interesting new results were derived for critical exponents in $\phi^4$ models using seven loop expansions \cite{seven-O(n),seven-O(n)-2} implementing these hypergeometric functions \cite{shalaby2020,shalaby2020critical}. And these results helped understand the decade-old discrepancy between theoretical predictions and experimental value in a model for superfluid helium-4 transition, known as "$\lambda$-point specific heat experimental anomaly" \cite{lambda}. The same issue could also be addressed using continued functions such as continued exponential (CE), a continued fraction (CF) and continued exponential fraction (CEF) from resummation, only using lower-order information \cite{abhignan2020continued}. Further, in the same work, critical exponents were derived for varied dimensions $>1$ in $O(n)-$symmetric $\phi^4$ models. Also, hypergeometric approximants could solve for precise estimates in this regime where perturbation parameter $>1$ \cite{shalaby2020critical}. These results are interesting since previous resummation studies of perturbative renormalization group functions could not predict reliable estimates for such systems due to the non-analyticity of the functions around their fixed points \cite{Eckmann1975,Orlov2000,Calabrese_2000}. Similarly, only using lower-order information, Borel-Leroy transformation was combined with continued exponential to implement a new resummation method \cite{abhignan2021}. Using continued functions interesting critical parameters were derived for modified $\phi^4$ models, such as $n$-vector model with cubic anisotropy \cite{ADZHEMYAN2019}, $O(n)\times O(m)$ spin models \cite{KOMPANIETS2020} and the weakly disordered Ising model \cite{RIM2021}. These forms of continuous iterative functions or self-similar approximants were first used for resummation by Yukalov \cite{Yukalov1991,Yukalov1992} which then developed into self-similar approximation theory \cite{physics3040053,YUKALOV2002,Yukalov2019}. Such self-similar approximants were also involved in using weak coupling information and strong coupling information to interpolate information from interesting problems of field theories \cite{PhysRevDself}. While approximation is one application of such resummation techniques, extrapolation also becomes interesting since information at large parameter limit (perturbation parameter $\rightarrow \infty$) is relevant when approximation is required at perturbation parameter $\geq1$. Recently implementing self-similar approximants, strong coupling information was extrapolated using the weak coupling information in Gell-Mann-Low functions of quantum field theories \cite{physrevdself2}.

 Here we implement hypergeometric functions, continued functions and Borel-Leroy transformation on recently derived perturbative expansions in $\phi^4$ \cite{six-O(n),adzhemyan2021sixloop,seven-O(n),seven-O(n)-2} and $\phi^3$ \cite{phi^3} models. Extrapolation and resummation of interesting field-theoretic quantities are performed. Increasing the precision of parameters like critical exponents in $\phi^4$ and $\phi^3$ models improves the description of the attributed physical systems. Also, in all these cases, the exact solutions are not known, and the reliability of the existing interpolated and extrapolated predictions are sustained only when different methods produce comparable results. We also test the predictions from extrapolated information of critical exponents by implementing it in an appropriate resummation procedure.
 
 The paper is organized as follows: We initially introduce the hypergeometric functions, continued functions and illustrate their applications on field theories in Sec. 2. We handle the Gell-Mann-Low functions from $O(n)-$symmetric $\phi^4$ model in Sec. 3.  We then handle the $\epsilon$ expansions from $\phi^4$ and $\phi^3$ models using hypergeometric functions, continued functions and Borel-Leroy transformation in Sec. 4 and 5, respectively.
\section{Continued functions, Borel-Leroy transformation and hypergeometric functions}
Physically relevant quantities are solved from perturbative methods of field theories in the form of\begin{equation}
    Q(\epsilon)\approx\sum^N_{i=0} q_i \epsilon^i
\end{equation}  where $\epsilon$ is the perturbation parameter. This expression with $\{q_i\}$ holds the weak coupling information ($\epsilon \rightarrow 0$), while the strong coupling information is given as \begin{equation}
    Q(\epsilon)\sim A\epsilon^{-a}+B\epsilon^{-b}+C\epsilon^{-c}+\cdots,\,\,\,\epsilon \rightarrow \infty.
\end{equation} Further, these series $Q(\epsilon)$ are divergent in nature and possess large-order asymptotic behaviour of the form \begin{equation}
    q_i \sim i! (\sigma)^i i^l \left(1+O\left(\frac{1}{i}\right)\right),\,\,\,i \rightarrow \infty,
\end{equation} 
where $\sigma$ is the large-order parameter.
\subsection{Resummation from continued exponential fraction, continued exponential and Borel-Leroy transformation; Parametrization with weak coupling information} 
Here we illustrate the convergence nature of CEF, CE and Borel-Leroy transformation by implementing only weak coupling information of quantities $Q(\epsilon)$ from light quantum chromodynamics $U(N_f) \times U(N_f)$ model \cite{calabrese2004,pisarski,agostino2003} for $N_f$ quark flavours \cite{PRBPISARKSI,JPAPISARSKI,PATERSON1981188}. The mean-field approach for this model predicts a second-order transition with an unbroken anomaly at critical temperature; however, flow of couplings in renormalization group theory of $U(n) \times U(m)$ model predict the absence of infrared stable fixed points for $n=m=N_f$. It was deduced that this finite-temperature transition in the limit of vanishing quark masses must be of a first-order kind by measuring the inequality for upper marginal dimensionality $n^+(m,3)>m$ which holds for physically relevant $N_f\geq2$ in three-dimensional systems. The recently derived six-loop approximate quantities of interest for $n^+(m,4-\epsilon)$ in minimal subtraction scheme of renormalization are of the form \cite{adzhemyan2021sixloop} \begin{subequations}
\begin{align}
    n^+(2,4-\epsilon) \approx 18.485 - 19.899 \epsilon + 2.926 \epsilon^2 + 4.619 \epsilon^3  - 0.718 \epsilon^4  - 1.766 \epsilon^5,\\
    n^+(3,4-\epsilon) \approx 28.856 - 30.083\epsilon + 6.557\epsilon^2 + 3.406\epsilon^3 - 0.796\epsilon^4 - 1.451\epsilon^5,\\
    n^+(4,4-\epsilon) \approx  38.975 - 40.239\epsilon + 9.609\epsilon^2 + 3.050\epsilon^3 - 0.616\epsilon^4 - 1.370\epsilon^5,\\
    n^+(5,4-\epsilon) \approx 49.000 - 50.375\epsilon + 12.49\epsilon^2 + 2.981\epsilon^3 - 0.463\epsilon^4 - 1.415\epsilon^5,\\
    n^+(6,4-\epsilon) \approx 58.983 - 60.501\epsilon + 15.29\epsilon^2 + 3.043\epsilon^3 - 0.342\epsilon^4 - 1.515\epsilon^5, 
\end{align}
\end{subequations} in $4-\epsilon$ dimensions $(\epsilon \rightarrow 0)$. Typically simple Taylor representation of such perturbation expansions do not converge to a meaningful value for $\epsilon=1$ (three-dimensional, in this case) due to a nonphysical singularity that determines the radius of convergence ($\epsilon<1$). The objective of the resummation method is to find a convergent representation for such series that can produce a meaningful sum \cite{bender1999advanced}. Due to the irregular structure of these quantities for $n^+(\{2,3,4,5,6\},4-\epsilon), $ resummation using Borel and Pad\'e-Borel methods lead to unstable and erroneous predictions \cite{calabrese2004}. However, Pad\'e and Pad\'e-Borel-Leroy methods can be implemented when only such lower-order information is given \cite{adzhemyan2021sixloop}.  \\ Similar to Pad\'e, we find that convergence can be obtained by converting quantities in general form of Eq. (1) into CF \begin{equation}
    Q(\epsilon) \sim \frac{h_0}{\frac{h_1\epsilon}{\frac{h_2\epsilon}{\frac{h_3\epsilon}{\frac{h_4 \epsilon}{\cdots}+1}+1}+1}+1}
\end{equation} or a CEF \cite{abhignan2020continued, abhignan2021} \begin{equation}
   Q(\epsilon) \sim
 c_0\exp\left(\frac{1}{1+c_1\epsilon\exp\left(\frac{1}{1+c_2\epsilon\exp\left(\frac{1}{1+c_3\epsilon\exp\left(\frac{1}{1+\cdots}\right)}\right)}\right)}\right)  \end{equation} or a CE \begin{equation}
    Q(\epsilon) \sim d_0\exp(d_1\epsilon \exp(d_2 \epsilon \exp(d_3 \epsilon \exp(d_4 \epsilon\exp(d_5 \epsilon\exp(d_6 \epsilon\exp(\cdots))))))) \end{equation} or continued exponential with Borel-Leroy transformation (CEBL) \begin{equation}
        Q(\epsilon) \sim \int_0^\infty \exp(-t) t^l e_0\exp(e_1\epsilon t\exp(e_2\epsilon t\exp(e_3\epsilon t\exp(\cdots)))) dt\; \; \; \hbox{for} \; \; \; (\epsilon \rightarrow 0),
    \end{equation} where $l$ is the Borel-Leroy parameter. CF are intimately related with the Pad\'e approximants and can be algebraically manipulated with ease \cite{Bultheel2001,Aptekarev_2011,LORENTZEN20101364}. CE were initially explored by Bender and Vinson \cite{contexp} followed by it was used for convergence in phase transition studies \cite{abhignan2020continued,abhignan2021,POLAND1998394}. CEF was devised by combining CE and CF \cite{abhignan2020continued}. As mentioned earlier, using such continued functions and their combinations were commonly developed by Yukalov (see, e.g. recent reviews \cite{Yukalov2019,physics3040053}). CEBL is based on Pad\'e-Borel-Leroy transformation  \cite{six-O(n),ADZHEMYAN2019,KOMPANIETS2020,kompaniets2020critical,adzhemyan2021sixloop,RIM2021,Hardy} \begin{equation}
     Q(\epsilon) = \int_0^\infty \exp(-t) t^l f(\epsilon t) dt ,\,\,\, f(y) = \sum_{i=0}^\infty \frac{q_i}{\Gamma(i+l+1)} y^i,
 \end{equation} which replaces Pad\'e with CE for better convergence. From Stirling's approximation for large $i$ the growth of  $\Gamma(i+l+1)$ can be determined as $i^l i!$ which can account for factorial growth of coefficients $q_i$ as in Eq. (3) \cite{kleinert}. These transformation of variables are affine and generally enlarge the radius of convergence for $Q(\epsilon)$ on a cut plane. \\ The convergence is observed by measuring quantities at successive resummation orders  \begin{equation}
      H_1\equiv\frac{h_0 }{h_1\epsilon+1},\,H_2\equiv\frac{h_0}{\frac{h_1\epsilon}{h_2\epsilon+1}+1},\,H_3\equiv\frac{h_0}{\frac{h_1\epsilon}{\frac{h_2\epsilon}{h_3\epsilon+1}+1}+1},\,H_4\equiv\frac{h_0}{\frac{h_1\epsilon}{\frac{h_2\epsilon}{\frac{h_3\epsilon}{h_4 \epsilon+1}+1}+1}+1},\cdots
  \end{equation} for CF,
\begin{multline}
     C_1 \equiv c_0\exp\left(\frac{1}{1+c_1\epsilon}\right),\,C_2 \equiv c_0\exp\left(\frac{1}{1+c_1\epsilon\exp\left(\frac{1}{1+c_2\epsilon}\right)}\right),\\ C_3 \equiv c_0\exp\left(\frac{1}{1+c_1\epsilon\exp\left(\frac{1}{1+c_2\epsilon\exp\left(\frac{1}{1+c_3\epsilon}\right)}\right)}\right),\cdots
 \end{multline} for CEF,
 \begin{equation}
    D_1 \equiv d_0\exp(d_1\epsilon),\,D_2 \equiv d_0\exp(d_1\epsilon\exp(d_2\epsilon)),\,D_3 \equiv d_0\exp(d_1\epsilon\exp(d_2\epsilon\exp(d_3\epsilon))),\,\cdots
\end{equation} for CE
and \begin{multline}
     E_1 \equiv \int_0^\infty \exp(-t) t^l e_0\exp(e_1\epsilon t) dt,\,E_2 \equiv \int_0^\infty \exp(-t) t^l e_0\exp(e_1\epsilon t\exp(e_2\epsilon t)) dt,\\ E_3 \equiv \int_0^\infty \exp(-t) t^l e_0\exp(e_1\epsilon t\exp(e_2\epsilon t\exp(e_3\epsilon t))) dt,\cdots
 \end{multline}
 for CEBL. Transformed variables $\{h_i\}$, $\{c_i\}$ and $\{d_i\}$ can be obtained from Taylor expansion of approximants at arbitary order and by relating them with coefficients $\{q_i\}$ from Eq. (1). \begin{equation}
     q_0=h_0,\,q_1=-h_0h_1,\,q_2=h_0h_1{\left(h_2+h_1 \right)},\,q_3=-h_0h_1 {\left(h_2 h_3 +{h_2}^2 +2h_1h_2 +{h_1}^2 \right)},\cdots
 \end{equation} for CF, \begin{equation}
    \begin{gathered}
    q_0 = c_0\exp(1),\,
    q_1 = -c_0c_1\exp(2),\,q_2 = c_0\exp(3){\left( c_1c_2 +\frac{3{c_1}^2}{2}\right)},\\
    q_3 = -c_0c_1\exp(4){\left(\frac{13{c_1 }^2}{6}+3c_1c_2+ \,c_2 \,c_3 +\frac{3{c_2 }^2 }{2} \, \right)},\, \cdots
\end{gathered}
\end{equation} for CEF and \begin{equation}
    \begin{gathered}
    q_0 = d_0,\,
    q_1 = d_0 d_1,\,q_2 = d_0{\left(d_1 d_2 +\frac{{d_1 }^2 }{2}\right)},
    q_3 = d_0 d_1 {\left( d_2 \,d_3 +\frac{{d_2 }^2 }{2} +d_1d_2+\frac{{d_1 }^2 }{6} \right)},\, \cdots
\end{gathered}
\end{equation} for CE. Similarly, solving these CE relations sequentially for coefficients of Borel-Leroy transformed series $f(y)$ in Eq. (9) provides transformed variables $\{ e_i \}$ for CEBL. \\ Error calculation is significant to obtain accuracy of the estimates produced from these continued functions. This is predicted from the principle of fastest apparent convergence where difference in estimates at successive orders is measured \cite{six-O(n),ERROR}. To obtain accelerated convergence and measure this error for sequences $\{C_i\},\{D_i\},\{E_i\}$, these representations are combined with Shanks transformation for any partial sum $\{A_i\}$ such as \cite{bender1999advanced} \begin{equation}
  S(A_i) = \frac{A_{i+1}A_{i-1}-A_i^2}{A_{i+1}+A_{i-1}-2A_i}.
 \end{equation} Shanks transformation can further be iterated $S^2(A_i) \equiv S(S(A_i))$ depending on improvement in convergence nature and availability of parameters $\{q_i\}$ from weak coupling information. Pad\'e approximants are equivalent to Shanks transformation in some aspects, however Shanks can be easily applied on sequences and its further iterations \cite{CALICETI20071,ANDREWS198670}. For a sequence $\{A_i\}$ the error is calculated from a relation \cite{shalaby2020critical} \begin{equation}
     (|A_{i+1} - A_{i}| + |A_{i+1} - S(A_{i})|)/2,
 \end{equation} when $S(A_{i})$ is taken as optimized estimate for $Q(\epsilon)$. Further the shift parameter $l$ which inhibits the convergence in Borel-Leroy transformation is obtained by minimzing this error. \\ 
 For instance, to tackle the quantity $n^+(2,4-\epsilon)$ with these resummation methods we consider CEF for $1/n^+$ while CE and CEBL are implemented directly. We empirically observe that CF is more reliable for region of convergence where $\epsilon>1$, which is further seen in Sec. 4.3 and Sec. 5.2. Numerical estimate of $n^+(2,4-\epsilon)$ using CEF sequences from Eq. (11), CE sequences from Eq.(12) and CEBL sequences from Eq. (13) ($l=7.08$) for relevant physical condition ($\epsilon=1$) are \begin{equation}
     C_3 = 2.397, C_4 = 5.349, C_5 = 4.391,D_3 = 4.208,\,D_4=4.513, D_5=4.362,\ \hbox{and},\ E_{3,4,5} =4.420,
 \end{equation} at four-loop ($A_3$), five-loop ($A_4$) and six-loop ($A_5$). Further Shanks transformation (Eq.(17)) is applied and estimates with errors (Eq.(18)) are obtained as $S(C_4)=4.62(59)$, $S(D_4)=4.41(10)$ and $S(E_4)=4.42032(14)$. We illustrate the convergence behaviours of these oscillating sequence of continued functions for $n^+(2,3)$ at consecutive orders in Fig. 1. These values show the effectiveness of CEBL procedure. Displayed in Figs. 2 and 3 are $S(E_4)$ (six-loop) and $S(E_3)$ (five-loop) with error bars, respectively showing their consecutive errors computed from relation as in Eq. (18) derived from CEBL sequence.
 \begin{figure}
    \centering
    \includegraphics[width=0.5\linewidth, height=6cm]{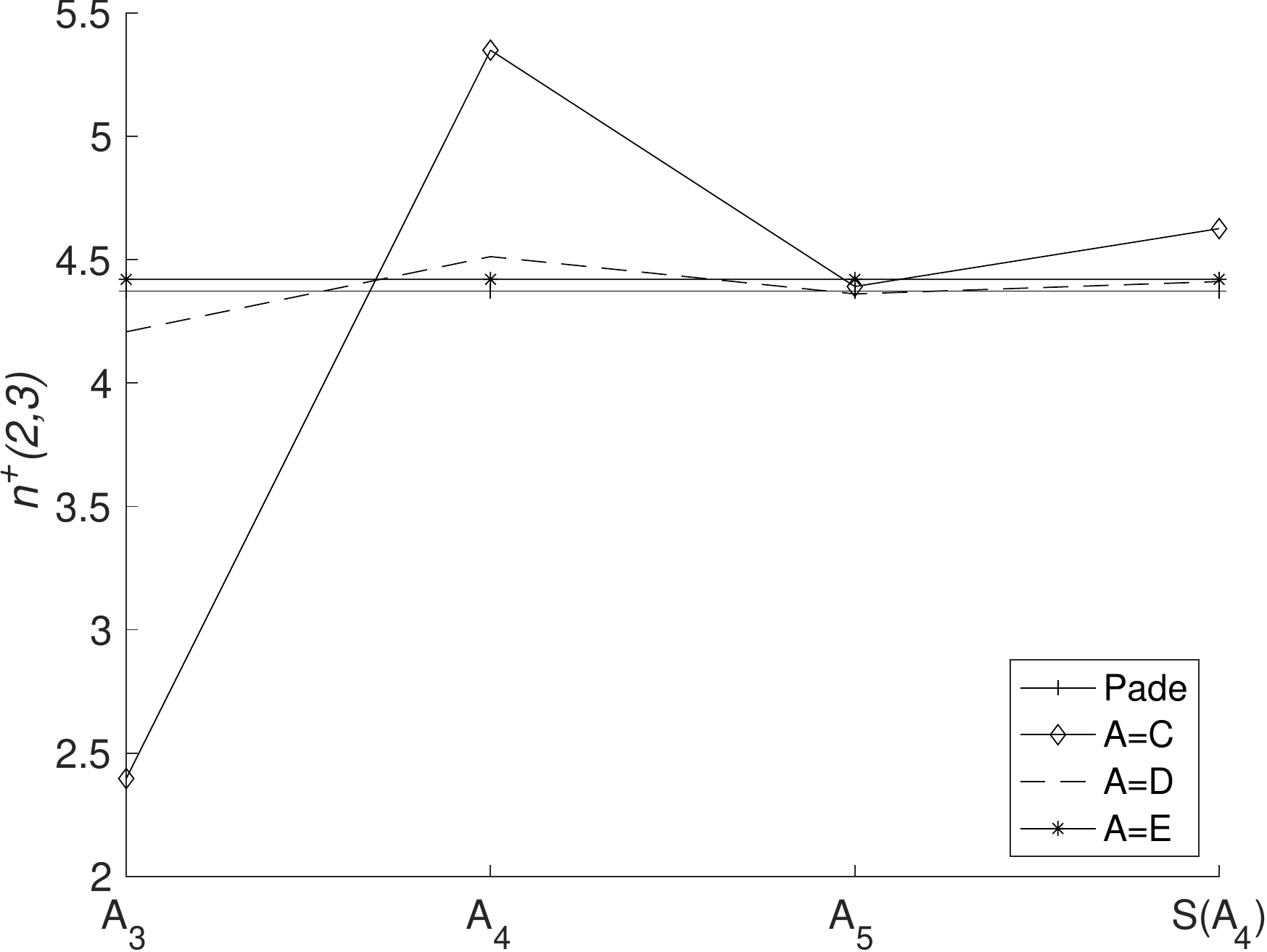}
    \caption{Estimate of $n^+(2,3)$ at consecutive orders compared with Pad\'e prediction \cite{adzhemyan2021sixloop}.}
\end{figure}
 \begin{figure}[ht]
\centering
\begin{subfigure}{0.46\textwidth}
\includegraphics[width=1\linewidth, height=5cm]{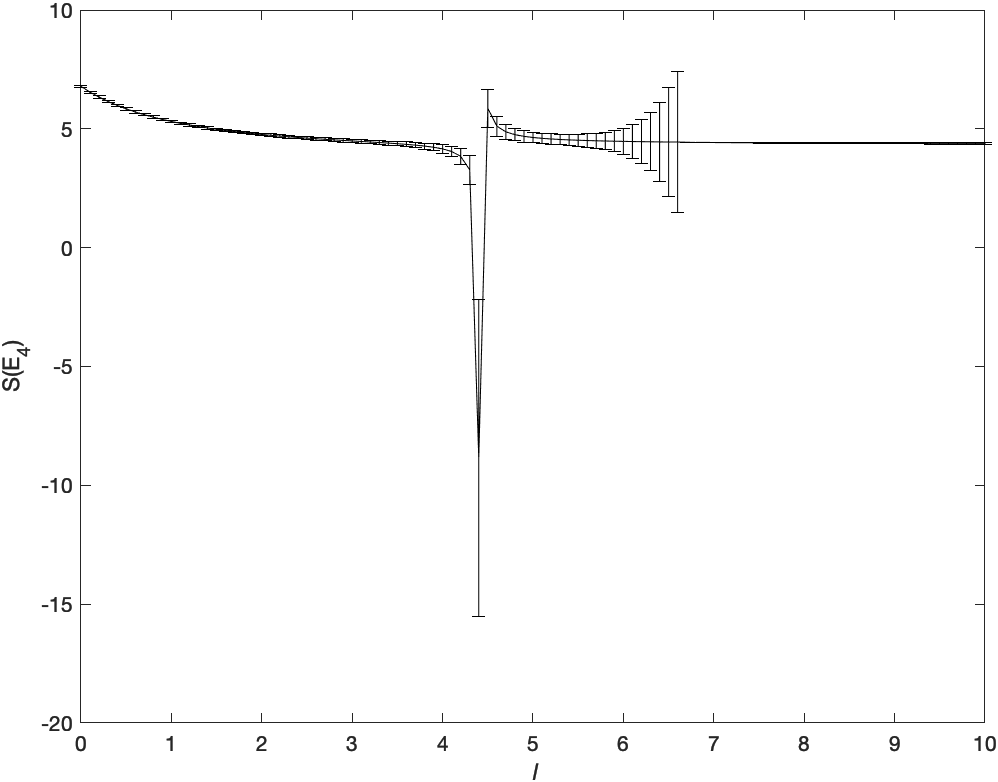} 
\caption{$S(E_4)$  vs $l$ for $l \in [0,10]$}

\end{subfigure}
\begin{subfigure}{0.46\textwidth}
\includegraphics[width=1\linewidth, height=5cm]{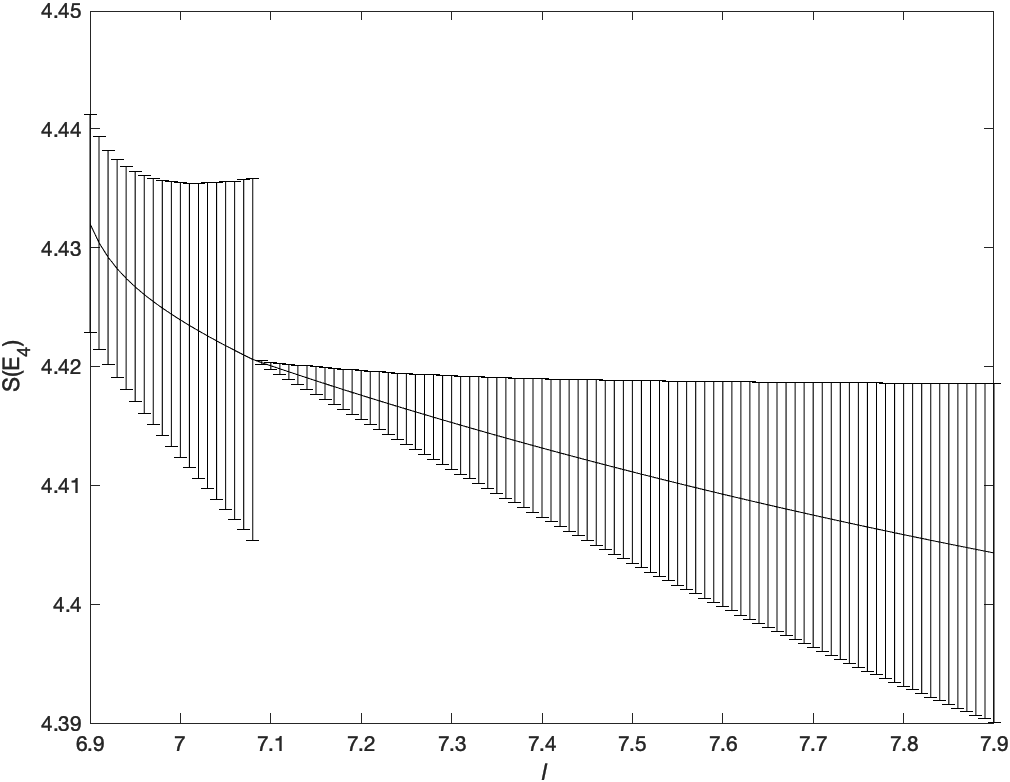}
\caption{$S(E_4)$  vs $l$ for $l \in [6.9,7.9]$}

\end{subfigure}
 
\caption{We plot the estimate of $n^+(2,3)$ derived from $S(E_4)$ vs shift parameter $l$, with the error bars showing the value of $(|E_{5} - E_{4}| + |E_{5} - S(E_{4})|)/2$.}

\end{figure}
\begin{figure}[ht]
\centering
\begin{subfigure}{0.46\textwidth}
\includegraphics[width=1\linewidth, height=5cm]{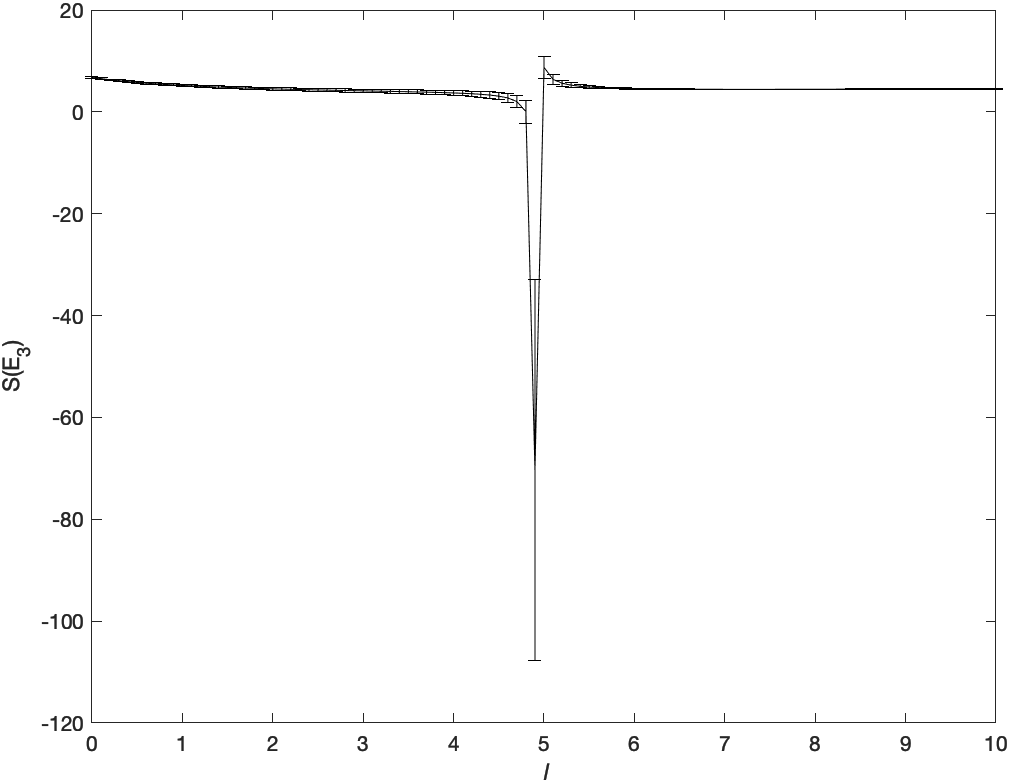} 
\caption{$S(E_3)$  vs $l$ for $l \in [0,10]$}

\end{subfigure}
\begin{subfigure}{0.46\textwidth}
\includegraphics[width=1\linewidth, height=5cm]{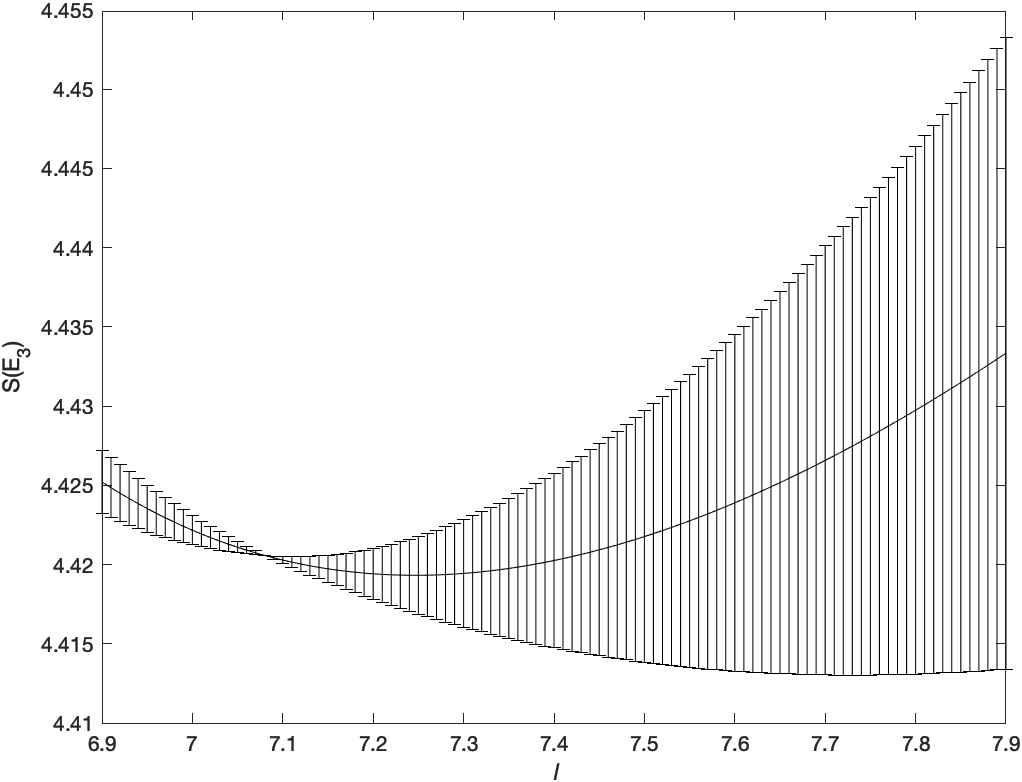}
\caption{$S(E_3)$  vs $l$ for $l \in [6.9,7.9]$}

\end{subfigure}
 
\caption{We plot the estimate of $n^+(2,3)$ derived from $S(E_3)$ vs shift parameter $l$, with the error bars showing the value of $(|E_{4} - E_{3}| + |E_{4} - S(E_{3})|)/2$.}

\end{figure}
It is to be noted here that convergence in CEBL is remarkable already at 5-loop to the predicted value and that by shifting the parameter $l$, the unstable nature of the approximants is removed. Also, the parameter $l=7.08$ remains the same for different orders, whereas this parameter changes for different orders in the Pad\'e-Borel-Leroy procedure since convergence is checked for the entire Pad\'e triangle and many unstable estimates are obtained \cite{adzhemyan2021sixloop}. CEBL estimate is obtained from fewer calculations analogous with only using the diagonal Pad\'e terms. To reiterate the convergence behaviour of the CEBL procedure, we display  $n^+(\{3,4,5,6\},3)$ vs shift parameter $l$ obtained from 5-loop expansions in Figs. 4 and 5. 
\begin{figure}[ht]
\centering
\begin{subfigure}{0.46\textwidth}
\includegraphics[width=1\linewidth, height=5cm]{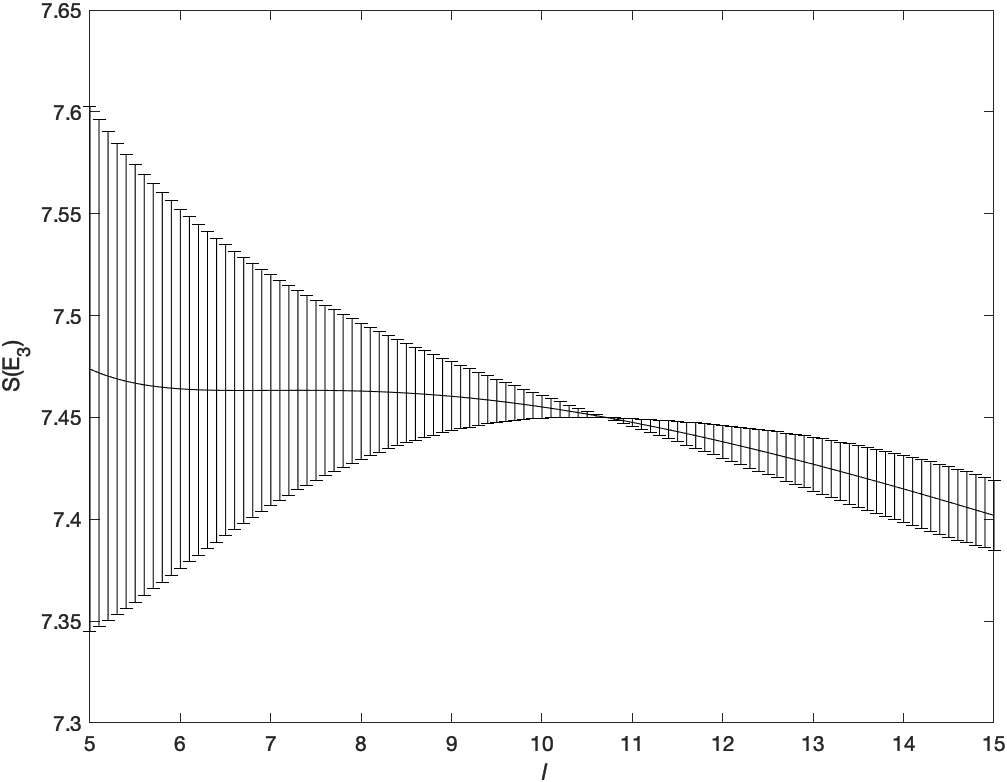} 
\caption{Estimate of $n^+(3,3)$ from $S(E_3)$}

\end{subfigure}
\begin{subfigure}{0.46\textwidth}
\includegraphics[width=1\linewidth, height=5cm]{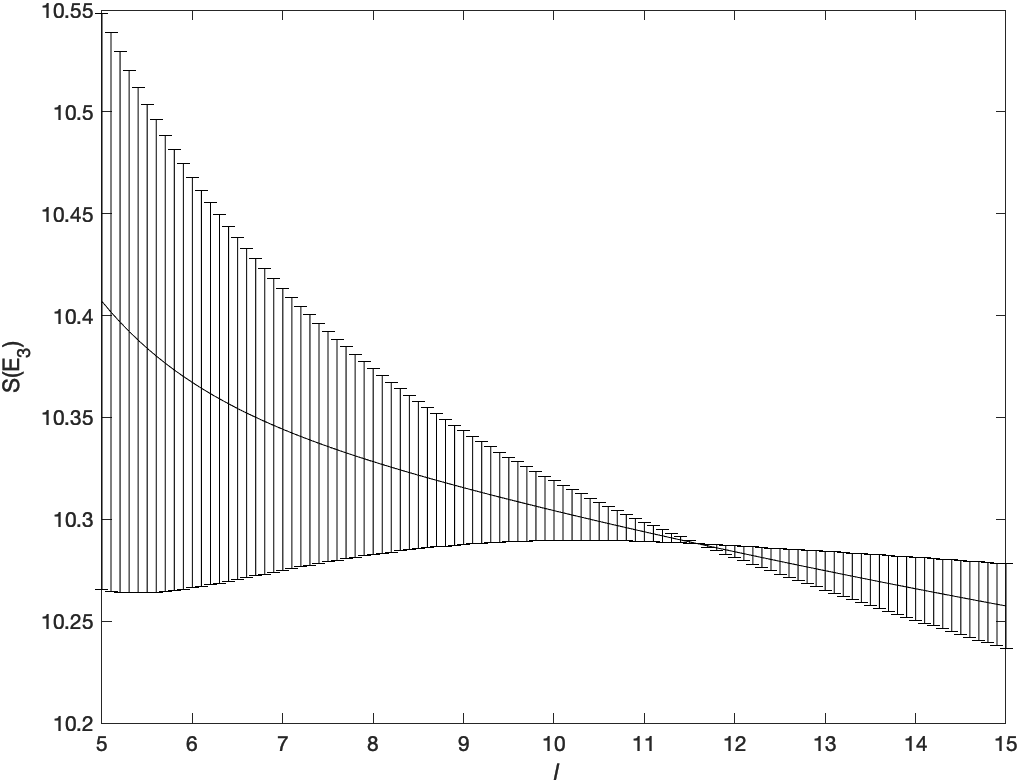}
\caption{Estimate of $n^+(4,3)$ from $S(E_3)$}

\end{subfigure}
 
\caption{We plot the estimate of $n^+(\{3,4\},3)$ derived from $S(E_3)$ vs shift parameter $l$ for $l \in [5,15]$, with the error bars showing the value of $(|E_{4} - E_{3}| + |E_{4} - S(E_{3})|)/2$.}

\end{figure}
\begin{figure}[ht]
\centering
\begin{subfigure}{0.46\textwidth}
\includegraphics[width=1\linewidth, height=5cm]{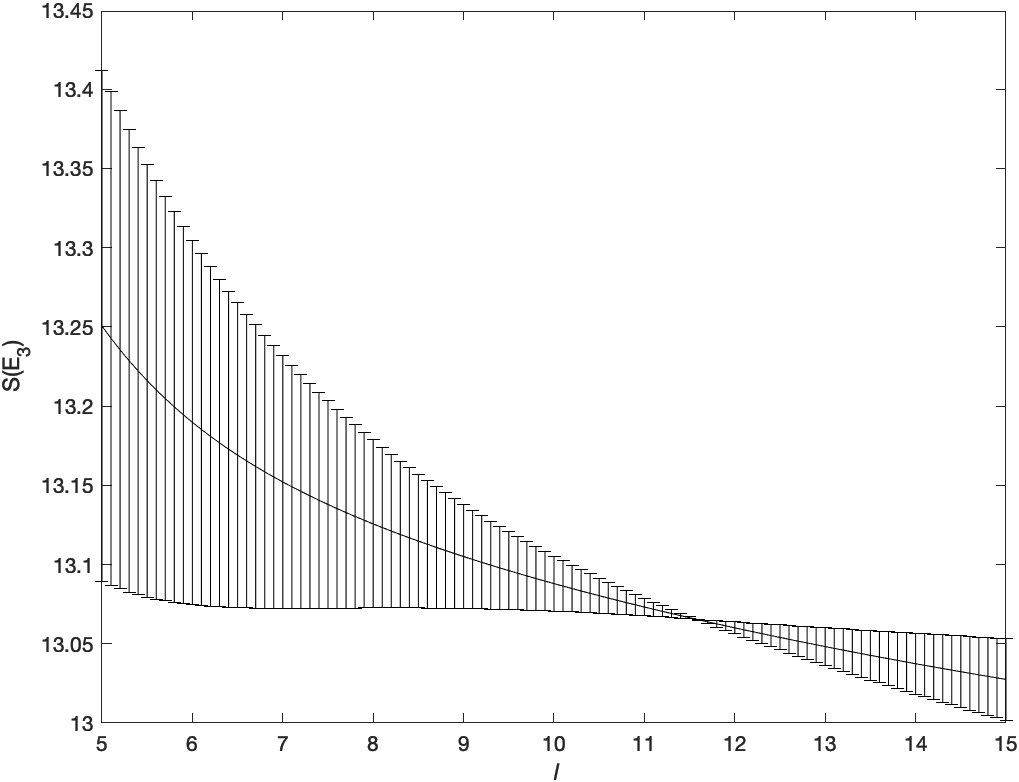} 
\caption{Estimate of $n^+(5,3)$ from $S(E_3)$}

\end{subfigure}
\begin{subfigure}{0.46\textwidth}
\includegraphics[width=1\linewidth, height=5cm]{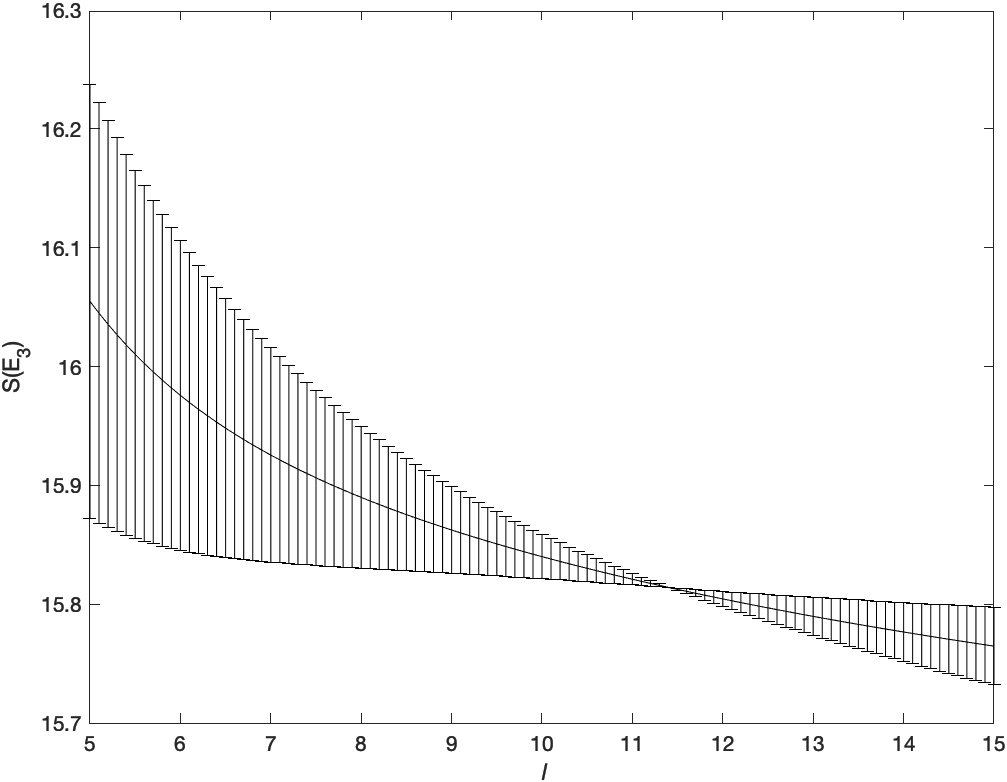}
\caption{Estimate of $n^+(6,3)$ from $S(E_3)$}

\end{subfigure}
 
\caption{We plot the estimate of $n^+(\{5,6\},3)$ derived from $S(E_3)$ vs shift parameter $l$ for $l \in [5,15]$, with the error bars showing the value of $(|E_{4} - E_{3}| + |E_{4} - S(E_{3})|)/2$.}

\end{figure}
Further, we compare the estimates from CEF ($S(C_4)$), CE ($S(D_4)$) and CEBL ($S(E_4)$) with existing predictions from Pad\'e based methods \cite{adzhemyan2021sixloop} in Table \ref{table:n^+}. The estimates of  $n^+(\{3,4,5,6\},3)$ at consecutive orders for continued functions are illustrated in Figs. 6 and 7. The estimates seem to be comparable with existing predictions, however not exactly compatible. Using constrained series where exact known solutions are used may shift these values \cite{calabrese2004, adzhemyan2021sixloop}, but we are only interested in illustrating the direct convergence behaviours of continued functions. 

\begin{figure}[ht]
\centering
\begin{subfigure}{0.46\textwidth}
\includegraphics[width=1\linewidth, height=6cm]{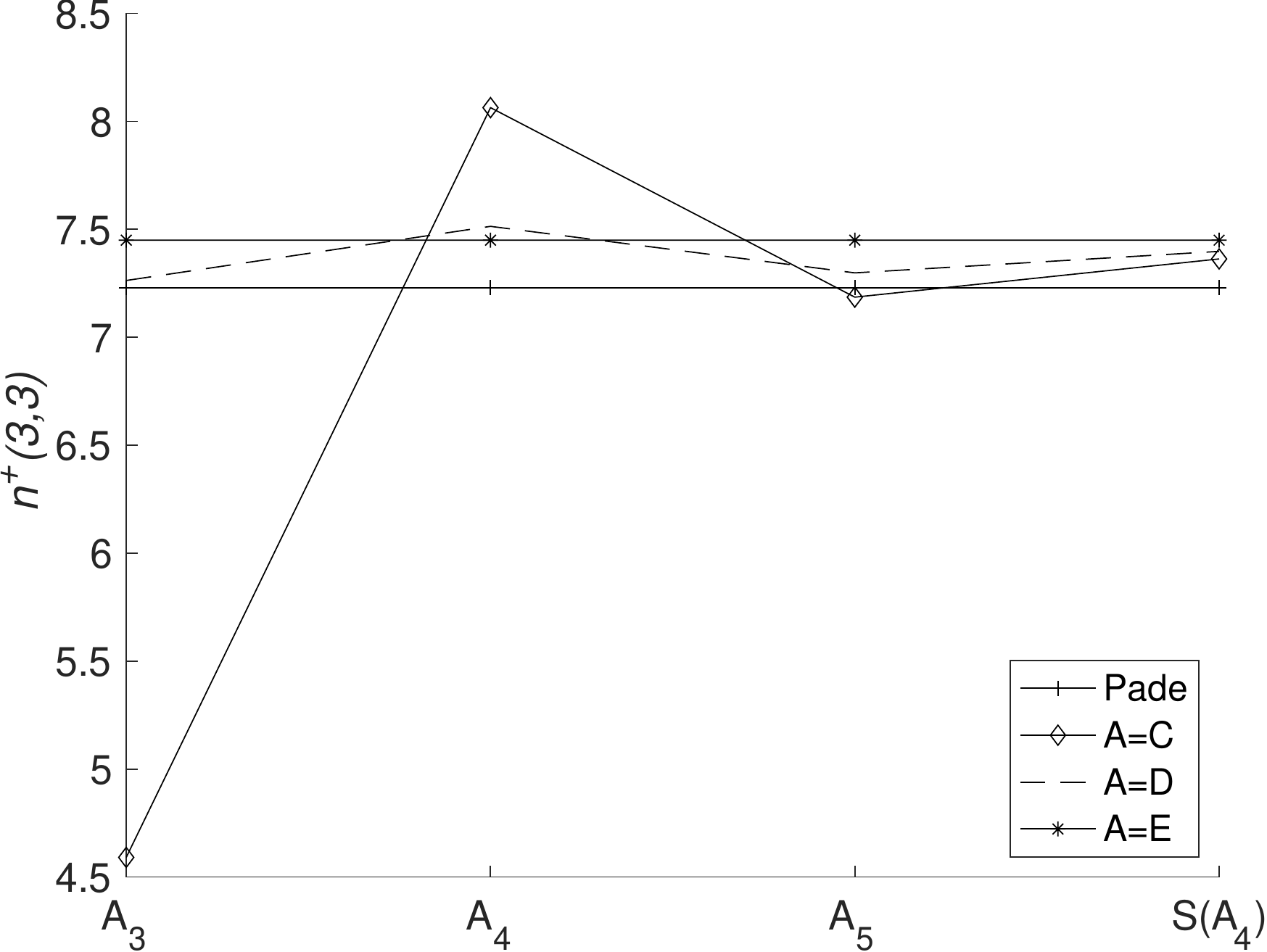} 
\caption{Estimate of $n^+(3,3)$}

\end{subfigure}
\begin{subfigure}{0.46\textwidth}
\includegraphics[width=1\linewidth, height=6cm]{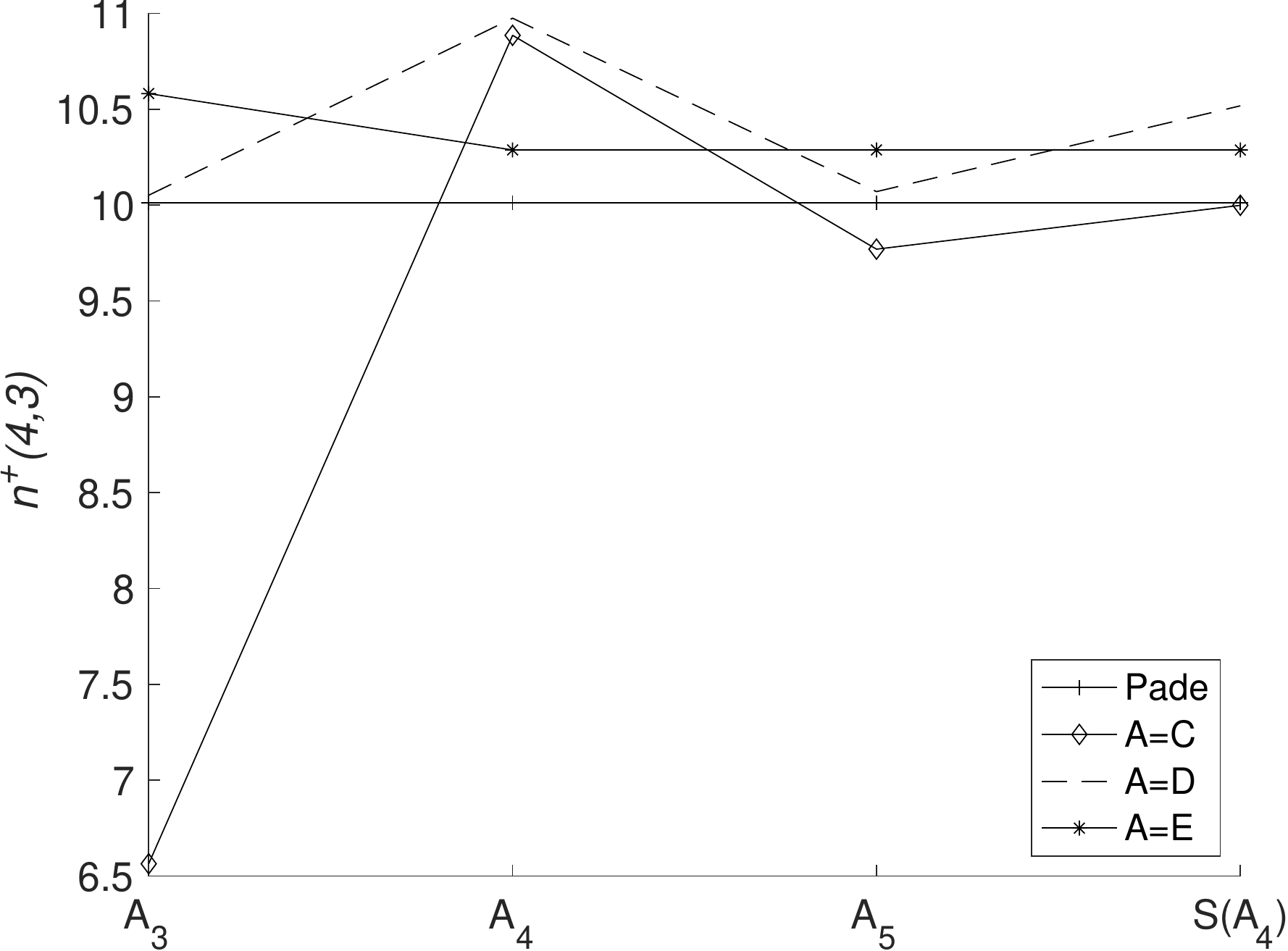}
\caption{Estimate of $n^+(4,3)$}

\end{subfigure}
 
\caption{Estimate of $n^+(\{3,4\},3)$ at consecutive orders compared with Pad\'e prediction \cite{adzhemyan2021sixloop}.}

\end{figure}
\begin{figure}[ht]
\centering
\begin{subfigure}{0.46\textwidth}
\includegraphics[width=1\linewidth, height=6cm]{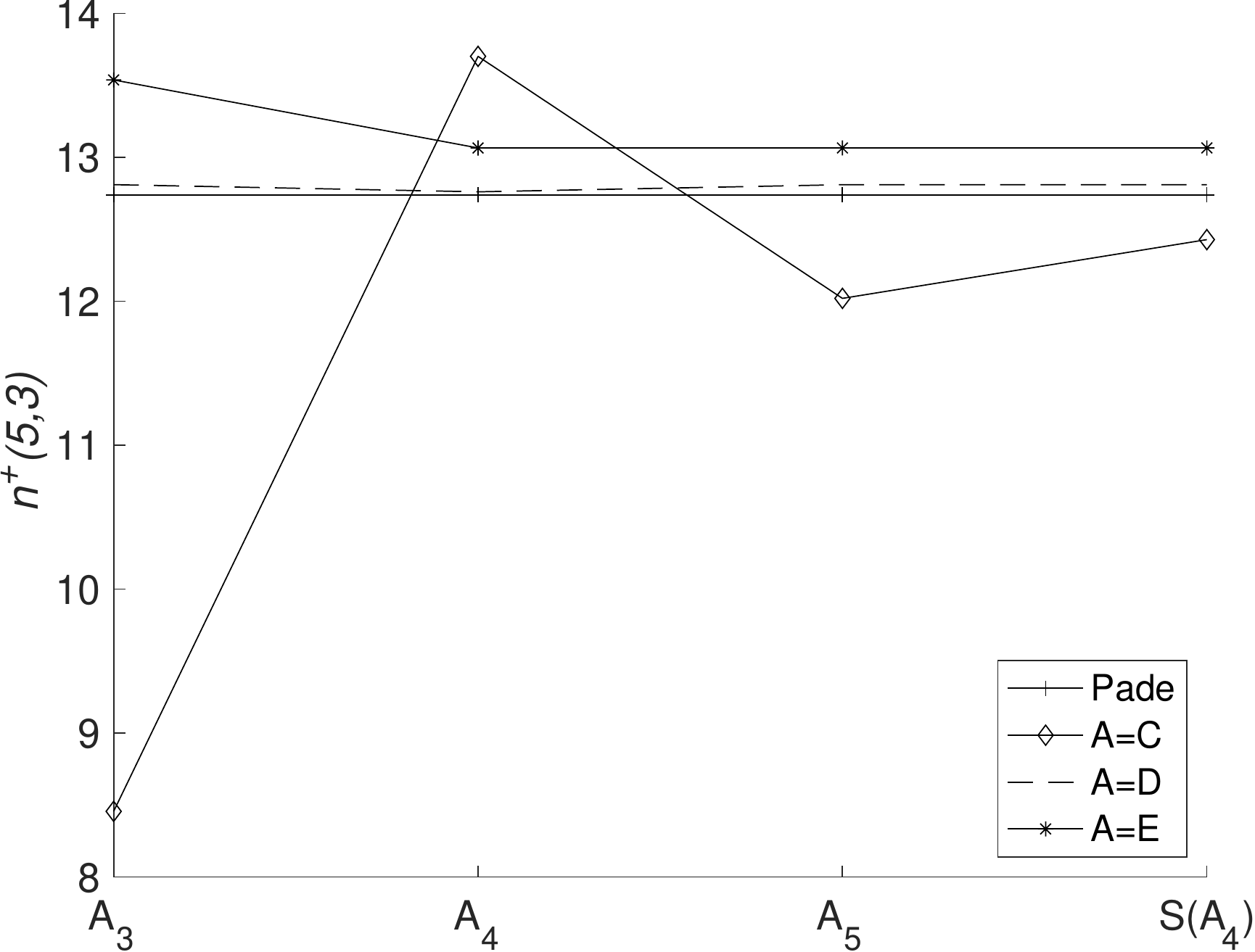} 
\caption{Estimate of $n^+(5,3)$}

\end{subfigure}
\begin{subfigure}{0.46\textwidth}
\includegraphics[width=1\linewidth, height=6cm]{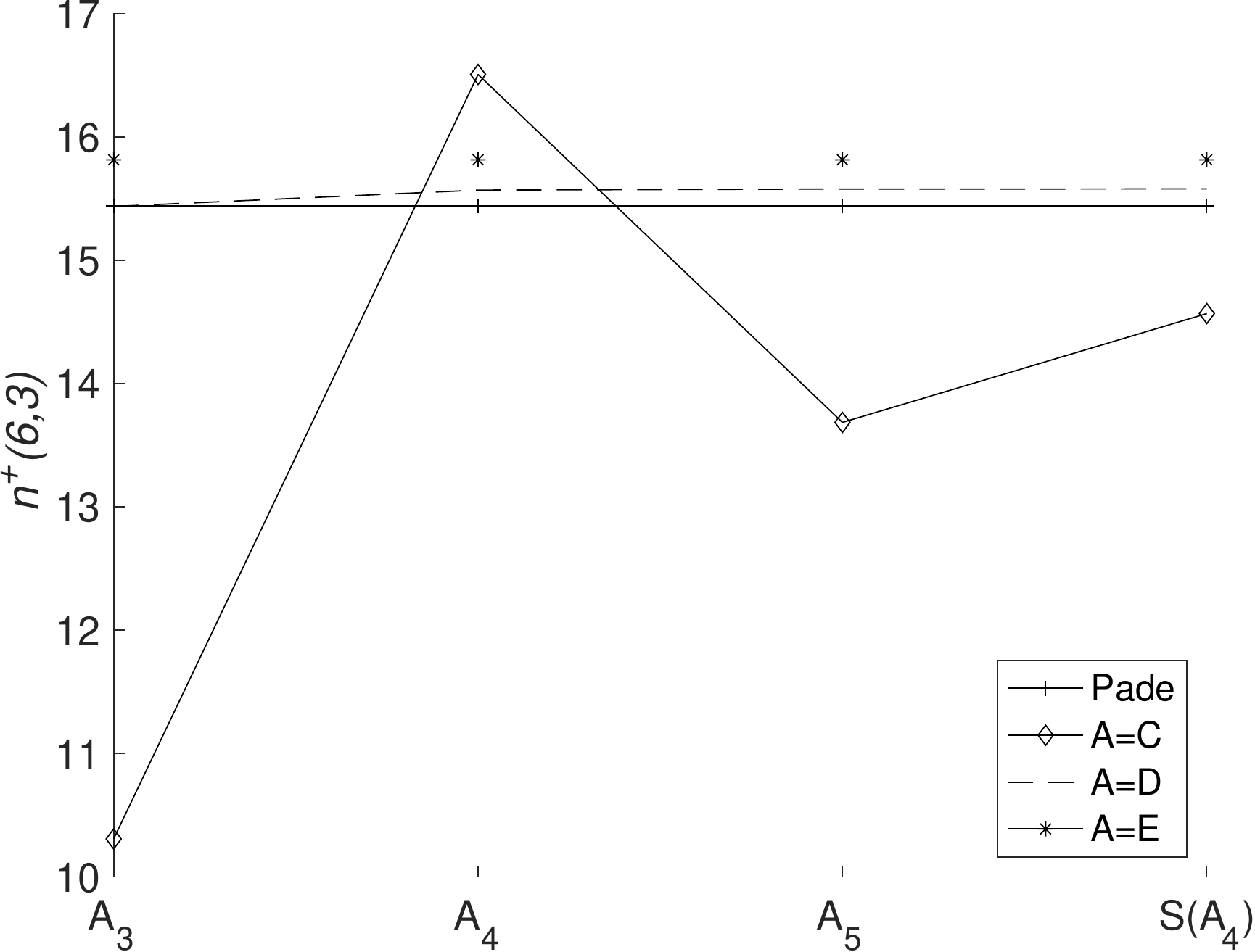}
\caption{Estimate of $n^+(5,3)$}

\end{subfigure}
 
\caption{Estimate of $n^+(\{5,6\},3)$ at consecutive orders compared with Pad\'e prediction \cite{adzhemyan2021sixloop}.}

\end{figure}
\begingroup
\setlength{\tabcolsep}{3pt} 
\renewcommand{\arraystretch}{1} 
\begin{table}[ht]

\begin{center}
\caption{Continued functions and Borel-Leroy transformation on upper marginal dimensionality $n^+(m,4-\epsilon)$ from light quantum chromodynamics $U(n) \times U(m)$ three-dimensional model.} 

 \begin{tabular}{||c c c c c||}
 
 \hline
$n^+(m,3)$ & $S(C_4)$ & $S(D_4)$ &  $S(E_4)$  & Pad\'e-based \cite{adzhemyan2021sixloop}  \\ [0.5ex] 
 \hline\hline
 
    $m=$ 2
   & 4.62(59)
   & 4.41(10)
   &  \begin{tabular}{c c}
        & 4.4203(1)  \\
        & $(l=7.08)$
   \end{tabular} 
   & 4.373(18)
   \\ 
 \hline
   $m=$ 3
   & 7.36(52)
   & 7.39(16)
     & \begin{tabular}{c c}
        & 7.4498  \\
        & $(l=10.72)$
   \end{tabular} 
  
   & 7.230(22)
   \\ 
 \hline
  $m=$ 4
   & 9.99(67)
   & 10.52(67)
     & 
 \begin{tabular}{c c}
        &10.2877(1)  \\
        &  $(l=11.61)$
   \end{tabular} 
   & 10.012(28)
   \\ 
 \hline
  $m=$ 5
   & 12(1)
   & 12.81
   &   \begin{tabular}{c c}
        & 13.0653   \\
        &  $(l=11.58)$
   \end{tabular} 
   & 12.74(05)
   \\ 
 \hline
   $m=$ 6
   & 15(1)
   & 15.57
   &    \begin{tabular}{c c}
        & 15.8137  \\
        &  $(l=11.42)$
   \end{tabular} 
   & 15.44(08)
   \\ 
 \hline
\end{tabular}
\label{table:n^+}
\end{center}
\end{table}
\subsection{Extrapolation from hypergeometric functions; Parametrization with weak coupling, strong coupling and large-order information} 
We implement hypergeometric function to estimate information of $(i+1)$th order in the hypergeometric series from $(i)$th order information. Shalaby \cite{shalaby2020,shalaby2020critical,SHALABY202010,Shalaby2022,Shalaby2022arxiv} suggested an appropriate hypergeometric approximant to get a convergent value of $Q(\epsilon)$ as \begin{equation}
    Q(\epsilon)\sim q_0\left({}_{j+2}F_j(a_1,a_2,\cdots,a_{j+2};b_1,b_2,\cdots,b_{j};\sigma \epsilon)\right)
\end{equation} The lower order information in Eq. (1) and large-order information in Eq. (3) can be accommodated to this approximant from its expansion by solving for $\{a_i\}$ and $\{b_i\}$ from $\{q_i\}$ such as \begin{equation}
     {}_{j+2}F_j(a_1,a_2,\cdots,a_{j+2};b_1,b_2,\cdots,b_{j};\sigma \epsilon) = \sum_{i=0} \frac{\frac{\Gamma(a_1+i)}{\Gamma(a_1)}\frac{\Gamma(a_2+i)}{\Gamma(a_2)}\cdots\frac{\Gamma(a_{j+2}+i)}{\Gamma(a_{j+2})}}{i!\frac{\Gamma(b_1+i)}{\Gamma(b_1)}\frac{\Gamma(b_2+i)}{\Gamma(b_2)}\cdots\frac{\Gamma(b_{j}+i)}{\Gamma(b_{j})}} (\sigma \epsilon)^i
\end{equation} where $i! = \Gamma(i+1)$(gamma function). The asymptotic large order behaviour for series expansion of this approximant can be derived to be in form of Eq. (3) with factorial growth. Similarly, the strong coupling behaviour for this approximant is given as \begin{equation}
     {}_{j+2}F_j(a_1,a_2,\cdots,a_{j+2};b_1,b_2,\cdots,b_{j};\sigma \epsilon) \sim A\epsilon^{-a_1}+B\epsilon^{-a_2}+C\epsilon^{-a_3}+\cdots,\,\,\,\epsilon \rightarrow \infty.
\end{equation} 
We primarily use approximants ${}_{2}F_0(a_1,a_2;\sigma \epsilon)$ and ${}_{3}F_1(a_1,a_2,a_3;b_1;\sigma \epsilon)$ for which hypergeometric variables $\{a_i,b_i,\sigma\}$ can be obtained from coefficients $\{q_i\}$ in Eq. (1). The relations for obtaining hypergeometric variables can be deduced from series expansion in Eq. (21) such as \begin{subequations}
\begin{align}
    a_1 a_2 \sigma &= \frac{q_1}{q_0},\\
    \frac{1}{2!} a_1(a_1+1) a_2(a_2+1) \sigma^2 &= \frac{q_2}{q_0},\\
    \frac{1}{3!} a_1(a_1+1)(a_1+2) a_2(a_2+1)(a_2+2) \sigma^3 &= \frac{q_3}{q_0},
\end{align}
\end{subequations} for ${}_{2}F_0(a_1,a_2;\sigma \epsilon)$ approximant and \begin{subequations}
\begin{align}
    \frac{a_1 a_2 a_3}{b_1} \sigma &= \frac{q_1}{q_0}\\
    \frac{a_1 (a_1+1) a_2 (a_2+1) a_3 (a_3+1)}{2! b_1 (b_1+1)} \sigma^2 &= \frac{q_2}{q_0}\\
    \frac{a_1 (a_1+1) (a_1+2) a_2 (a_2+1) (a_2+2) a_3 (a_3+1) (a_3+2)}{3! b_1 (b_1+1) (b_1+2)} \sigma^3 &= \frac{q_3}{q_0}\\ \frac{a_1 (a_1+1) (a_1+2) (a_1+3) a_2 (a_2+1) (a_2+2) (a_2+3) a_3 (a_3+1) (a_3+2) (a_3+3)}{4! b_1 (b_1+1) (b_1+2) (b_1+3)} \sigma^4 &= \frac{q_4}{q_0}\\ \frac{a_1 (a_1+1) \cdots (a_1+4) a_2 (a_2+1) \cdots (a_2+4) a_3 (a_3+1) \cdots (a_3+4)}{5! b_1 (b_1+1) (b_1+2) (b_1+3) (b_1+4)} \sigma^5 &= \frac{q_5}{q_0}
\end{align}
\end{subequations} for ${}_{3}F_1(a_1,a_2,a_3;b_1;\sigma \epsilon)$ approximant. \subsubsection{Ground-state energy of anharmonic oscillator} To illustrate the extrapolation nature of hypergeometric approximant, it is initially tested on the well known ground-state energy of one-dimensional anharmonic oscillator \cite{anharmonic}. The ground-state energy $E$ at weak coupling limit $g \rightarrow0$ is known as \begin{equation}
    E\approx0.5+0.75g-2.625g^2+20.8125g^3-241.289g^4+3580.98g^5-63982.8g^6,
\end{equation} where $g$ is the coupling constant. Initially we approximate this energy with ${}_{2}F_0$ approximant at third order, by implementing $q_0=0.5,q_1=0.75,q_2=-2.625,q_3=20.8125$ in Eqs. (23) to identify hypergeometric parameters $a_1=-0.27326,a_2=1.32499,\sigma=-4.1429$. The approximant is \begin{equation}
    E\sim 0.5 {}_{2}F_0(-0.27326,1.32499;-4.1429 g)
\end{equation} at third-order. The leading strong coupling behaviour from this approximant is $E\sim g^{0.27326},\,g\rightarrow\infty$ from Eq. (22). The ground-state energy can also be approximated by ${}_{3}F_1$ approximant at fifth-order by utilising $q_0=0.5,q_1=0.75,q_2=-2.625,q_3=20.8125,q_4=-241.289,q_5=3580.98$ in Eqs. (24) to obtain $a_1=-0.335298,a_2=0.36446,a_3=5.89769,b_1=0.22952,\sigma=-2.41998$, and the approximant is \begin{equation}
E\sim  0.5 {}_{3}F_1(-0.335298,0.36446,5.89769;0.22952;-2.41998 g). \end{equation} The prediction for strong coupling behaviour from this approximant is $E\sim g^{0.335298},\,g\rightarrow\infty$. The strong coupling behaviours of these approximants ${}_{2}F_0\sim g^{0.27326}$ and ${}_{3}F_1\sim g^{0.335298}$ seem to approach the exact limiting behaviour $E\sim g^{1/3}$ with percentage error of $0.6\%$ \cite{scah}. And, the asymptotic large order parameter $\sigma=-4.14$ for ${}_{2}F_0$ approximant, $\sigma=-2.42$ for ${}_{3}F_1$ approximant seem to approach exact behaviour $\sigma=-3$ with percentage error of $-20\%$\cite{abah}. \\ This prediction can be improved using higher hypergeometric approximants, but here, implementing this strong coupling information we are interested in estimating $(i+1)$th order information from $(i)$th order information. The quantity can also be approximated such as $E \sim 0.5+0.75g\,{}_{3}F_1(a_1,a_2,a_3;b_1;\sigma x)$. But, in this case we fix the strong coupling information such as $a_1=(1-0.335298)$ (Since for $\,g\rightarrow\infty$, here $E \sim g\,{}_{3}F_1 \sim g\,g^{-1+0.335298}$) and implement $q_0=0.75,q_1=-2.625,q_2=20.8125,q_3=-241.289,q_4=3580.98$ in Eqs. (24a),(24b),(24c),(24d) to identify hypergeometric parameters $a_2=0.9034,a_3=8.034,b_1=3.229,\sigma=-2.343$. Using all these parameters in Eq. (24e), the estimate of $q_5$ at sixth-order can be predicted as $-63887.5$, which shows compatibility with actual value $-63982.8$ in Eq. (25). This value at sixth-order can also be predicted fixing the large order behaviour $\sigma=-3$, by similarly implementing $q_0=0.75,q_1=-2.625,q_2=20.8125,q_3=-241.289,q_4=3580.98$ in Eqs. (24a),(24b),(24c),(24d) to identify hypergeometric parameters $a_1=0.4205,a_2=1.228,a_3=2.779,b_1=-1.125$. Using all these parameters in Eq. (24e), the estimate of $q_5$ can be predicted as $-64430.9$, which again shows compatibility with actual value $-63982.8$.
\subsubsection{Energy of massive Schwinger model}
Further, here we illustrate the properties of hypergeometric approximant by handling the divergent expressions of the massive Schwinger model \cite{Schwinger,LOWENSTEIN1971172}. This model is generally used as testing ground for numerical methods \cite{PhysRevDself,physrevdself2} since it has properties of quantum electrodynamics in (1+1) dimensions and also of quantum chromodynamics \cite{COLEMAN1975267,COLEMAN1976239,Prds,PRDS2,PRDS3}. In the lattice formulation of this model \cite{PRDSE}, energy of the excited-"vector state" can be perturbatively expressed from the series \begin{equation}
    f(z)\approx1+2z-10z^2+78.667z^3-736.22z^4+7572.9z^5-82736.7z^6,\,z \rightarrow 0
\end{equation} where $z\equiv1/(g a)^4$, with coupling parameter $g$ and lattice spacing $a$. Initially, we approximate this energy at third-order with ${}_{2}F_0$ approximant, by implementing $q_0=1,q_1=2,q_2=-10,q_3=78.667$ in Eqs. (23) to identify $a_1=-0.17635,a_2=14.176,\sigma=-0.8$, and the approximant is \begin{equation}
    f(z)\sim{}_{2}F_0(-0.17635,14.176;-0.8 z).
\end{equation} The leading strong coupling behaviour from this approximant is $f(z)\sim z^{0.17635},\,z\rightarrow\infty$ from Eq. (22). Similarly, the expression for $f(z)$ can be approximated with ${}_{3}F_1$ approximant by utilising $q_0=1,q_1=2,q_2=-10,q_3=78.667,q_4=-736.22,q_5=7572.9$ in Eqs. (24) to obtain $a_1=-0.21969,a_2=0.15373,a_3=208.16,b_1=0.22952,\sigma=-0.06529$, and the approximant is \begin{equation}
f(z)\sim{}_{3}F_1(-0.21969,0.15373,208.16;0.22952;-0.06529 z) \end{equation} at fifth-order. The prediction for strong coupling behaviour from this approximant is $f(z)\sim z^{0.21969},\,z\rightarrow\infty$. The strong coupling behaviours of these approximants ${}_{2}F_0\sim z^{0.17635}$ and ${}_{3}F_1\sim z^{0.21969}$ seem to approach the exact limiting behaviour $f(z)\sim z^{1/4}$ with percentage error of $-12\%$. \\The quantity can also be approximated such as $f(z) \sim 1+2z\,{}_{3}F_1(a_1,a_2,a_3;b_1;\sigma x)$. Again, in this case we fix the strong coupling information such as $a_1=(1-0.21969)$ (Since for $\,z\rightarrow\infty$, here $f(z) \sim z\,{}_{3}F_1 \sim z\,z^{-1+0.21969}$) and implement $q_0=2,q_1=-10,q_2=78.667,q_3=-736.22,q_4=7572.9$ in Eqs. (24a),(24b),(24c),(24d) to identify hypergeometric parameters $a_2=0.96014,a_3=228.17,b_1=2.0546,\sigma=-0.06009$. Using all these parameters in Eq. (24e), the estimate of $q_5$ can be predicted as $-82755.4$, which shows compatibility with actual value $-82736.7$ in Eq. (28). It can also be observed from this approximant that large-order behaviour  $\sigma=-0.06009$ seems to be compatible with previous prediction $\sigma=-0.06529$ in Eq. (30) which could have also been used to estimate the same value. \\ The ground state energy $(E/g)$ of the "vector state" in this model was solved perturbatively in presence of a fermion with mass $m$. At the continuum limit ($a\rightarrow0$) of this nontrivial gauge theory, considering $x\equiv m/g$ it was found that \begin{equation}
    \frac{E}{g} \approx 
    0.5642-0.219x+0.1907x^2,\,x \rightarrow 0 \,\,\hbox{and}\,\, \frac{E}{g} \approx 
    0.6418x^{-1/3},\,x \rightarrow \infty
\end{equation} at weak \cite{PRDS2,PRDS4,PRDS6} and strong coupling limits \cite{PRDSE,PRDS6,PRDS8}, respectively. This information $q_0=0.5642,q_1=-0.219,q_2=0.1907$ and strong coupling information $a_1=1/3$ is used in Eqs.(23a), (23b) to find the appropriate hypergeometric approximant \begin{equation}
   \frac{E}{g} \sim 
    0.5642\,\,{}_{2}F_0\left(1/3,8.219;-0.1416 x\right).
\end{equation}
Using this information $a_1=1/3,a_2=8.219,\sigma=-0.1416$ in Eq.(23c), $q_3$ can be estimated as $-0.2147$. With this estimate at third order, the energy in Eq. (31) from continuum limit \begin{equation}
    \frac{E}{g} \approx 
    0.5642-0.219x+0.1907x^2-0.2147x^3
    \end{equation} is compatible with the ground state energy estimated from a different approach in lattice theory \cite{PRDSE}  at weak coupling limit ($x \rightarrow 0$) \begin{equation}
    \frac{E}{g} \approx 
    0.56-0.2x+0.16x^2-0.22x^3.
\end{equation}

 These examples illustrate that the coefficients of the series in the regime of low perturbation parameter possess hidden information regarding the entire regime, at finite values and infinite values of the perturbation parameter. We exploit the properties of hypergeometric functions which can efficiently find the relation between these coefficients. \\ So far we have only considered approximate instances. We now implement information from an exact prediction for $SU(N)$ circular Wilson loop ($\langle W \rangle$) of $\mathcal{N}=4$ supersymmetric Yang–Mills theory \cite{ERICKSON2000155,Drukker2001}. $\mathcal{N}=4$ supersymmetric Yang–Mills theory in four dimensions is dual to type $\RomanNumeralCaps{2}$B string theory on an $AdS_5 \times S^5$ background based on the AdS/CFT conjecture \cite{Maldacena1999,AHARONY2000183}. This conjecture makes it applicable to study the correlation functions for large gauge group of rank $N$ and large $g^2N$, at the ’t Hooft limit. Here $\lambda'^2 \equiv g^2N$ is the ’t Hooft coupling, where $g^2$ is the string coupling. The exact result for expectation value of circular Wilson loop is deduced as \cite{ERICKSON2000155} \begin{equation}
    \langle W \rangle = \frac{2}{\lambda'}I_1\left(\lambda'\right)
\end{equation} 
where $I_1$ is the modified Bessel function of the first kind. The AdS/CFT correspondence \cite{PRL_maldacena} predicts this quantity can be approximated as \cite{PhysRevDself} \begin{equation}
    \langle W \rangle = e^{\lambda'}\left(1-\lambda'+\frac{5}{8}\lambda'^2-\frac{7}{24}\lambda'^3+\frac{7}{64}\lambda'^4-\frac{11}{320}\lambda'^5+\frac{143}{15360}\lambda'^6\right),\,\lambda' \rightarrow 0.
\end{equation} where expansion of Bessel function is taken up to sixth-order. Using this information at weak-coupling limit up to fifth-order we construct a ${}_3F_1$ hypergeometric function such as \begin{equation}
    \langle W \rangle \sim e^{\lambda'} \,{}_{3}F_1\left(1.499,215.3,-225.8;2.995;0.00004 \lambda'\right).
\end{equation} This hypergeometric function predicts the leading strong coupling behaviour as $\langle W \rangle \sim e^{\lambda'} \lambda'^{-1.499} $, which is exactly compatible with the limiting value $\langle W \rangle \sim e^{\lambda'} \lambda'^{-3/2}, \,\lambda' \rightarrow \infty $ \cite{ERICKSON2000155}. Using this strong coupling information we predict the sixth-order term as $0.0091$ which when compared with exact value $143/15360$ gives $-2\%$ deviation. These illustrate the ability of hypergeometric functions to predict strong coupling information and next-order corrections in typical perturbative formulations. These next-order information can be roughly estimated either using strong-coupling information or large-order information. However, we do not study the convergence nature of these ${}_{j+2}F_j$ hypergeometric functions, which have been studied exclusively through Meijer G-function representation \cite{shalaby2020,shalaby2020critical,SHALABY202010,Shalaby2022,Shalaby2022arxiv}. Further, we exploit these properties of hypergeometric functions to predict the leading strong coupling behaviours and the next order perturbative corrections from known orders of information in Gell-Mann-Low $\beta$ functions in $O(n)-$symmetric $\phi^4$ theory. 

\section{Gell-Mann-Low $\beta$ function in $O(n)-$symmetric $\phi^4$ field theory} 
 One can also effectively extrapolate strong coupling information from known asymptotic large order behaviour and weak coupling information. Here using such information, the divergent Gell-Mann-Low $\beta$ functions for $O(n)-$symmetric $\phi^4$ model are considered. The strong coupling information for this function was recently extrapolated using self-similar approximants \cite{physrevdself2}. Renormalization group $\beta$-functions are of interest in field theories which determine the flow of coupling and are defined as \begin{equation}
    \beta(g) = \mu \frac{\partial g}{\partial \mu}
\end{equation} where $\mu$ is the renormalization scale and $g$ is the coupling parameter. For $n$-component field the seven-loop approximation in minimal subtraction scheme of renormalization \cite{shalaby2020critical,seven-O(n),seven-O(n)-2} produced \begin{equation}
    \beta(g) \approx -g+\sum_{i=1}^7 B_i(n)\,\,g^{(i+1)}
\end{equation} in three-dimensional systems. We implement $\sigma = -0.14777423 (9/n+8)$ \cite{kleinert} asymptotic large order information and weak coupling information from Eq. (39) to determine hypergeometric approximants ${}_{3}F_1$ for $0\leq n \leq 4$ at successive orders (second-order phase transitions on interesting systems; $n=0$: dilute polymer solutions, $n=1$: Ising model, $n=2$: superconductivity and superfluid helium-4 transition, $n=3$: Heisenberg ferromagnets, $n=4$: some models of quark-gluon plasma). Initially, we consider the $\beta$ function in three-loop order such as \begin{equation}
    \beta(g) \approx -g+\sum_{i=1}^3 B_i(n)\,g^{(i+1)} \sim {}_{3}F_1(a_1,a_2,a_3;b_1;\sigma g)-1.
\end{equation} We consider $q_0=1$, $q_1=-1$, $q_2=B_1(n)$, $q_3=B_2(n)$, $q_4=B_3(n)$ and $\sigma = -0.14777423 (9/n+8)$ in Eqs. (24a), (24b), (24c), (24d) to obtain hypergeometric parameters $\{a_1,a_2,a_3,b_1\}$ which can provide the strong coupling information and obtain its leading behaviour in Table \ref{table:O(n)}. \begingroup
\setlength{\tabcolsep}{5pt} 
\renewcommand{\arraystretch}{1} 
\begin{table}[ht]

\begin{center}
\caption{Hypergeometric parameters of Gell-Mann-Low $\beta$ function (three-loop approximation) for \\ $O(n)-$symmetric $\phi^4$ model.} 

 \begin{tabular}{||c c c c||}
 
 \hline
$n$ & $(B_1,B_2,B_3)$  & $(a_1,a_2,a_3,b_1)$  &   \begin{tabular}{c c}
   & $_{3}F_1 \sim$ \\
     &   $(g \rightarrow \infty)$
   \end{tabular} \\ [0.5ex] 
 \hline\hline
 
     0
   & (2.66667,-4.66667,25.4571)
   & (-2.147,0.2223,29.29,-2.324)
   &$g^{2.147}$ \\ 
 \hline
    1
   & (3.00000,-5.66667,32.5497) 
   & (-2.117,0.1928,37.54,-2.265)
   &$g^{2.117}$\\ 
 \hline
    2
   & (3.33333,-6.66667,39.9478)
   & (-2.089,0.1710,46.46,-2.207)
   &$g^{2.089}$\\
 \hline
   3
   & (3.66667,-7.66667,47.6514)
   & (-2.064,0.1542,55.97,-2.153)
   &$g^{2.064}$\\
 \hline
   4
   & (4.00000,-8.66667,55.6606)
   & (-2.042,0.1408,66.03,-2.104)
   &$g^{2.042}$\\
 \hline
\end{tabular}
\label{table:O(n)}
\end{center}
\end{table} \\Next, we consider the $\beta$ function in four-loop order, such as  \begin{equation}
    \beta(g) \approx g\left(-1+\sum_{i=1}^4 B_i(n)\,\,g^i\right) \sim -g\,{}_{3}F_1(a_1,a_2,a_3;b_1;\sigma g).
\end{equation} We now consider $q_0=-1$, $q_1=B_1(n)$, $q_2=B_2(n)$, $q_3=B_3(n)$, $q_4=B_4(n)$ and $\sigma = -0.14777 (9/n+8)$ in Eqs. (24a), (24b), (24c), (24d) to obtain hypergeometric parameters $\{a_1,a_2,a_3,b_1\}$ which can provide the strong coupling information and compare its leading behaviour with existing predictions in Table \ref{table:O(n)2}.\begingroup
\setlength{\tabcolsep}{5pt} 
\renewcommand{\arraystretch}{1} 
\begin{table}[ht]

\begin{center}
\caption{Hypergeometric parameters of Gell-Mann-Low $\beta$ function (four-loop approximation) for \\ $O(n)-$symmetric $\phi^4$ model.} 

 \begin{tabular}{||c c c c c||}
 
 \hline
$n$ & $B_4$  & $(a_1,a_2,a_3,b_1)$  &   \begin{tabular}{c c}
   & $g\,_{3}F_1 \sim$ \\
     &   $(g \rightarrow \infty)$
   \end{tabular} &  \begin{tabular}{c c}
   & $\beta \sim$ \\
     &   $(g \rightarrow \infty)$ \cite{physrevdself2}
   \end{tabular}  \\ [0.5ex] 
 \hline\hline
 
     0
   & -200.926
   & (-0.580,-0.176,97.6,0.622)
   &$g^{1.580}$
   &$g^{1.789}$ \\ 
 \hline
    1
   & -271.606
   & (-0.525,-0.232,120,0.719)
   &$g^{1.525}$
   &$g^{1.770}$ \\ 
 \hline
    2
   & -350.515
   & (-0.403,-0.364,145,0.849)
   &$g^{1.403}$
   &$g^{1.772}$ \\
 \hline
   3
   & -437.646
   & (-0.426,-0.352,168,0.830)
   &$g^{1.426}$
   &$g^{1.765}$ \\
 \hline
   4
   & -532.991
   & (-0.433,-0.358,192,0.824)
   &$g^{1.433}$
   &$g^{1.759}$ \\
 \hline
\end{tabular}
\label{table:O(n)2}
\end{center}
\end{table}The extrapolated strong coupling information from hypergeometric approximants have qualitatively similar behaviour with self-similar approximant for varying $n$  \cite{physrevdself2}, and their values lie between our prediction from third-loop (Table 2) and four-loop (Table 3) approximation with deviation of $\pm 20\%$ for leading behaviour. \\ Further, we approximate the $\beta$ function in five-loop order such as  \begin{multline}
    \beta(g) \approx -g+B_1(n)g^{2}-g^2\left(\sum_{i=2}^5 B_i(n)g^{(i-1)}\right) \\ \sim -g+B_1(n)\,g^{2}-g^2\left({}_{3}F_1(a_1,a_2,a_3;b_1;\sigma g)-1\right).
\end{multline}
Here we consider $q_0=1$, $q_1=-B_2(n)$, $q_2=-B_3(n)$, $q_3=-B_4(n)$, $q_4=-B_5(n)$ and $\sigma = -0.14777 (9/n+8)$ in Eqs. (24a), (24b), (24c), (24d) to obtain parameters $\{a_1,a_2,a_3,b_1\}$. We implement these parameters in Eq. (24e) to predict an estimate for $q_5$ at six-loop order as displayed in Table \ref{table:O(n)3}. Another approach to predict this estimate is to fix the strong coupling information such as $a_1=(2-s)$ where $s=1.580$, $1.525$, $1.403$, $1.426$, $1.433$ (Table 3) for $n=0,1,2,3,4$, respectively (Since $g^2\left({}_{3}F_1(a_1,a_2,a_3;b_1;\sigma g)-1\right)\sim g^2g^{(-2+s)}$ for $g\rightarrow\infty$ in Eq.(42)). Then we solve for hypergeometric parameters $\{a_2,a_3,b_1,\sigma\}$  and implement them in Eq. (24e) to predict an estimate for $q_5$ at six-loop order as displayed in Table \ref{table:O(n)3}. The two predictions for the six-loop order in Table \ref{table:O(n)3} seem to provide an upper bound and lower bound with a deviation of $\pm4\%$ when compared with the actual value in Table \ref{table:O(n)4}. 
\begingroup
\setlength{\tabcolsep}{5pt} 
\renewcommand{\arraystretch}{1} 
\begin{table}[ht]

\begin{center}
\caption{Hypergeometric parameters of Gell-Mann-Low $\beta$ function (five-loop approximation) for $O(n)-$symmetric $\phi^4$ model. Prediction for six-loop in column three, from fixing the large-order behaviour $\sigma$ with hypergeometric parameters $(a_1,a_2,a_3,b_1)$. Prediction for six-loop in column four, from fixing the strong coupling behaviour $a_1$ with hypergeometric parameters $(a_2,a_3,b_1,\sigma)$.} 

 \begin{tabular}{||c c c c||}
 
 \hline
$n$ & $B_5$  &  \begin{tabular}{c c}
   & $(a_1,a_2,a_3,b_1)$ \\& Predicted $B_6$ \\ \end{tabular} & \begin{tabular}{c c}
   & $(a_2,a_3,b_1,\sigma)$  \\& Predicted $B_6$ \\ \end{tabular} \\ [0.5ex] 
 \hline\hline
 
     0
   & 2003.98
   & \begin{tabular}{c c}
   & (-1.5269,-0.31116,88.259,-1.4938) \\& -22375.9 \\ \end{tabular} 
   & \begin{tabular}{c c}
   & (-0.06603,3.8485,0.04037,-1.7650) \\&  -23894.6 \\ \end{tabular}  \\
 \hline
    1
   & 2848.57
   & \begin{tabular}{c c}
   & (-1.3592,-0.35034,106.52,-1.3227) \\& -33497.4 \\ \end{tabular} 
   & \begin{tabular}{c c}
   & (-0.10307,4.2429,0.06464,-1.7633) \\&  -35527.2 \\ \end{tabular}  \\
 \hline
   2
   & 3844.51
   & \begin{tabular}{c c}
   & (-1.2037,-0.38987,125.45,-1.1744) \\& -47292.7 \\ \end{tabular} 
   & \begin{tabular}{c c}
   & (-1.0962,64.554,-1.0793,-0.25230) \\&  -47559.4 \\ \end{tabular}  \\
 \hline
    3
   & 4998.62
   & \begin{tabular}{c c}
   & (-1.0500,-0.43337,144.96,-1.0403) \\& -64006.3 \\ \end{tabular} 
   & \begin{tabular}{c c}
   & (-0.18005,5.4805,0.12334,-1.6695) \\&  -67050.9 \\ \end{tabular}  \\
 \hline
    4
   & 6317.66
   & \begin{tabular}{c c}
   & (-0.88834,-0.48997,165.01,-0.91842) \\& -83884.1 \\ \end{tabular} 
   & \begin{tabular}{c c}
   & (-0.20607,6.3649,0.13674,-1.5935) \\&  -87411.1 \\ \end{tabular}  \\
 \hline
\end{tabular}
\label{table:O(n)3}
\end{center}
\end{table}
\\ Next, we approximate the $\beta$ function in six-loop order such as  \begin{multline}
    \beta(g) \approx -g+\sum_{i=1}^2 B_i(n)\,g^{(i+1)}+B_3(n)g^3\left(g+\sum_{i=4}^6 \frac{B_i(n)}{B_3(n)}\,g^{(i-2)}\right) \\ \sim -g+\sum_{i=1}^2 B_i(n)\,g^{(i+1)}+B_3(n)g^3\left({}_{3}F_1(a_1,a_2,a_3;b_1;\sigma g)-1\right).
\end{multline}
Here we consider $q_0=1$, $q_1=1$, $q_2=B_4(n)/B_3(n)$, $q_3=B_5(n)/B_3(n)$, $q_4=B_6(n)/B_3(n)$ and $\sigma = -0.14777 (9/n+8)$ in Eqs. (24a), (24b), (24c), (24d) to obtain hypergeometric parameters $\{a_1,a_2,a_3,b_1\}$. We implement these parameters in Eq. (24e) to predict an estimate for $q_5$ at seven-loop order as displayed in Table \ref{table:O(n)4}. Another approach to predict this estimate is to fix the strong coupling information such as $a_1=(3-s)$ where $s=1.580$, $1.525$, $1.403$, $1.426$, $1.433$ for $n=0,1,2,3,4$, respectively (Since $g^3\left({}_{3}F_1(a_1,a_2,a_3;b_1;\sigma g)-1\right)\sim g^3g^{(-3+s)}$ for $g\rightarrow\infty$ in Eq.(43)). Then we solve for hypergeometric parameters $\{a_2,a_3,b_1,\sigma\}$  and implement them in Eq. (24e) to predict an estimate for $q_5$ at seven-loop order as displayed in Table \ref{table:O(n)4}. Similar to previous order, the two predictions for the seven-loop order in Table \ref{table:O(n)4} seem to provide an upper bound and lower bound with a deviation of $\pm3\%$ when compared with the actual value in Table \ref{table:O(n)5}.
\begingroup
\setlength{\tabcolsep}{5pt} 
\renewcommand{\arraystretch}{1} 
\begin{table}[ht]

\begin{center}
\caption{Hypergeometric parameters of Gell-Mann-Low $\beta$ function (six-loop approximation) for $O(n)-$symmetric $\phi^4$ model. Prediction for seven-loop in column three, from fixing the large-order behaviour $\sigma$ with hypergeometric parameters $(a_1,a_2,a_3,b_1)$. Prediction for seven-loop in column four, from fixing the strong coupling behaviour $a_1$ with hypergeometric parameters $(a_2,a_3,b_1,\sigma)$.} 

 \begin{tabular}{||c c c c||}
 
 \hline
$n$ & $B_6$  &  \begin{tabular}{c c}
   & $(a_1,a_2,a_3,b_1)$ \\& Predicted $B_7$ \\ \end{tabular} & \begin{tabular}{c c}
   & $(a_2,a_3,b_1,\sigma)$  \\& Predicted $B_7$ \\ \end{tabular} \\ [0.5ex] 
 \hline\hline
 
     0
   & -23314.7
   & \begin{tabular}{c c}
   & (-1.2749,-0.063459,93.376,-1.2559) \\& 295093 \\ \end{tabular} 
   & \begin{tabular}{c c}
   & (-0.02875,4.8271,0.29382,-1.4912) \\&  307677 \\ \end{tabular}  \\
 \hline
    1
   & -34776.1
   & \begin{tabular}{c c}
   & (-1.2665,-0.060257,110.62,-1.2475) \\& 460935 \\ \end{tabular} 
   & \begin{tabular}{c c}
   & (-0.02680,4.7188,0.29201,-1.5654) \\&  480844 \\ \end{tabular}  \\
 \hline
   2
   & -48999.1
   & \begin{tabular}{c c}
   & (-1.2647,-0.057513,128.73,-1.2452) \\& 677213 \\ \end{tabular} 
   & \begin{tabular}{c c}
   & (-0.02527,4.4386,0.29588,-1.6518) \\&  707399 \\ \end{tabular}  \\
 \hline
    3
   & -66242.7
   & \begin{tabular}{c c}
   & (-1.2565,-0.055183,147.55,-1.2369) \\& 950862 \\ \end{tabular} 
   & \begin{tabular}{c c}
   & (-0.02309,4.4367,0.27658,-1.7153) \\&  993764 \\ \end{tabular}  \\
 \hline
    4
   & -86768.4
   & \begin{tabular}{c c}
   & (-1.2475,-0.053155,167.09,-1.2279) \\& 1.2896$\times10^6$ \\ \end{tabular} 
   & \begin{tabular}{c c}
   & (-0.02129,4.4112,0.26159,-1.7772) \\&  1.3483$\times10^6$ \\ \end{tabular}  \\
 \hline
\end{tabular}
\label{table:O(n)4}
\end{center}
\end{table} 
\\ Using a similar approach, we approximate the $\beta$ function at seven-loop order such as  \begin{multline}
    \beta(g) \approx -g+\sum_{i=1}^2 B_i(n)\,g^{(i+1)}+g^4\left(B_3(n)+\sum_{i=4}^7 B_i(n)\,g^{(i-3)}\right) \\ \sim -g+\sum_{i=1}^2 B_i(n)\,g^{(i+1)}+B_3(n)g^4\left({}_{3}F_1(a_1,a_2,a_3;b_1;\sigma g)\right).
\end{multline}
Here we consider $q_0=B_3(n)$, $q_1=B_4(n)$, $q_2=B_5(n)$, $q_3=B_6(n)$, $q_4=B_7(n)$ and $\sigma = -0.14777 (9/n+8)$ in Eqs. (24a), (24b), (24c), (24d) to obtain parameters $\{a_1,a_2,a_3,b_1\}$. We implement these parameters in Eq. (24e) to predict an estimate for $q_5$ at eight-loop order as displayed in Table \ref{table:O(n)5}. Another approach to predict this estimate is to fix the strong coupling information such as $a_1=(4-s)$ where $s=1.580$, $1.525$, $1.403$, $1.426$, $1.433$ for $n=0,1,2,3,4$, respectively (Since $g^4\,{}_{3}F_1(a_1,a_2,a_3;b_1;\sigma g)\sim g^4g^{(-4+s)}$ for $g\rightarrow\infty$ in Eq.(44)). Then we solve for hypergeometric parameters $\{a_2,a_3,b_1,\sigma\}$  and implement them in Eq. (24e) to predict an estimate for $q_5$ at eight-loop order as displayed in Table \ref{table:O(n)5}. Very similar to predictions from previous two orders, the two predictions for the eight-loop order seem to vary between an upper bound and lower bound.
\begingroup
\setlength{\tabcolsep}{5pt} 
\renewcommand{\arraystretch}{1} 
\begin{table}[ht]

\begin{center}
\caption{Hypergeometric parameters of Gell-Mann-Low $\beta$ function (seven-loop approximation) for $O(n)-$symmetric $\phi^4$ model. Prediction for eight-loop in column three, from fixing the large-order behaviour $\sigma$ with hypergeometric parameters $(a_1,a_2,a_3,b_1)$. Prediction for eight-loop in column four, from fixing the strong coupling behaviour $a_1$ with hypergeometric parameters $(a_2,a_3,b_1,\sigma)$.} 

 \begin{tabular}{||c c c c||}
 
 \hline
$n$ & $B_7$  &  \begin{tabular}{c c}
   & $(a_1,a_2,a_3,b_1)$ \\& Predicted $B_8$ \\ \end{tabular} & \begin{tabular}{c c}
   & $(a_2,a_3,b_1,\sigma)$  \\& Predicted $B_8$ \\ \end{tabular} \\ [0.5ex] 
 \hline\hline
 
     0
   & 303869
   & \begin{tabular}{c c}
   & (0.94424,2.8511,93.376,9.3317) \\& -4.3273$\times10^6$ \\ \end{tabular} 
   & \begin{tabular}{c c}
   & (0.97191,42.187,6.1953,-0.49280) \\& -4.3308$\times10^6$ \\ \end{tabular}  \\
 \hline
    1
   & 474651
   & \begin{tabular}{c c}
   & (0.96533,2.7105,188.32,8.7264) \\& -7.0637$\times10^6$ \\ \end{tabular} 
   & \begin{tabular}{c c}
   & (0.98211,67.174,6.9427,-0.35480) \\& -7.0671$\times10^6$ \\ \end{tabular}  \\
 \hline
   2
   & 696998
   & \begin{tabular}{c c}
   & (0.99283,2.5509,211.99,8.1383) \\& -1.0791$\times10^7$ \\ \end{tabular} 
   & \begin{tabular}{c c}
   & (0.98922,354.59,8.5325,-0.08219) \\& -1.0789$\times10^7$ \\ \end{tabular}  \\
 \hline
    3
   & 978330
   & \begin{tabular}{c c}
   & (1.0233,2.4012,236.37,7.6458) \\& -1.5701$\times10^7$ \\ \end{tabular} 
   & \begin{tabular}{c c}
   & (1.0073,-354.6,9.1459,0.09136) \\& -1.5694$\times10^7$ \\ \end{tabular}  \\
 \hline
    4
   & 1.326$\times10^6$
   & \begin{tabular}{c c}
   & (1.0635,2.2185,258.95,7.0712) \\& -2.1981$\times10^7$ \\ \end{tabular} 
   & \begin{tabular}{c c}
   & (1.0242,-98.474,10.198,0.37719) \\& -2.1961$\times10^7$ \\ \end{tabular}  \\
 \hline
\end{tabular}
\label{table:O(n)5}
\end{center}
\end{table}
\section{Critical exponents $\nu$ and $\omega$ from $O(n)-$symmetric $\phi^4$ models}
\subsection{Eight-loop prediction for critical exponents $\nu$ and $\omega$}
In $\phi^4$ models, fluctuations become important at the point of transition which are governed by correlation functions. Further at the critical temperature ($T = T_c$), fluctuations on one part of the physical system influences on other parts of the system based on a characteristic length scale called as the correlation length, $\xi$. This is controlled in the vicinity of critical point, $T_c$ as \begin{equation}
\xi(T) \sim |T-T_c|^{-\nu}(1+\hbox{const}.|T-T_c|^{\omega \nu}+\cdots). \end{equation} In this expression for $\xi$, $\nu$ is the leading exponent and $\omega$ is the subleading exponent. Similar to previous section we approximate these exponent with hypergeometric approximants for extrapolation of eight-loop predictions using exact seven-loop information \cite{seven-O(n),seven-O(n)-2}. We consider $\nu$ at seven-loop order based on recent minimal subtraction renormalization scheme such as \cite{shalaby2020critical}
\begin{subequations}
    \begin{align}
        1/\nu \approx 2-0.25000\epsilon-0.08594\epsilon^2+0.11443\epsilon^3-0.28751\epsilon^4+0.95613\epsilon^5-3.8558\epsilon^6+17.784\epsilon^7, \\ 1/\nu \approx 2-0.33333\epsilon-0.11728\epsilon^2 + 0.12453\epsilon^3- 0.30685\epsilon^4 + 0.95124\epsilon^5- 3.5726\epsilon^6 + 15.287\epsilon^7,\\
        1/\nu \approx 2 - 0.40000\epsilon -0.14000\epsilon^2 + 0.12244\epsilon^3 -0.30473\epsilon^4
+ 0.87924\epsilon^5 - 3.1030\epsilon^6 + 12.419\epsilon^7, \\ 1/\nu \approx 2 - 0.45455\epsilon - 0.15590\epsilon^2 + 0.11507\epsilon^3 - 0.29360\epsilon^4
+ 0.78994\epsilon^5 - 2.6392\epsilon^6 + 9.9452\epsilon^7, \\ 1/\nu \approx 2 - 0.50000\epsilon - 0.16667\epsilon^2 + 0.10586\epsilon^3 -0.27866\epsilon^4 + 0.70217\epsilon^5 - 2.2337\epsilon^6 + 7.9701\epsilon^7,
    \end{align}
\end{subequations} for $n=0,1,2,3,4$, respectively in $4-\epsilon$ dimensions ($\epsilon \rightarrow 0$). Initially, this leading exponent is approximated at six-loop order with large-order behaviour $\sigma=-3/(n+8)$ to obtain hypergeometric approximants \begin{subequations}
    \begin{align}
        1/\nu &\sim 2+\cdots-0.08594\epsilon^2+\epsilon^2({}_{3}F_1(-0.026719,-0.97399,11.449;-0.97649;-3/8 \epsilon)-1) , \\ 1/\nu &\sim 2+\cdots-0.11728\epsilon^2+\epsilon^2({}_{3}F_1(-0.030597,-1.01204,12.190;-1.01041;-3/9 \epsilon)-1),\\
        1/\nu &\sim 2+\cdots -0.14000\epsilon^2+\epsilon^2({}_{3}F_1(-0.030638,-1.13045,13.045;-1.1070;-3/10 \epsilon)-1), \\ 1/\nu &\sim 2+\cdots- 0.15590\epsilon^2+\epsilon^2({}_{3}F_1(-0.029064,-1.21489,13.939;-1.1666;-3/11\epsilon)-1), \\ 1/\nu &\sim 2+\cdots- 0.16667\epsilon^2+\epsilon^2({}_{3}F_1(-0.026931,-1.26785,14.831;-1.1959;-3/12 \epsilon)-1),
    \end{align}
\end{subequations} for $n=0,1,2,3,4$, respectively. Implementing these hypergeometric parameters $\{a_1,a_2,a_3,b_1\}$ and $\sigma$ in Eq. (24e), estimate for seven-loop order can be predicted as $17.766\epsilon^7$, $15.298\epsilon^7$, $12.495\epsilon^7$, $10.080\epsilon^7$, $8.141\epsilon^7$ for $n=0,1,2,3,4$, respectively. These estimates seem to deviate $-0.1\%$, $-0.1\%$, $0.6\%$, $1\%$, $2\%$, respectively from their actual values in Eqs. (46). The exponent $\nu$ is now approximated at seven-loop order with large-order behaviour $\sigma=-3/(n+8)$ to obtain hypergeometric approximants \begin{subequations}
    \begin{align}
        1/\nu &\sim 2+\cdots+0.11443\epsilon^3+\epsilon^3({}_{3}F_1(-0.10504,0.15710,11.783;-0.25363;-3/8 \epsilon)-1) , \\ 1/\nu &\sim 2+\cdots+0.12453\epsilon^3+\epsilon^3({}_{3}F_1(-0.082758,0.19289,12.357;-0.21427;-3/9 \epsilon)-1),\\
        1/\nu &\sim 2+\cdots+0.12244\epsilon^3+\epsilon^3({}_{3}F_1(-0.0023981,0.39308,12.678;-0.011765;-3/10 \epsilon)-1), \\ 1/\nu &\sim 2+\cdots+0.11507\epsilon^3+\epsilon^3({}_{3}F_1(0.038248,1.0876,12.5003;0.483007;-3/11\epsilon)-1), \\ 1/\nu &\sim 2+\cdots+0.10586\epsilon^3+\epsilon^3({}_{3}F_1(0.0517109,3.4722,10.6125;1.7095;-3/12 \epsilon)-1),
    \end{align}
\end{subequations} for $n=0,1,2,3,4$, respectively. Similarly, we implement these parameters $\{a_1,a_2,a_3,b_1\}$ and $\sigma$ in Eq. (24e), to obtain an estimate for eight-loop order such as $-90.985\epsilon^8$, $-72.323\epsilon^8$, $-54.724\epsilon^8$, $-41.020\epsilon^8$, $-30.878\epsilon^8$ for $n=0,1,2,3,4$, respectively. 
\\Then we handle the subleading exponent $\omega$ at seven-loop order based on recent minimal subtraction renormalization scheme such as \cite{shalaby2020critical} \begin{subequations}
\begin{align}
    \omega &\approx \epsilon-0.65625\epsilon^2+1.8236\epsilon^3-6.2854\epsilon^4+26.873\epsilon^5-130.01\epsilon^6+692.10\epsilon^7, \\
    \omega &\approx \epsilon-0.62963\epsilon^2+1.6182\epsilon^3-5.2351\epsilon^4+20.750\epsilon^5-93.111\epsilon^6+458.74\epsilon^7, \\
    \omega &\approx \epsilon-0.60000\epsilon^2+1.4372\epsilon^3-4.4203\epsilon^4+16.374\epsilon^5-68.777\epsilon^6+316.48\epsilon^7, \\
    \omega &\approx \epsilon-0.57025\epsilon^2+1.2829\epsilon^3-3.7811\epsilon^4+13.182\epsilon^5-52.204\epsilon^6+226.02\epsilon^7, \\
    \omega &\approx \epsilon - 0.54167\epsilon^2 + 1.1526\epsilon^3 - 3.2719\epsilon^4 + 10.802\epsilon^5 - 40.567\epsilon^6 + 166.26\epsilon^7,
    \end{align}
\end{subequations} for $n=0,1,2,3,4$, respectively ($\epsilon \rightarrow 0$). Initially, this subleading exponent is approximated at six-loop order with large-order behaviour $\sigma=-3/(n+8)$ to obtain hypergeometric approximants
\begin{subequations}
\begin{align}
    \omega &\sim \epsilon-0.65625\epsilon^2-\epsilon^2({}_{3}F_1(-1.7444,0.40044,12.156;-1.7461;-3/8 \epsilon)-1), \\
    \omega &\sim \epsilon-0.62963\epsilon^2-\epsilon^2({}_{3}F_1(0.34825,1.3526,12.603;1.2229;-3/9 \epsilon)-1), \\
    \omega &\sim \epsilon-0.60000\epsilon^2-\epsilon^2({}_{3}F_1(-0.78826,0.35419,13.645;-0.79520;-3/10 \epsilon)-1), \\
    \omega &\sim \epsilon-0.57025\epsilon^2-\epsilon^2({}_{3}F_1(-0.94139,0.32700,14.432;-0.94447;-3/11 \epsilon)-1), \\
    \omega &\sim \epsilon - 0.54167\epsilon^2-\epsilon^2({}_{3}F_1(-0.98713,0.30402,15.179;-0.98804;-3/12 \epsilon)-1),
    \end{align}
\end{subequations} for $n=0,1,2,3,4$, respectively. Implementing these hypergeometric parameters $\{a_1,a_2,a_3,b_1\}$ and $\sigma$ in Eq. (24e), estimate for seven-loop order can be predicted as $693.72\epsilon^7$, $459.27\epsilon^7$, $317.73\epsilon^7$, $227.33\epsilon^7$, $167.48\epsilon^7$ for $n=0,1,2,3,4$, respectively. These estimates seem to deviate $0.2\%$, $0.1\%$, $0.4\%$, $0.6\%$, $1\%$, respectively from their actual values in Eqs. (49). The exponent $\omega$ is now approximated at seven-loop order with large-order behaviour $\sigma=-3/(n+8)$ to obtain hypergeometric approximants \begin{subequations}
\begin{align}
    \omega &\sim \epsilon+\cdots+1.8236\epsilon^3+\epsilon^3({}_{3}F_1(0.17478,0.53401,12.536;0.069810;-3/8 \epsilon)-1), \\
    \omega &\sim \epsilon+\cdots+1.6182\epsilon^3+\epsilon^3({}_{3}F_1(0.12565,0.58406,13.157;0.061478;-3/9 \epsilon)-1), \\
    \omega &\sim \epsilon+\cdots+1.4372\epsilon^3+\epsilon^3({}_{3}F_1(0.03842,0.65560,13.701;0.023421;-3/10 \epsilon)-1), \\
    \omega &\sim \epsilon+\cdots+1.2829\epsilon^3+\epsilon^3({}_{3}F_1(-0.056268,0.70761,14.217;-0.040829;-3/11 \epsilon)-1), \\
    \omega &\sim \epsilon+\cdots+1.1526\epsilon^3+\epsilon^3({}_{3}F_1(-0.15687,0.73296,14.736;-0.12946;-3/12\epsilon)-1),
    \end{align}
\end{subequations} for $n=0,1,2,3,4$, respectively. Implementing these parameters $\{a_1,a_2,a_3,b_1\}$ and $\sigma$ in Eq. (24e), estimate for eight-loop order can be predicted as $-3992.2\epsilon^8$, $-2443.3\epsilon^8$, $-1570.7\epsilon^8$, $-1053.1\epsilon^8$, $-731.91\epsilon^8$ for $n=0,1,2,3,4$, respectively.
\subsection{Resummation of critical exponents $\nu$ and $\omega$ with eight-loop estimates in three dimensions}
Similar to Sec. 2.1 and previous works \cite{abhignan2020continued,abhignan2021}, we perform resummation of these critical exponents with the estimated eight-loop predictions implementing continued functions for three-dimensional systems $(\epsilon=1)$. Numerical estimate for $1/\nu$ at successive orders are obtained from sequences of CEF ($\{C_i\}$), CE ($\{D_i\}$), CEBL ($\{E_i\}$) from Eqs. (11), (12), (13), respectively, inverted to obtain $\nu$ and their final estimate is interpolated from iterated Shanks $S^2$ defined below Eq. (17). The oscillating convergence nature of these estimates are illustrated in Figs. (8), (9), (10) at four-loop ($A_4$), five-loop ($A_5$), six-loop ($A_6$), seven-loop ($A_7$), eight-loop ($A_8$) and compared with the most reliable results from other field-theoretic approaches such as Monte Carlo simulations (MC) and Conformal Bootstrap calculations (CB). Similarly, estimates for $\omega/\epsilon$ at successive orders are obtained from sequences of CE ($\{D_i\}$), CEBL ($\{E_i\}$) from Eqs. (12), (13), respectively and their final estimate is interpolated from Shanks in Eq. (17). In this case, estimates are illustrated in Figs. (10), (11), (12) at five-loop ($A_4$, since we consider $Q/\epsilon$ in Eq. (1)), six-loop ($A_5$), seven-loop ($A_6$) and eight-loop ($A_7$) to be compared with other results. We observe empirically that reliable estimates in this case of $\epsilon=1$ are obtained for $\nu$ from CE, CEF and CEBL, while for $\omega$ only from CEBL procedure based on their numerical structure \cite{abhignan2020continued,abhignan2021}. CEF estimates seem to be most compatible in case of $\nu$, while CEBL estimates seem to be most compatible in case of $\omega$.
\begin{figure}[ht]
\centering
\begin{subfigure}{0.495\textwidth}
\includegraphics[width=1\linewidth, height=6cm]{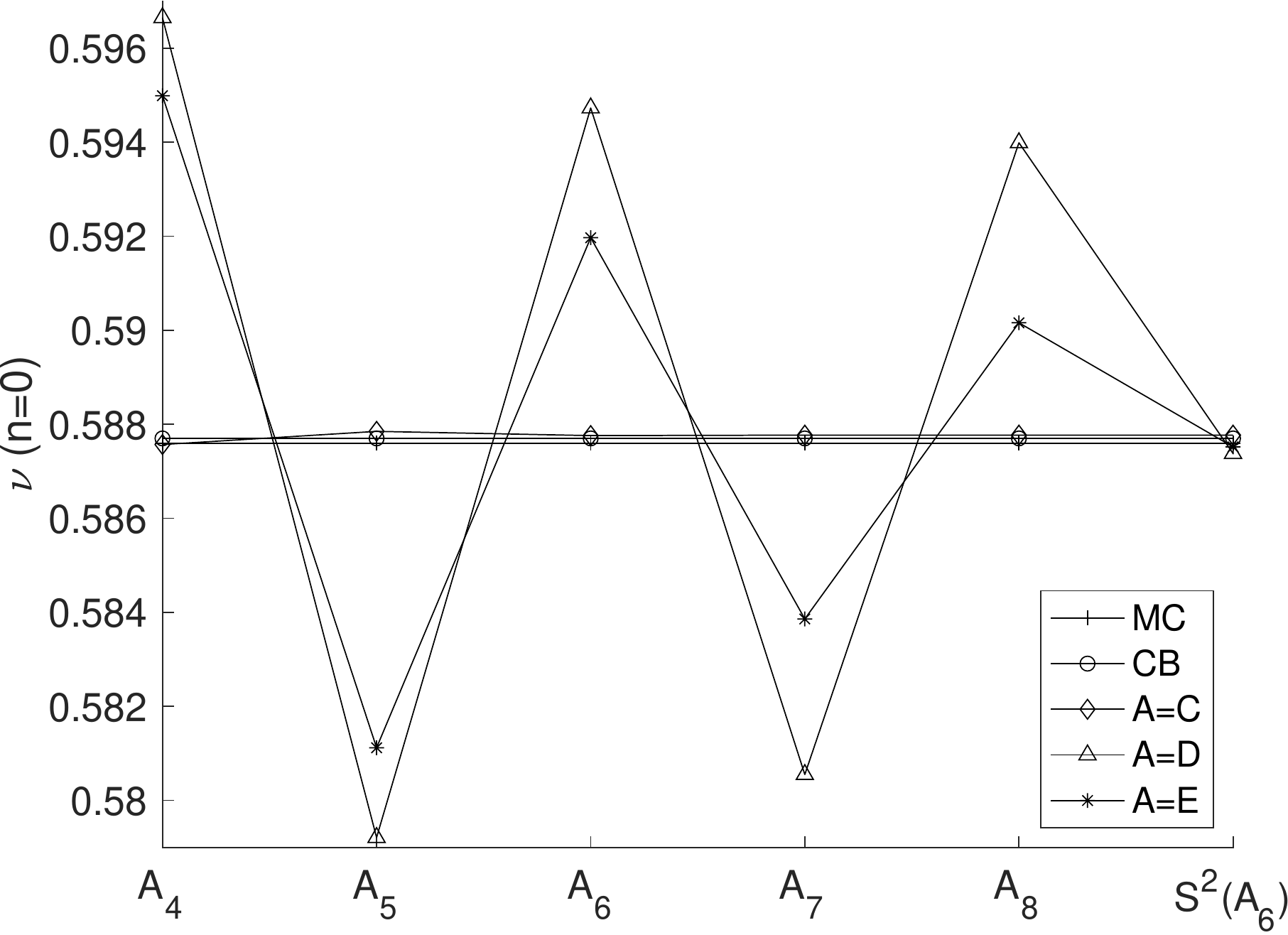} 
\caption{Estimates of self-avoiding walks model $\nu$ $(n=0)$ compared with MC result $\nu=0.5875970$ \cite{Clisby2016} and CB result $\nu=0.5877$ \cite{Shimada2016}.}

\end{subfigure}
\begin{subfigure}{0.495\textwidth}
\includegraphics[width=1\linewidth, height=6cm]{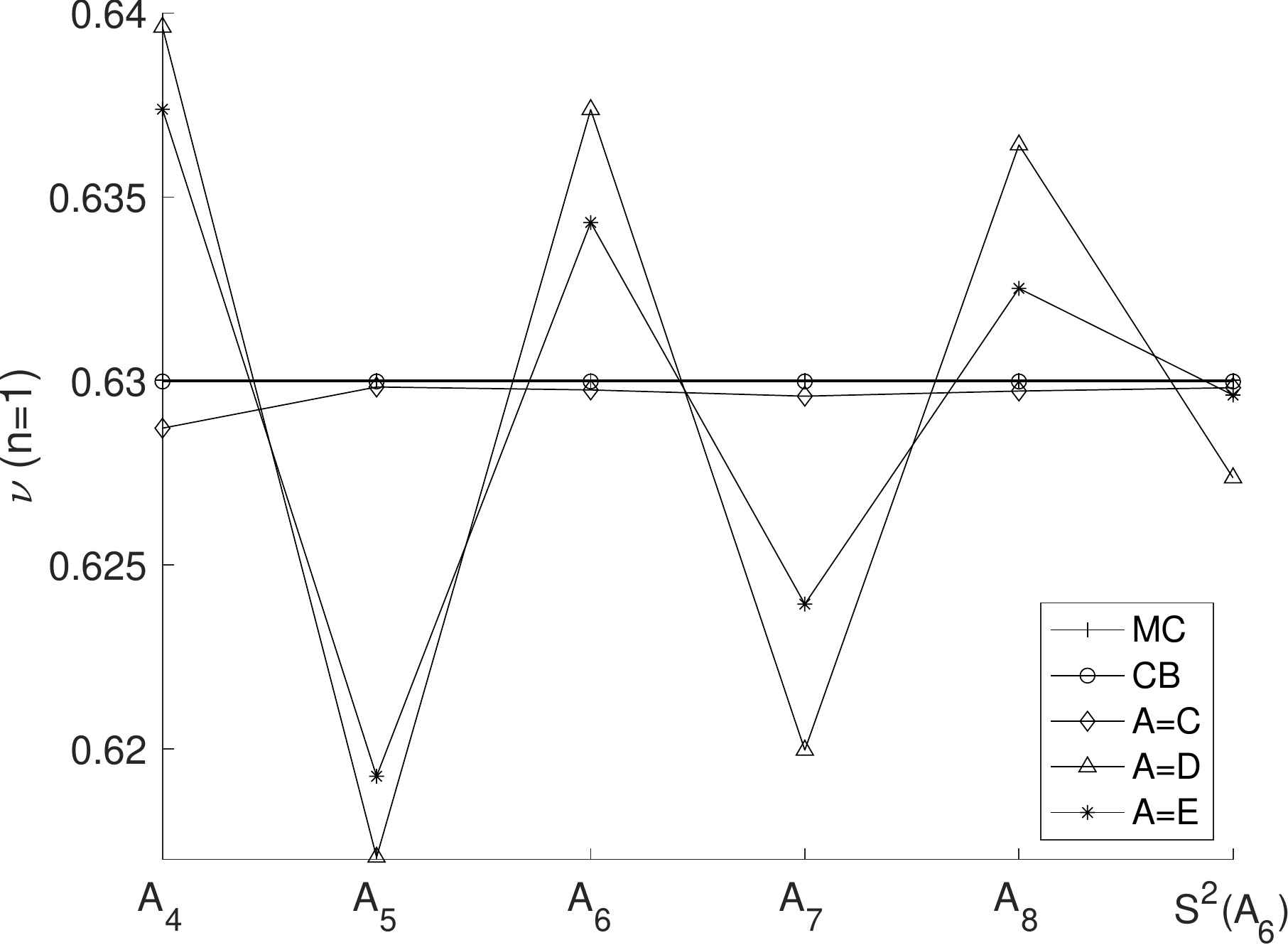}
\caption{Estimates of Ising model $\nu$ $(n=1)$ compared with MC result $\nu=0.63002$ \cite{Hasenbusch2010} and CB result $\nu=0.62999$ \cite{Showk2014}.}

\end{subfigure}

\caption{Estimates of $\nu$ at successive orders compared with MC results and CB results.}
\end{figure} 

\begin{figure}[ht]
\centering
\begin{subfigure}{0.495\textwidth}
\includegraphics[width=1\linewidth, height=6cm]{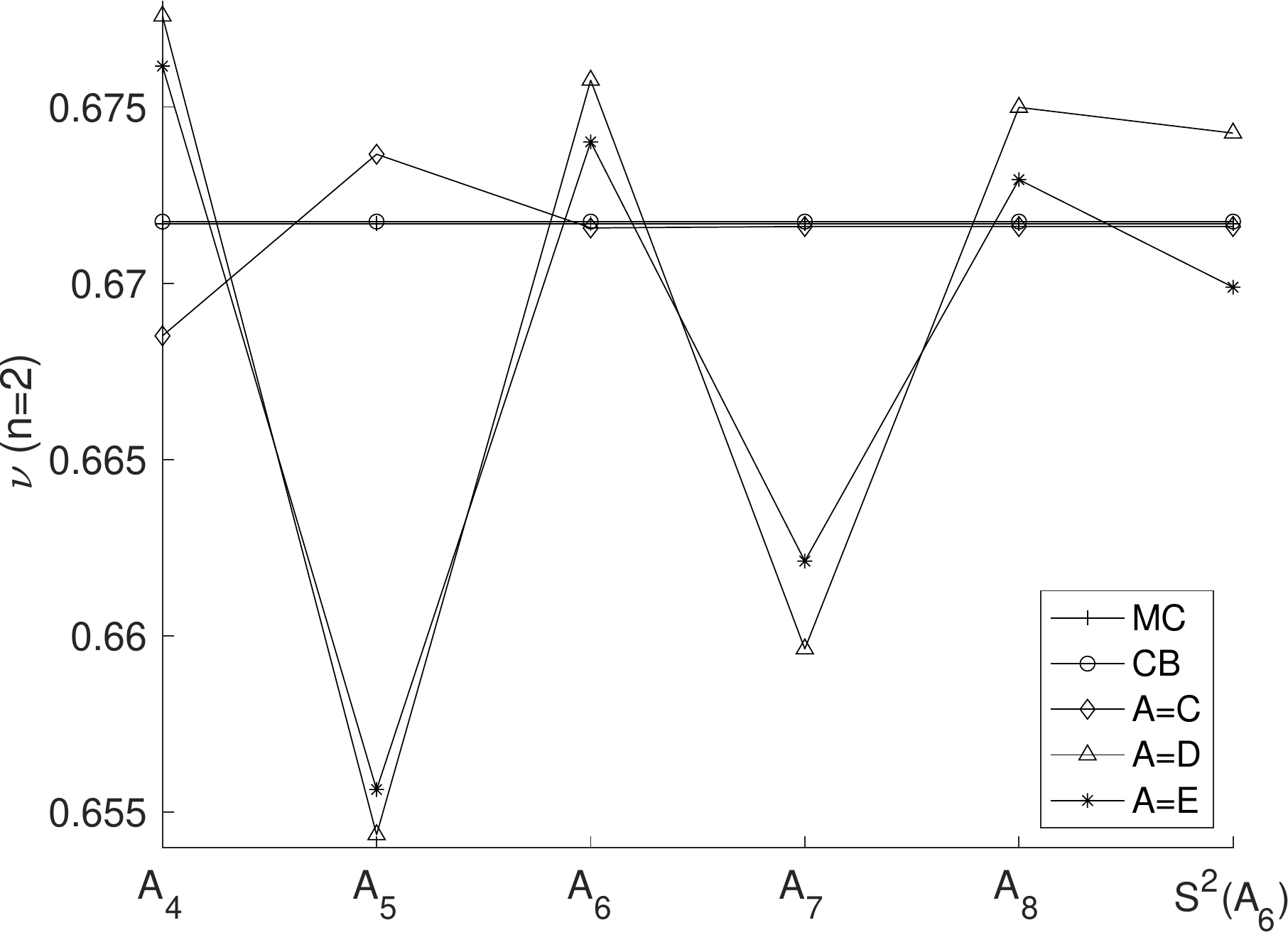} 
\caption{Estimates of $XY$ universality class $\nu$ $(n=2)$ compared with MC result $\nu=0.67169$ \cite{mcn=2} and CB result $\nu=0.67175$ \cite{Chester2020}.}

\end{subfigure}
\begin{subfigure}{0.495\textwidth}
\includegraphics[width=1\linewidth, height=6cm]{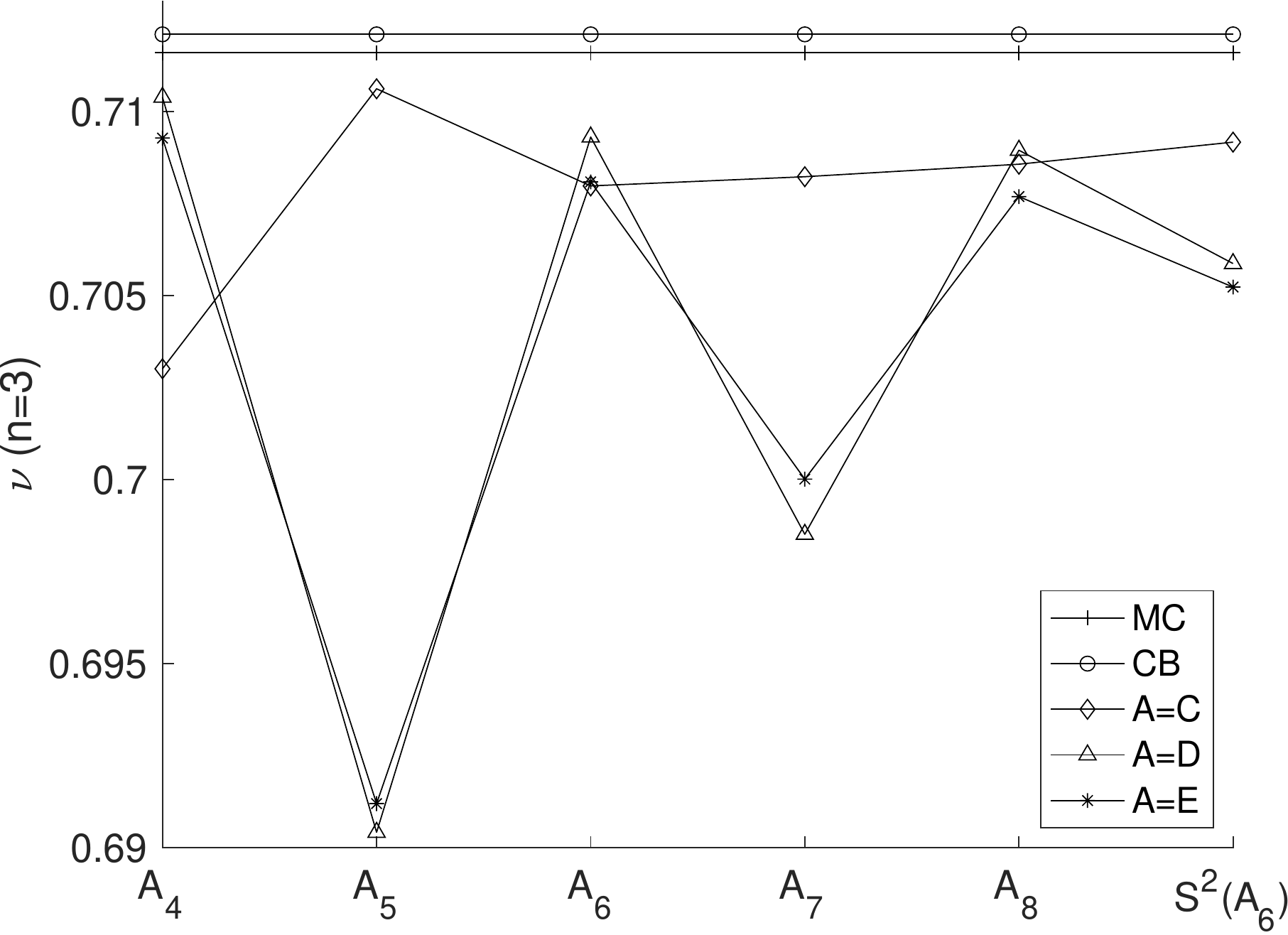}
\caption{Estimates of Heisenberg model $\nu$ $(n=3)$ compared with MC result $\nu=0.7116$ \cite{Hasenbusch2011} and CB result $\nu=0.7121$ \cite{Kos2016}.}

\end{subfigure}

\caption{Estimates of $\nu$ at successive orders compared with MC results and CB results.}
\end{figure} 

\begin{figure}[ht]
\centering
\begin{subfigure}{0.495\textwidth}
\includegraphics[width=1\linewidth, height=6cm]{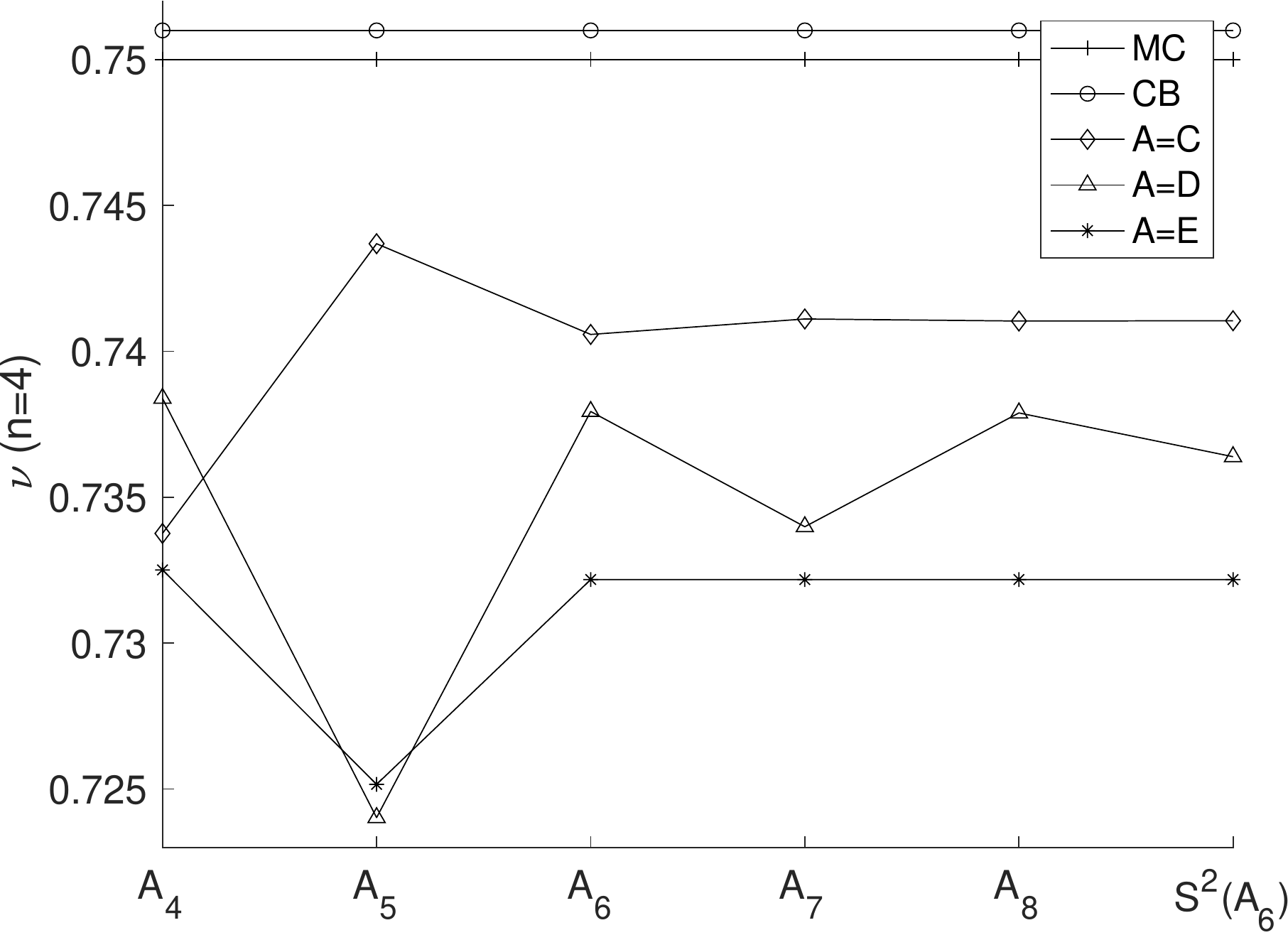} 
\caption{Estimates of $\nu$ $(n=4)$ compared with MC result $\nu=0.750$ \cite{Hasenbusch2011} and CB result $\nu=0.751$ \cite{Echeverri2016}.}

\end{subfigure}
\begin{subfigure}{0.495\textwidth}
\includegraphics[width=1\linewidth, height=6cm]{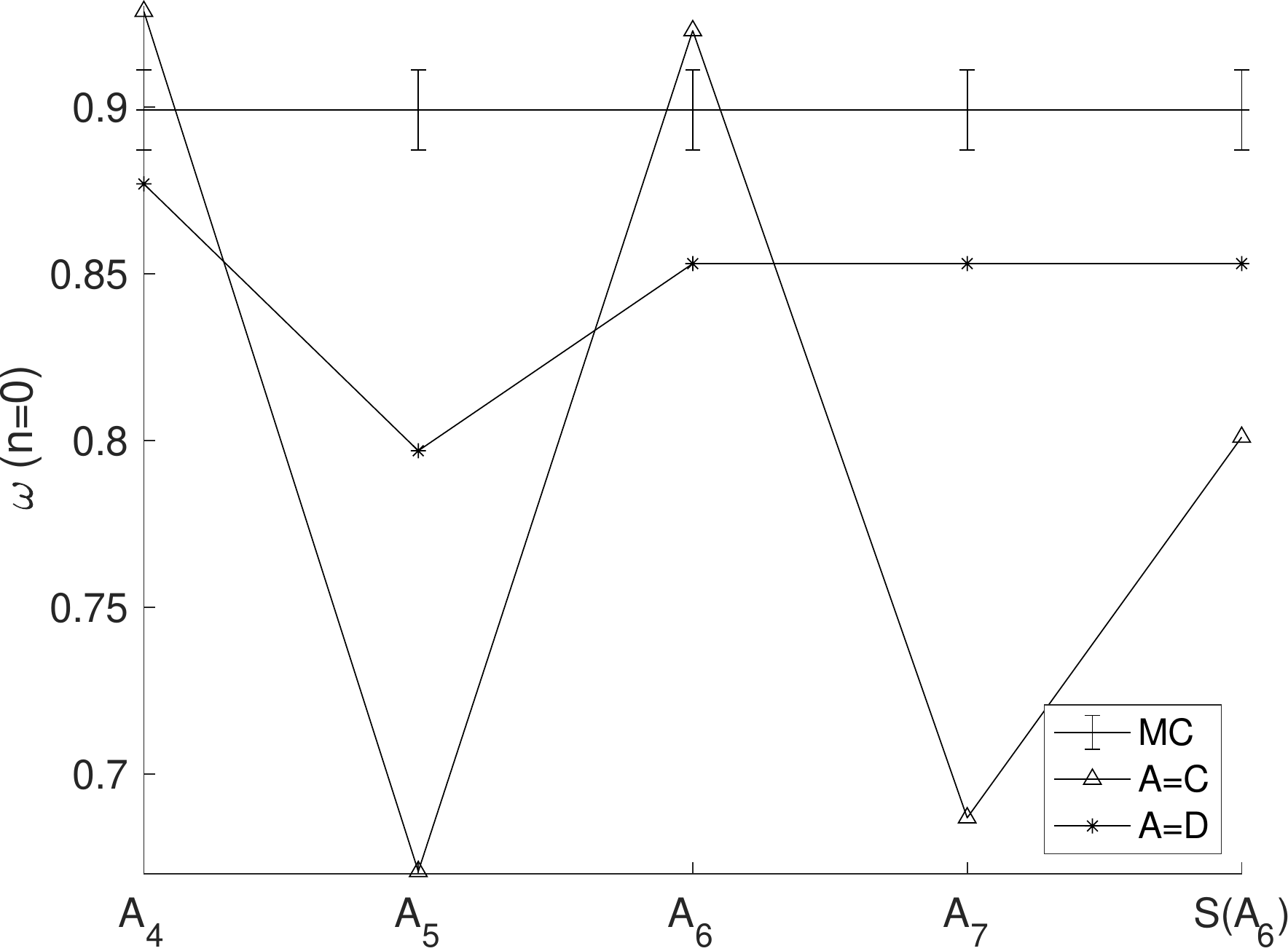}
\caption{Estimates of self-avoiding walks model $\omega$ $(n=0)$ compared with MC result $\omega= 0.899(12)$ \cite{Clisby2016}.}

\end{subfigure}

\caption{Estimates of $\nu$ and $\omega$ at successive orders compared with MC results and CB results.}
\end{figure} 

\begin{figure}[ht]
\centering
\begin{subfigure}{0.495\textwidth}
\includegraphics[width=1\linewidth, height=6cm]{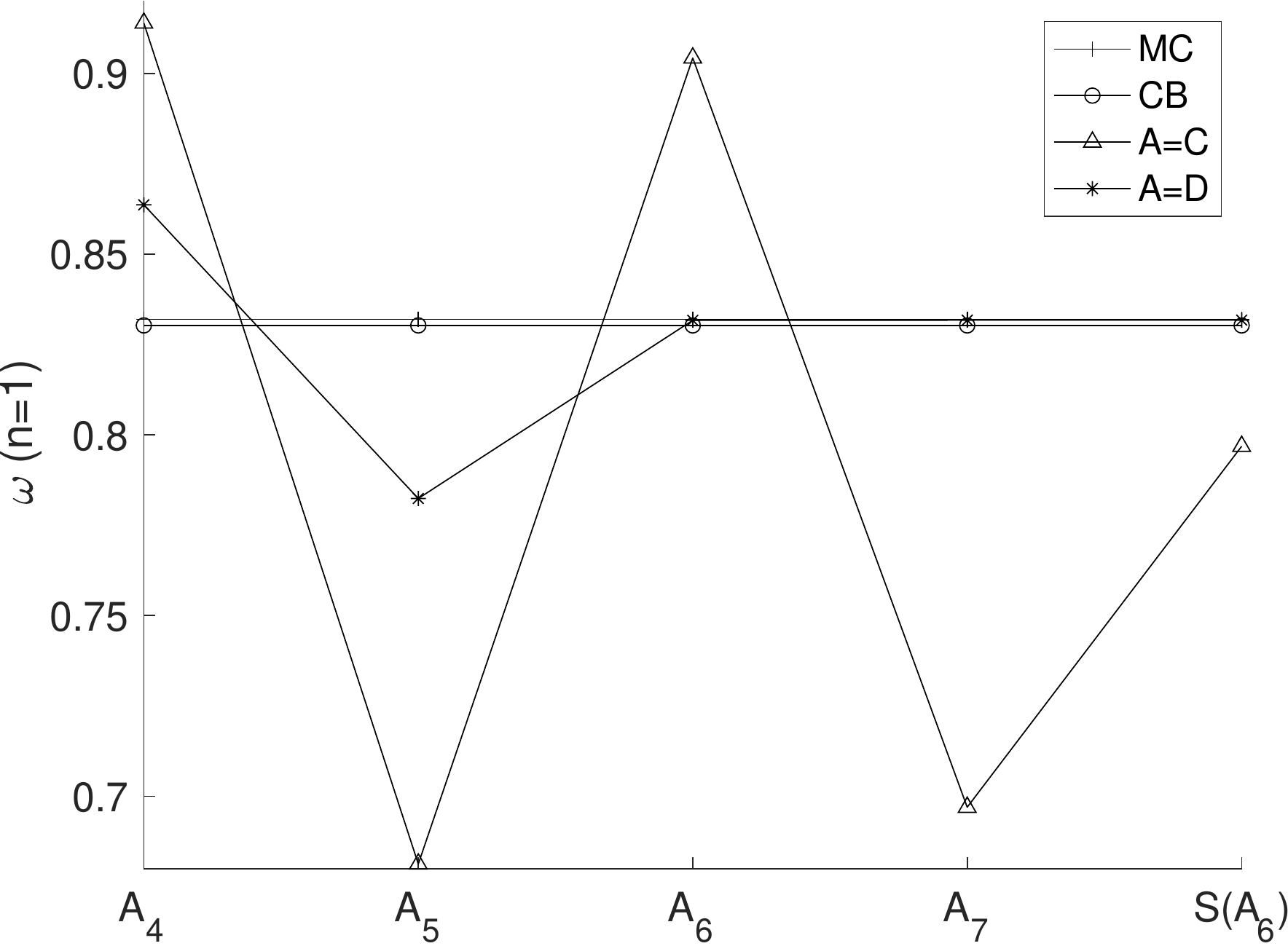} 
\caption{Estimates of Ising model $\omega$ $(n=1)$ compared with MC result $\omega=0.832$ \cite{Hasenbusch2010} and CB result $\omega=0.8303$ \cite{Showk2014}.}

\end{subfigure}
\begin{subfigure}{0.495\textwidth}
\includegraphics[width=1\linewidth, height=6cm]{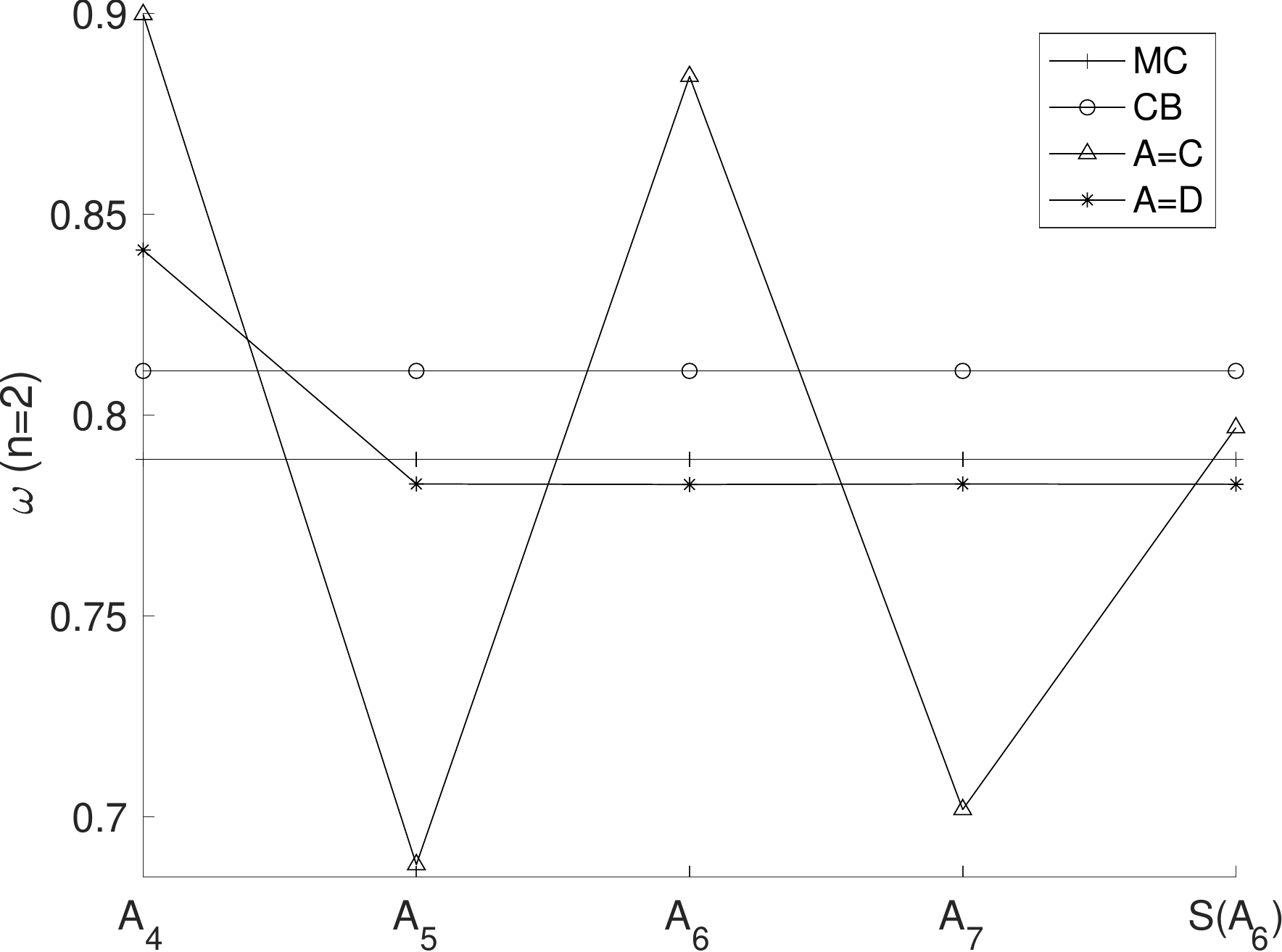}
\caption{Estimates of $XY$ universality class $\omega$ $(n=2)$ compared with MC result $\omega= 0.789$ \cite{mcn=2} and CB result $\omega=0.811$ \cite{Echeverri2016}.}

\end{subfigure}

\caption{Estimates of $\omega$ at successive orders compared with MC results and CB results.}
\end{figure} 

\begin{figure}[ht]
\centering
\begin{subfigure}{0.495\textwidth}
\includegraphics[width=1\linewidth, height=6cm]{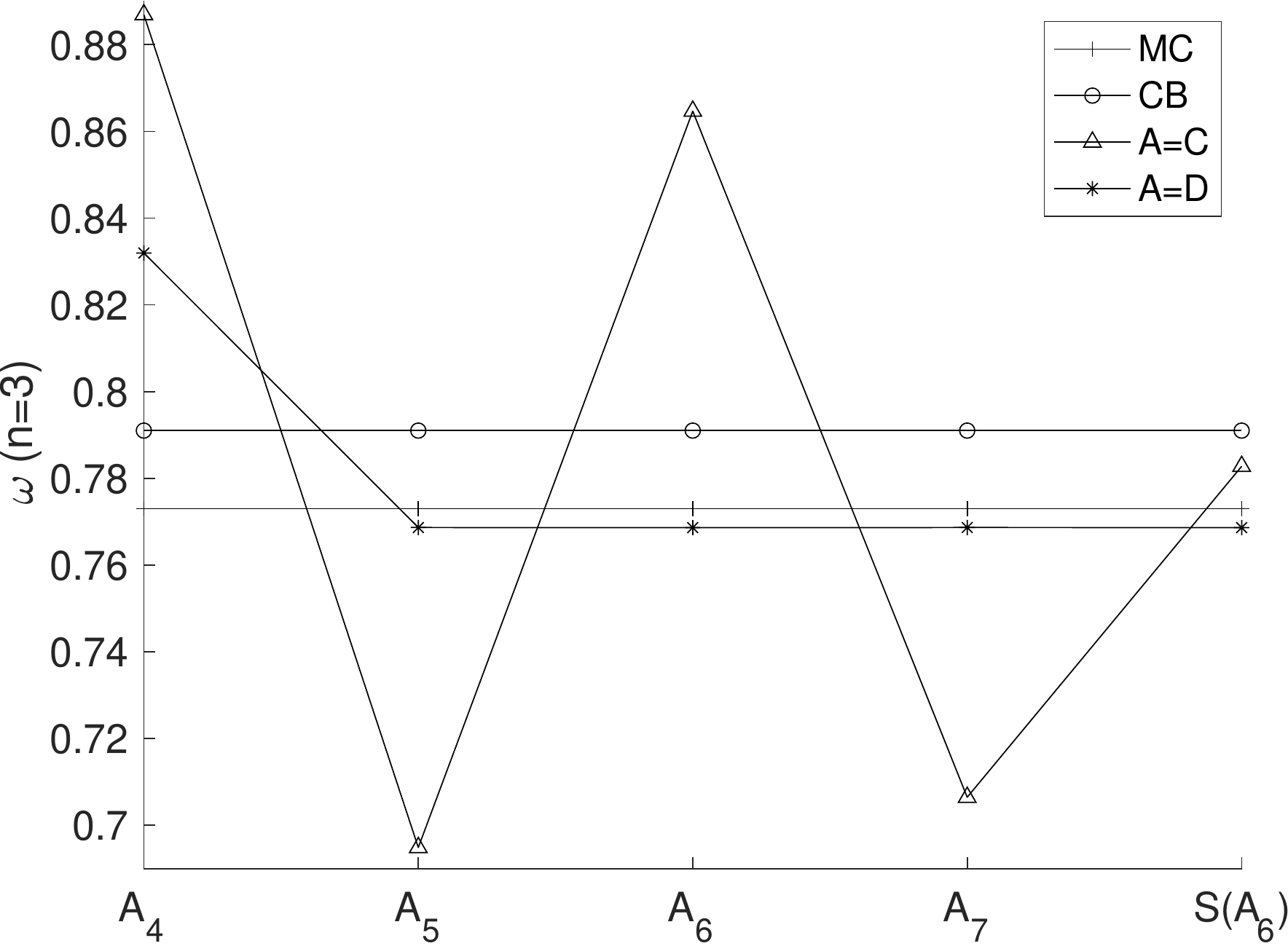} 
\caption{Estimates of Heisenberg model $\omega$ $(n=3)$ compared with MC result $\omega= 0.773$ \cite{Hasenbusch_2001} and CB result $\omega=0.791$ \cite{Echeverri2016}.}

\end{subfigure}
\begin{subfigure}{0.495\textwidth}
\includegraphics[width=1\linewidth, height=6cm]{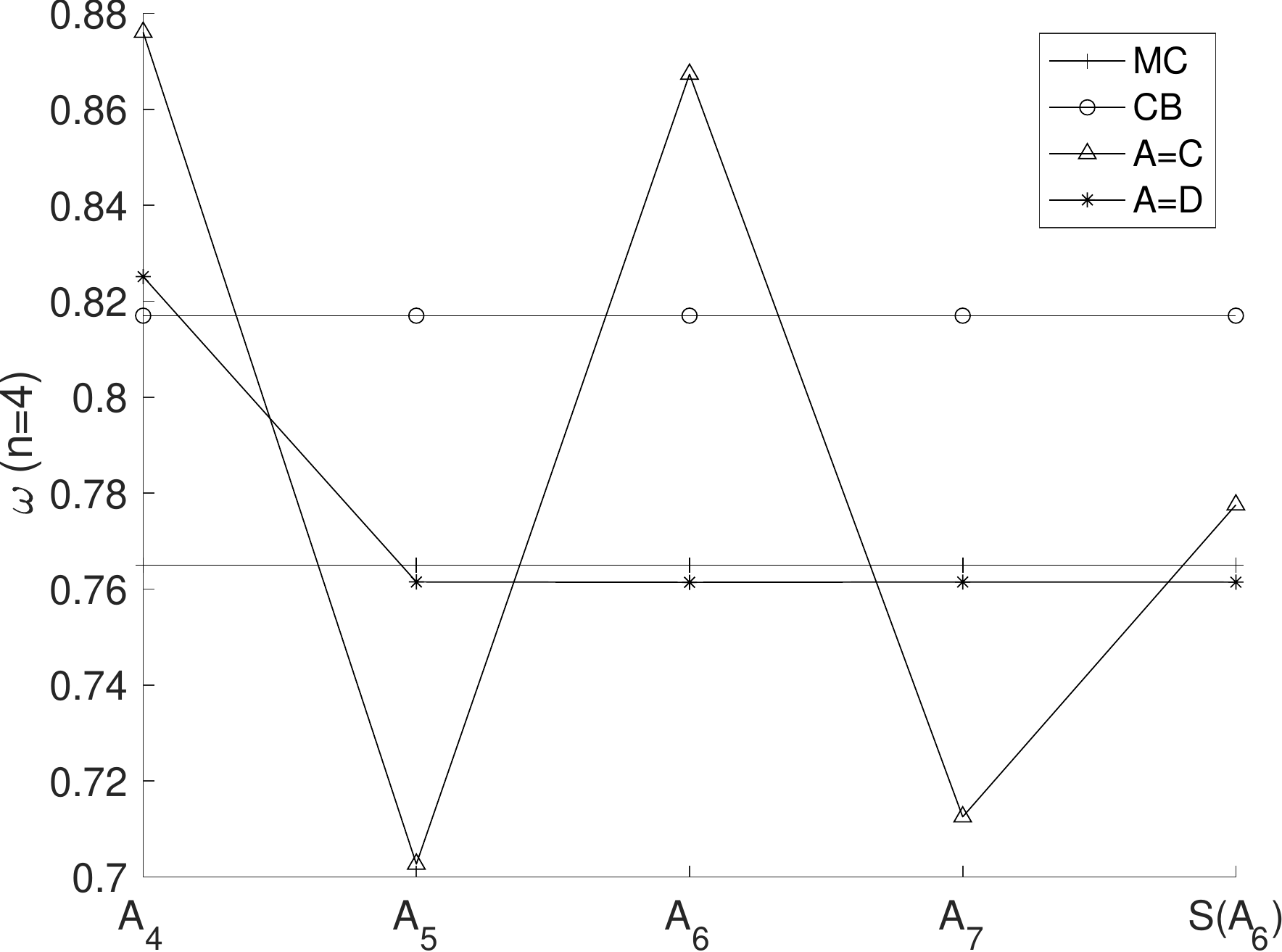}
\caption{Estimates of $\omega$ $(n=4)$ compared with MC result $\omega=0.765$ \cite{Hasenbusch_2001} and CB result $\omega=0.817$ \cite{Echeverri2016}.}

\end{subfigure}

\caption{Estimates of $\omega$ at successive orders compared with MC results and CB results.}
\end{figure} 

The final estimates of $\nu$ and $\omega$ along with their predictions for eight-loop from previous section are tabulated in Table \ref{table:O(n):exponents} with their error calculated from Eq. (18). They are compared with MC, CB results, predictions from non-perturbative renormalization group \cite{nprg} and from seven-loop resummation with hypergeometric functions \cite{shalaby2020critical}. For $XY$-universality classs ($n=2$) $\nu=0.67161(0)$ is obtained using CEF. This estimate can be compared with other predictions from Monte Carlo simulations $\nu=0.67169(7)$ \cite{mcn=2}, conformal bootstrap calculation $\nu=0.67175(10)$ \cite{Chester2020} and non-perturbative renormalization group value $\nu=0.6716(6)$ \cite{nprg}. We predict that at this order results from perturbative renormalization group are completely compatible with other theoretical predictions in contrast to experimental value $\nu=0.6709(1)$ \cite{exp3} and seven-loop results $\nu=0.67076(38)$ \cite{shalaby2020critical}, $\nu=0.67070(73)$ (CEF) \cite{abhignan2020continued} thus complementing the "$\lambda$-point specific heat experimental anomaly" \cite{lambda}. Further, in this case we obtain $\nu=0.6699(16)$ using CEBL and $\nu=0.6743(38)$ using CE, which are consistent up to two decimal places. These results may become more significant once actual calculations from eight-loop approximation of minimal subtraction scheme are solved \cite{seven-O(n)-2}, since the value is small and errors seem to be significant from different approaches of resummation. We observe all the estimates to be compatible with existing predictions while only deduced from lower-order information. For instance, for ($n=0$) self-avoiding walks model CE estimate for $\nu=0.58738(10)$, CEF estimate  $\nu=0.58777$ and CEBL estimate $\nu=0.58752(9)$ are compatible with unprecedented prediction from Monte Carlo simulations $\nu=0.5875970(4)$ \cite{Clisby2016}. 
    \begingroup
\setlength{\tabcolsep}{0.0pt} 
\renewcommand{\arraystretch}{1} 
\begin{table}[htbp]

\begin{center}
\caption{Predictions for eight-loop approximation and estimates for critical exponents $\nu$ and $\omega$ compared with existing predictions (Monte Carlo-MC, Conformal bootsrap-CB, Seven-loop hypergeometric approximant-HM, Non-perturbative renormalization group-NPRG) in $O(n)-$symmetric $\phi^4$ three-dimensional models.} 

 \begin{tabular}{||c| c c c c||}
 
 \hline
$n$ & \begin{tabular}{c c}
     & Hypergeometric 8-loop \\
     & prediction of $\nu$ 
\end{tabular}& $\nu$ &  \begin{tabular}{c c}
     & Hypergeometric 8-loop \\
     & prediction of $\omega$ 
\end{tabular} & $\omega$ \\ [0.5ex] 
 \hline\hline
 
     0
   & -90.985$\epsilon^8$
   & \begin{tabular}{c c c c c c}
        & 0.58777 (CEF)  \\
        & 0.58738(10) (CE) \\
        & 0.58752(9) (CEBL) \\
        & 0.5875970(4) (MC \cite{Clisby2016}) \\
        & 0.5877(12) (CB \cite{Shimada2016}) \\
        & 0.58770(17) (HM \cite{shalaby2020critical})
   \end{tabular}
   & -3992.2$\epsilon^8$
   & \begin{tabular}{c c c c c}
        & 0.80(18)  (CE) \\
        & 0.85299  (CEBL) \\
        & 0.899(12) (MC \cite{Clisby2016}) \\
        & 0.8484(17) (HM \cite{shalaby2020critical})
   \end{tabular}  \\ 
 \hline
    1
   & -72.323$\epsilon^8$
   &  \begin{tabular}{c c c c c}
        & 0.62983(21) (CEF) \\
        & 0.62809(17) (CE) \\
        & 0.62962(72) (CEBL) \\
        & 0.63002(10) (MC \cite{Hasenbusch2010}) \\
        & 0.62999(5) (CB \cite{Showk2014}) \\
        & 0.62977(22) (HM \cite{shalaby2020critical})\\
        & 0.63012(16) (NPRG \cite{nprg})
   \end{tabular}
   & -2443.3$\epsilon^8$
   & \begin{tabular}{c c c c c}
        & 0.80(15)  (CE) \\
        & 0.83176(2)  (CEBL) \\
        & 0.832(6) (MC \cite{Hasenbusch2010}) \\
        & 0.8303(18) (CB \cite{Showk2014}) \\
        & 0.8231(5) (HM \cite{shalaby2020critical})\\
        & 0.832(14) (NPRG \cite{nprg})
   \end{tabular} \\ 
 \hline
    2
   & -54.724$\epsilon^8$
   & \begin{tabular}{c c c c c}
        & 0.67161 (CEF) \\
        & 0.6743(38) (CE) \\
        & 0.6699(16) (CEBL) \\
        & 0.67169(7) (MC \cite{mcn=2}) \\
        & 0.67175(10) (CB \cite{Chester2020}) \\
        & 0.67076(38) (HM \cite{shalaby2020critical})\\
        & 0.6716(6)(16) (NPRG \cite{nprg})
   \end{tabular}
   & -1570.7$\epsilon^8$
   & \begin{tabular}{c c c c c}
        & 0.79(14)  (CE) \\
        & 0.78278(9)  (CEBL) \\
        & 0.789(4) (MC \cite{mcn=2}) \\
        & 0.811(10) (CB \cite{Echeverri2016}) \\
        & 0.789(13) (HM \cite{shalaby2020critical})\\
        & 0.791(8) (NPRG \cite{nprg})
   \end{tabular} \\
 \hline
   3
   & -41.020$\epsilon^8$
   & \begin{tabular}{c c c c c}
        & 0.7092(13) (CEF) \\
        & 0.7059(17) (CE) \\
        & 0.7052(13) (CEBL) \\
        & 0.7116(10) (MC \cite{Hasenbusch2011}) \\
        & 0.7121(28) (CB \cite{Kos2016}) \\
        & 0.70906(18) (HM \cite{shalaby2020critical})\\
        & 0.7114(9) (NPRG \cite{nprg})
   \end{tabular}
   & -1053.1$\epsilon^8$
   & \begin{tabular}{c c c c c}
        & 0.78(12)  (CE) \\
        & 0.76864(2)  (CEBL) \\
        & 0.773 (MC \cite{Hasenbusch_2001}) \\
        & 0.791(22) (CB \cite{Echeverri2016}) \\
        & 0.764(18) (HM \cite{shalaby2020critical})\\
        & 0.769(11) (NPRG \cite{nprg})
   \end{tabular}\\
 \hline
   4
   & -30.878$\epsilon^8$
   & \begin{tabular}{c c c c c}
        & 0.74105(1) (CEF) \\
        & 0.7364(8) (CE) \\
        & 0.73218 (CEBL) \\
        & 0.750(2) (MC \cite{Hasenbusch2011}) \\
        & 0.751(3) (CB \cite{Echeverri2016}) \\
        & 0.74425(32) (HM \cite{shalaby2020critical})\\
        & 0.7478(9) (NPRG \cite{nprg})
   \end{tabular}
   & -731.91$\epsilon^8$
   & \begin{tabular}{c c c c c}
        & 0.78(9)  (CE) \\
        & 0.76146(6)  (CEBL) \\
        & 0.765(30)(MC \cite{Hasenbusch_2001}) \\
        & 0.817(30) (CB \cite{Echeverri2016}) \\
        & 0.7519(13) (HM \cite{shalaby2020critical})\\
        & 0.761(12) (NPRG \cite{nprg})
   \end{tabular} \\
 \hline
\end{tabular}
\label{table:O(n):exponents}
\end{center}
\end{table}
\subsection{Two-dimensional resummation of $\nu$ and $\omega$}
More interesting resummation results are in the case of $\epsilon=2$, two-dimensional systems where exact solutions for critical exponents $\nu$ and $\omega$ are known in $n=0$, self-avoiding walks model \cite{Bernard1982,Caracciolo2005} and $n=1$, Ising model \cite{Onsager1944, Calabrese_2000}. These exact results are helpful to check the reliable nature of a resummation methods for larger parameter, $\epsilon>1$. Similar to previous section we construct CF (Eq.(10)) and CE (Eq.(11)) for $\nu,\omega$ of $n=0,1$. We obtain estimates for predicted eight-loop approximation of $\nu$, $n=0,1$ and $\omega$, $n=0,1$ in two dimensions $(\epsilon=2)$ to compare their convergence behaviours with exact values in Figs. (13) and (14), respectively. When compared with the exact values we observe slow oscillating convergence nature of CE when compared with CF. \begin{figure}[ht]
\centering
\begin{subfigure}{0.495\textwidth}
\includegraphics[width=1\linewidth, height=6cm]{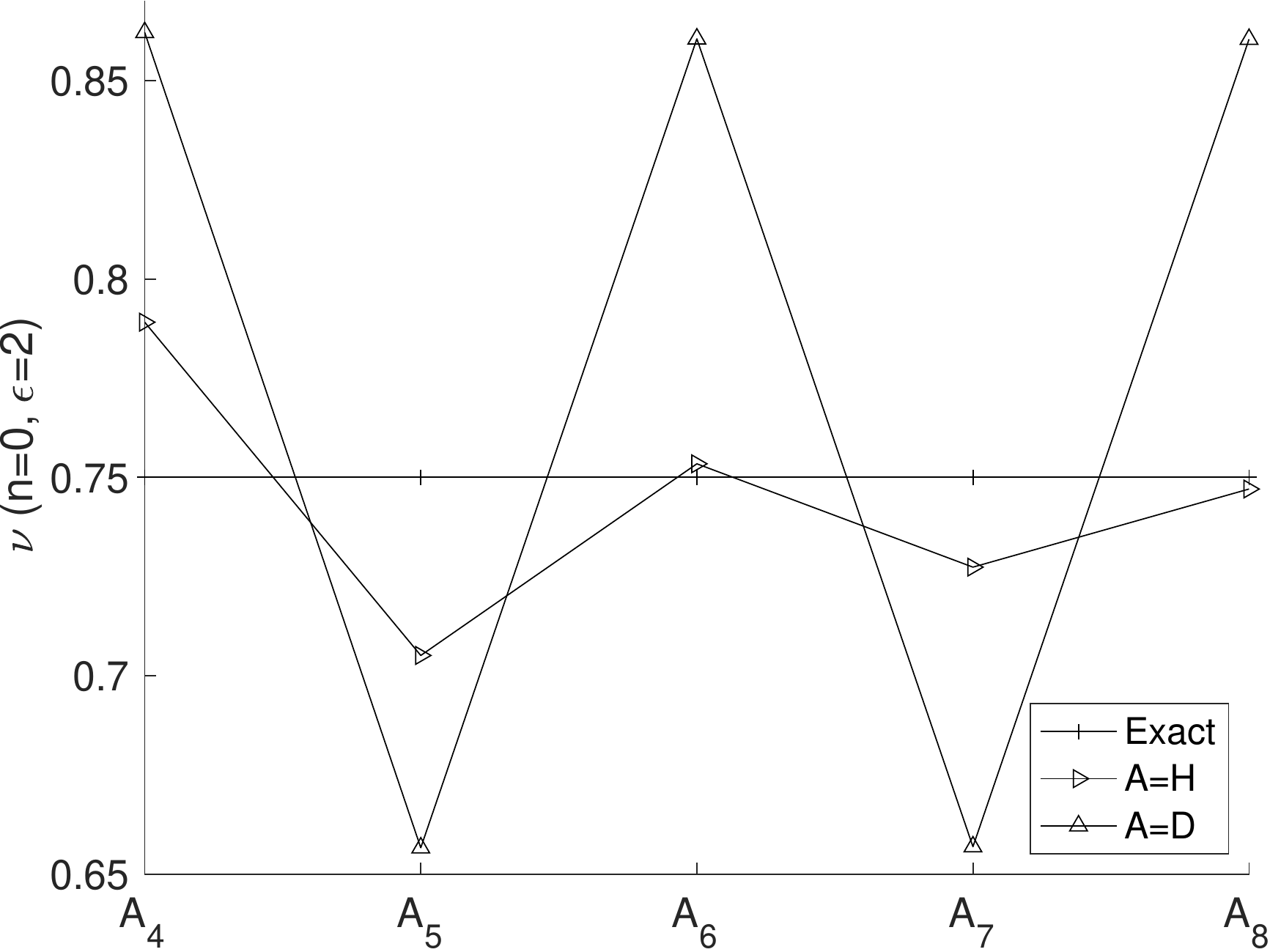} 
\caption{Estimates of two-dimensional $(n=0)$ self-avoiding walks model $\nu$ compared with exact result $\nu=0.75$ \cite{Bernard1982}.}

\end{subfigure}
\begin{subfigure}{0.495\textwidth}
\includegraphics[width=1\linewidth, height=6cm]{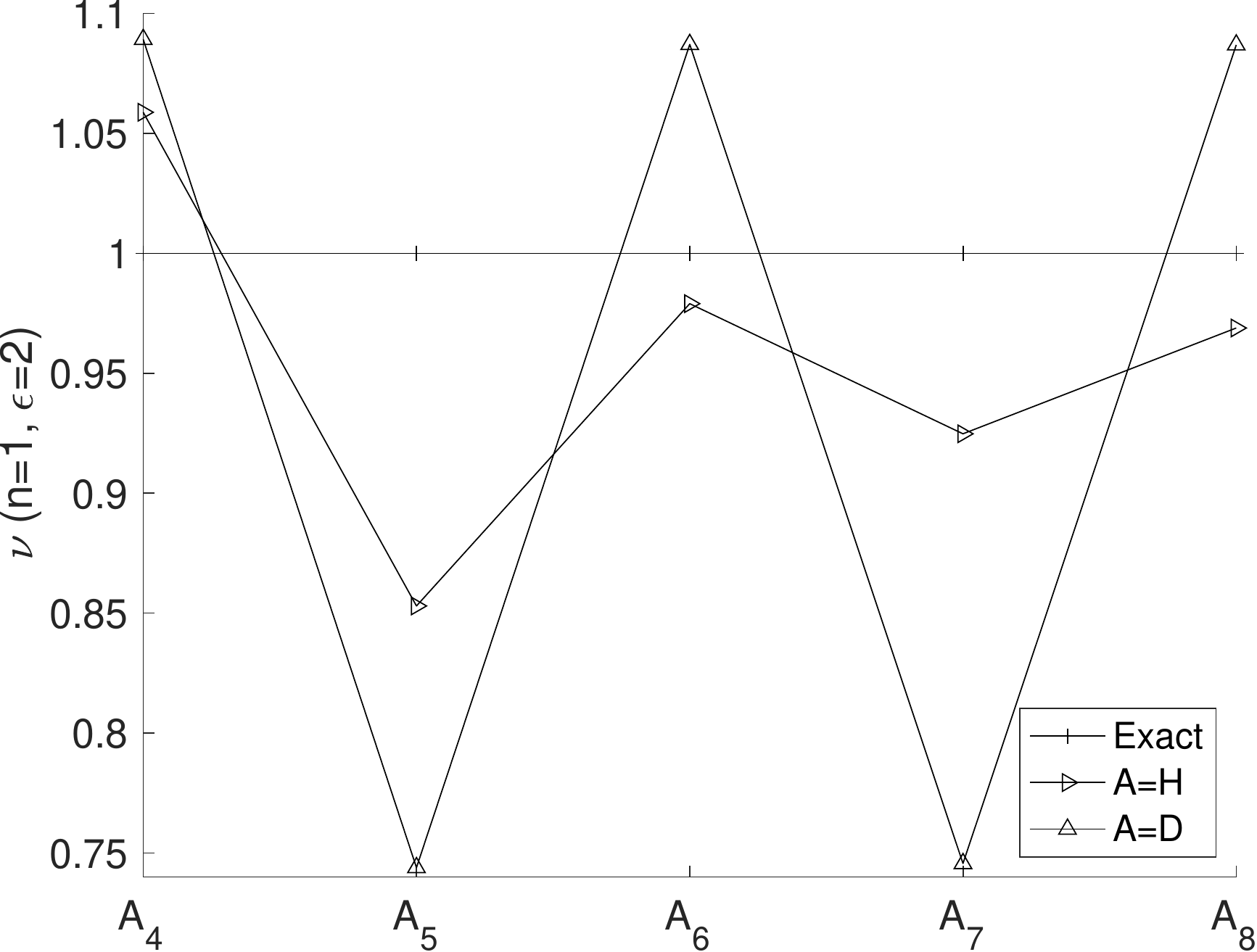}
\caption{Estimates of two-dimensional Ising model $\nu$ $(n=1)$ compared with exact lattice result $\nu=1$ \cite{Onsager1944}.}

\end{subfigure}

\caption{Estimates of two-dimensional $\nu$ at successive orders compared with exact results.}
\end{figure} 
 \begin{figure}[ht]
\centering
\begin{subfigure}{0.495\textwidth}
\includegraphics[width=1\linewidth, height=6cm]{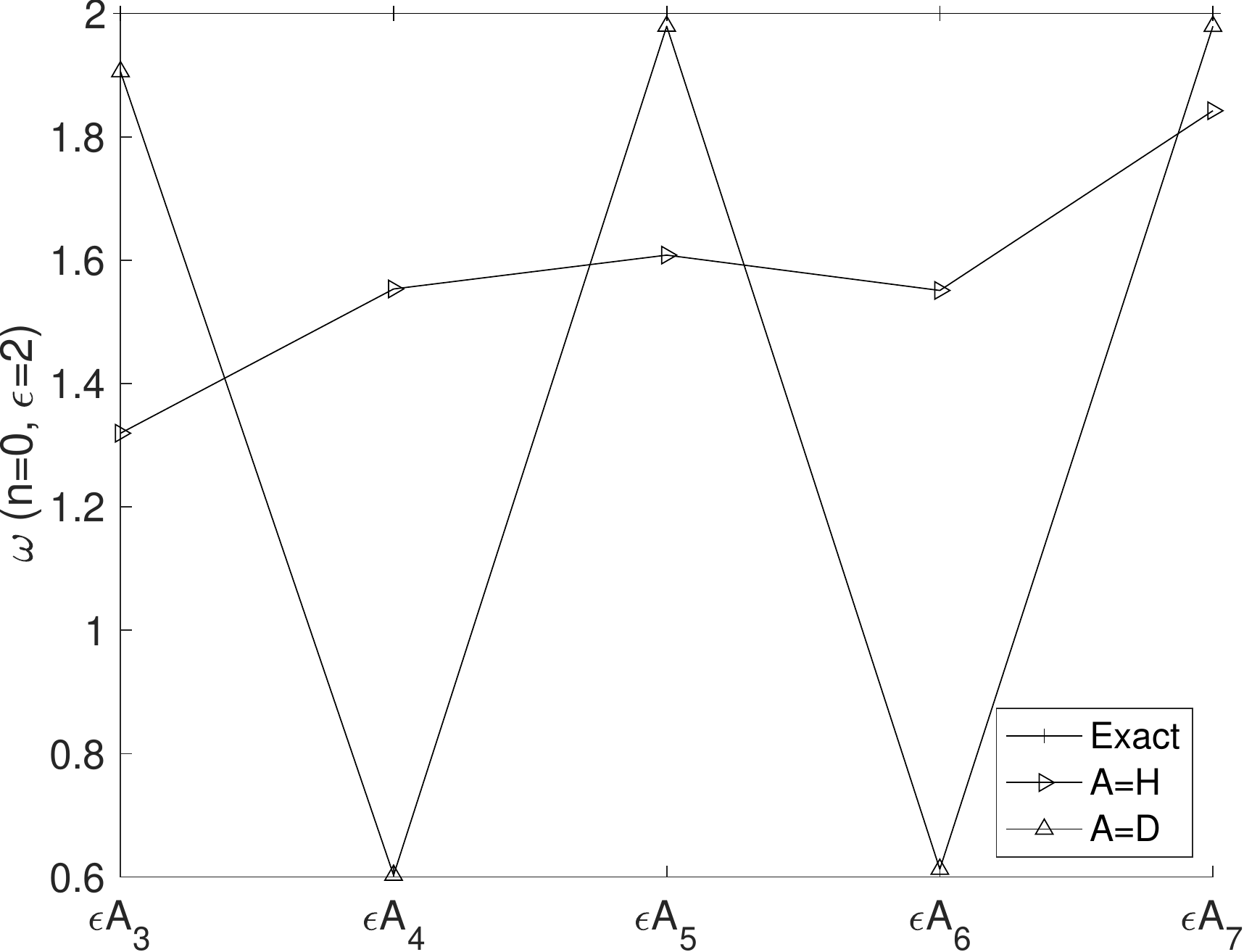} 
\caption{Estimates of two-dimensional $(n=0)$ self-avoiding walks model $\omega$ compared with exact result $\omega=2$ \cite{Caracciolo2005}.}

\end{subfigure}
\begin{subfigure}{0.495\textwidth}
\includegraphics[width=1\linewidth, height=6cm]{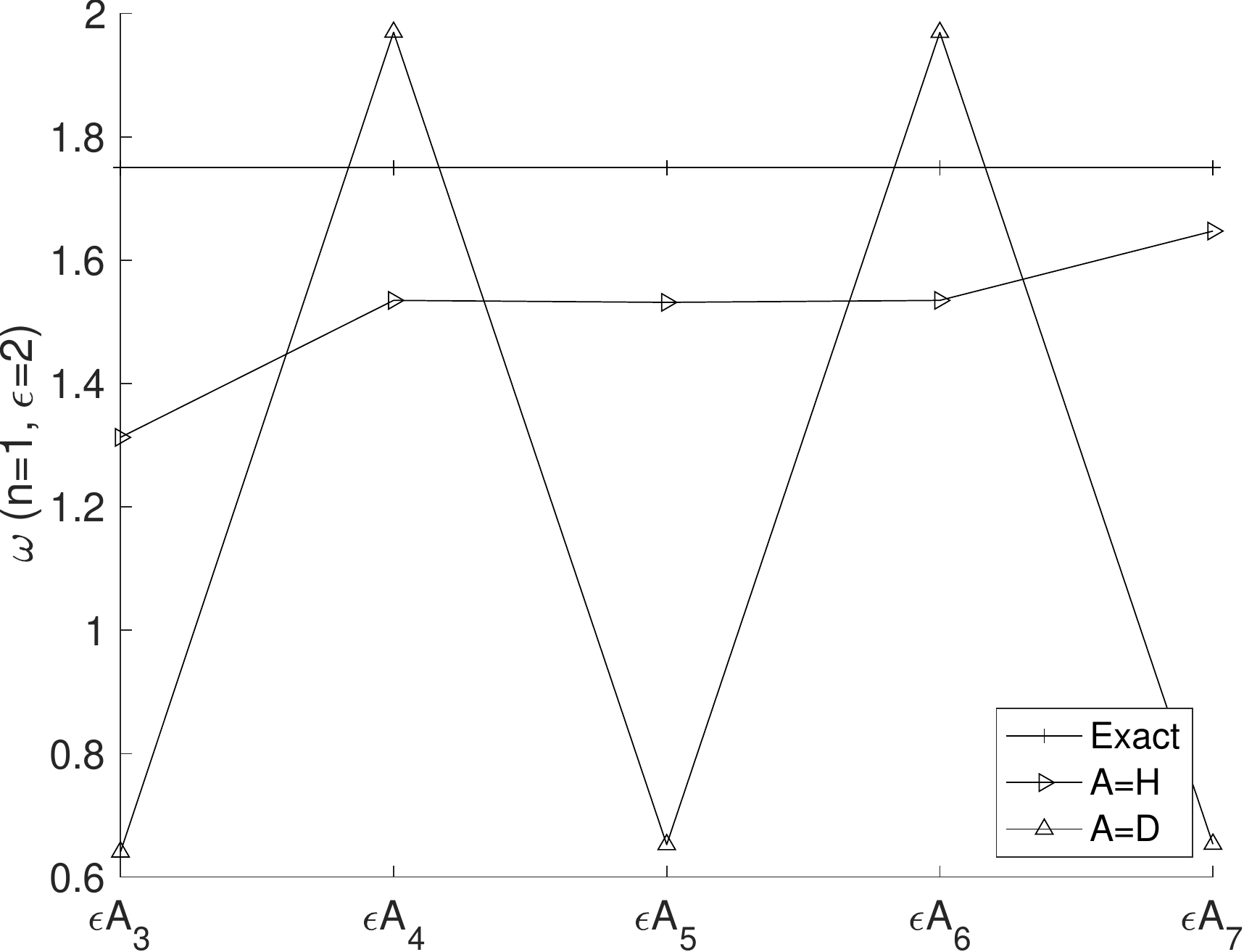}
\caption{Estimates of two-dimensional Ising model $\omega$ $(n=1)$ compared with exact result $\omega=1.75$ \cite{Calabrese_2000}.}

\end{subfigure}

\caption{Estimates of two-dimensional $\omega$ at successive orders compared with exact results.}
\end{figure} 

Exhibiting these convergence behaviours and comparing with their exact values, one-sided convergence is observed, where the estimates $H_2, H_4, H_6, H_8$ in case of $\nu$ and $\epsilon H_1, \epsilon H_3,\epsilon H_5,\epsilon H_7$ in case of $\omega$ approach faster towards the exact value (previously observed in \cite{abhignan2020continued}). We further deduce that reliable estimate in this case of $\epsilon>1$ using CF are obtained from modified Shanks $S^*(B_i)$ for partial sums $\{B_i\}$ such as \cite{abhignan2020continued} \begin{equation}
     S^*(B_i) = \frac{B_{i+2}B_{i-2}-B_i^2}{B_{i+2}+B_{i-2}-2B_i}.
 \end{equation} Using this we obtain estimates for $\nu$, $S^*(H_4)=0.752(18)$, $S^*(H_6)=0.745(4)$ and $\omega$, $S^*(\epsilon H_3)=2.11(40)$ which are comparable with the exact values $\nu=0.75$ \cite{Bernard1982} and $\omega=2$ \cite{Caracciolo2005} for self-avoiding walks model. Similarly, we obtain estimates for $\nu$, $S^*(H_4)=0.98(4)$, $S^*(H_6)=0.967(6)$ and $\omega$, $S^*(\epsilon H_3)=1.76(22)$, $S^*(\epsilon H_5)=1.77(12)$ which are comparable with the exact values $\nu=1$ \cite{Onsager1944} and $\omega=1.75$ \cite{calabrese2004} for Ising model. Further, we also observe one-sided convergence behaviour in the oscillating sequence of CF for large values of $\epsilon$, which can be seen in our analysis for critical exponents of $\phi^3$ models in next section. We empirically observe that CF is more reliable for producing estimates in case of $\epsilon>1$.
\section{Critical exponents from $\phi^3$ models for Lee-Yang edge singularity and percolation theory}
Scalar $\phi^3$ models can describe phase transitions related to Lee-Yang edge singularity and percolation problems \cite{PHI^3-1980,PHI^3-1981}. The non-unitary version of $\phi^3$ model explains the Lee-Yang edge singularity which is important in lattice gauge theory studies of quantum chromodynamics \cite{PhysRevLett1978}. For instance, this theory produces exponent $\sigma^{(LY)}$ which defines the analytic behaviour of the partition function of the chiral symmetry crossover in case of a nonzero chemical potential \cite{CONNELLY2021121834,PhysRevLett2020}. Phase transitions on percolation models have interesting consequences in condensed matter physics where the scaling properties of a percolating medium are discussed \cite{potts_1952}. Significance of critical exponents in these universality classes has led to development of other field-theoretic approaches to solve for them such as functional renormalization group \cite{11frg,12frg} and conformal bootstrap technique  \cite{10cb,13cb,15cb}. Other significant noncontinuum field approaches such as Monte Carlo simulations and series methods produced unprecedented precision in these exponents which can be further seen in our comparison. However, the most range of predictions for these physically relevant critical exponents in varied dimensions from continuum field-theoretic approach are derived recently from resummation of perturbative renormalization group functions of five-loop approximation \cite{phi^3} which show significant improvement over four-loop approximation \cite{four-loop-phi^3}. Similarly we handle these recently derived five-loop $\epsilon$ expansions \cite{phi^3} of percolation exponents \begin{subequations}
    \begin{align}
        \eta^{(P)}&\approx-0.047619\epsilon -0.022244\epsilon^2 + 0.031263\epsilon^3 -0.061922\epsilon^4 + 0.20454\epsilon^5,\\
        \eta_O^{(P)}&\approx-0.28571\epsilon - 0.057499\epsilon^2 + 0.080517\epsilon^3 - 0.18935\epsilon^4 + 0.63749\epsilon^5,\\
        1/\nu^{(P)}&\approx2 - 0.238095\epsilon - 0.035255\epsilon^2 + 0.049249\epsilon^3 - 0.12744\epsilon^4 + 0.43287\epsilon^5,\\
        \omega^{(P)}&\approx \epsilon - 0.76077\epsilon^2 + 2.0089\epsilon^3 - 7.0413\epsilon^4 + 30.216\epsilon^5,
    \end{align} 
\end{subequations} and Lee-Yang exponents \begin{subequations}
    \begin{align}
        \eta^{(LY)}&\approx-0.11111\epsilon - 0.058985\epsilon^2 + 0.043693\epsilon^3 - 0.078951\epsilon^4 + 0.20843\epsilon^5,\\
\eta_O^{(LY)}&\approx-0.66667\epsilon -0.088477\epsilon^2 + 0.065543\epsilon^3 - 0.11843\epsilon^4 + 0.31241\epsilon^5,\\
1/\nu^{(LY)}&\approx2 - 0.55556\epsilon - 0.029493\epsilon^2 + 0.021845\epsilon^3 - 0.039477\epsilon^4 + 0.10413\epsilon^5,\\\omega^{(LY)}&\approx\epsilon - 0.77160\epsilon^2 + 1.5907\epsilon^3 - 4.5329\epsilon^4 + 15.440\epsilon^5,\\ \sigma^{(LY)}&\approx0.50000 -0.083333\epsilon -0.020319\epsilon^2 + 0.0065494\epsilon^3 -0.014381\epsilon^4 + 0.038131\epsilon^5
\end{align}
\end{subequations} in $6-\epsilon$ dimensions ($\epsilon \rightarrow 0$). Similar to previous sections we approximate these with hypergeometric approximants for extrapolation of six-loop predictions. \subsection{Six-loop prediction for critical exponents}
These exponents are approximated at five-loop order with hypergoemetric functions 
\begin{subequations}
    \begin{align}
        \eta^{(P)}&\sim-0.047619\epsilon+\epsilon({}_{3}F_1(-2.3643,0.0096214,4.4247;-2.4241;-15/28\epsilon)-1),\\
        \eta_O^{(P)}&\sim-0.28571\epsilon +\epsilon({}_{3}F_1(-5.6338,0.027955,3.4359;-5.0418;-15/28\epsilon)-1),\\
        1/\nu^{(P)}&\sim2 - 0.238095\epsilon+\epsilon({}_{3}F_1(-0.60957,0.0088023,5.8863;-0.47992;-15/28\epsilon)-1),\\
        \omega^{(P)}&\sim \epsilon +\epsilon({}_{3}F_1(-1.4798,0.19969,7.0704;-1.4713;-15/28\epsilon)-1),
    \end{align} 
\end{subequations} with large-order parameter $\sigma=-15/28$ \cite{PhysRevB.17.2956} and \begin{subequations}
    \begin{align}
        \eta^{(LY)}&\approx-0.11111\epsilon +\epsilon({}_{3}F_1(-0.26576,-0.064564,18.159;1.4674;-5/18\epsilon)-1),\\
\eta_O^{(LY)}&\approx-0.66667\epsilon+\epsilon({}_{3}F_1(-0.24714,-0.10099,17.989;1.4097;-5/18\epsilon)-1),\\
1/\nu^{(LY)}&\approx2 - 0.55556\epsilon+\epsilon({}_{3}F_1(-0.28941,-0.029743,18.062;1.4644;-5/18\epsilon)-1),\\\omega^{(LY)}&\approx\epsilon+\epsilon({}_{3}F_1(-0.032594,0.09269,12.846;-0.013971;-5/18\epsilon)-1),\\ \sigma^{(LY)}&\approx0.50000 -0.083333\epsilon+\epsilon({}_{3}F_1(-1.0059,0.0078136,9.5591;-1.0271;-5/18\epsilon)-1),
\end{align}
\end{subequations} with large-order parameter $\sigma=-5/18$ \cite{KALAGOV2014672,MCKANE1979166,KIRKHAM1979}. Using these hypergeometric parameters $\{a_1,a_2,a_3,b_1\}$, $\sigma$ in Eq. (24e) we predict the estimates for six-loop order as $-0.76835\epsilon^6$, $-3.2083\epsilon^6$, $-1.7704\epsilon^6$, $-150.01\epsilon^6$ for $\eta^{(P)}$, $\eta_O^{(P)}$, $1/\nu^{(P)}$, $\omega^{(P)}$, respectively and $-0.68971\epsilon^6$, $-1.0323\epsilon^6$, $-0.34409\epsilon^6$, $-58.443\epsilon^6$, $-0.11594\epsilon^6$ for $\eta^{(LY)}$, $\eta_O^{(LY)}$, $1/\nu^{(LY)}$, $\omega^{(LY)}$, $\sigma^{(LY)}$, respectively.
\subsection{Resummation of critical exponents with six-loop predictions} We initially handle these predicted expansions in $6-\epsilon$ dimensions without using constrained series to check their actual convergence behaviours. Further, we determine constrained series $c(\epsilon)$ where two-dimensional ($\epsilon_{bc}=4$) and one-dimensional ($\epsilon_{bc}=5$) exact solutions are used by simple transformation \cite{phi^3,Guida_1998} \begin{equation}
    Q(\epsilon)\approx\sum^N_{i=0} q_i \epsilon^i = f(\epsilon_{bc})+(\epsilon_{bc}-\epsilon) c(\epsilon)
\end{equation} where $bc$ is boundary condition. Here $f(\epsilon_{bc})$ is known to be the exact value of critical exponent, which we assume has sufficient smoothness for varying $\epsilon$. We approximate the value of $c(\epsilon)$ using continued fraction and substitute it in above equation to determine $Q(\epsilon)$. Due to large values of $\epsilon$ in this case constrained series are more favourable for producing precise estimates and we restrict our resummation analysis to CF.  \\ Initially, we handle exponents $\eta^{(LY)}$, $\nu^{(LY)}$, $\sigma^{(LY)}$ for one-dimensional ($\epsilon=5$) and two-dimensional case ($\epsilon=4$) using CF similar to previous sections. The convergence nature of these estimates are illustrated in Figs. (15), (16), (17) at four-loop (4-L), five-loop (5-L), six-loop (6-L) and their refined Shanks values are compared with exact results. Constrained series estimates are denoted by their value of $\epsilon_{bc}$ and direct estimates are denoted by CF. \begin{figure}[ht]
\centering
\begin{subfigure}{0.495\textwidth}
\includegraphics[width=1\linewidth, height=6cm]{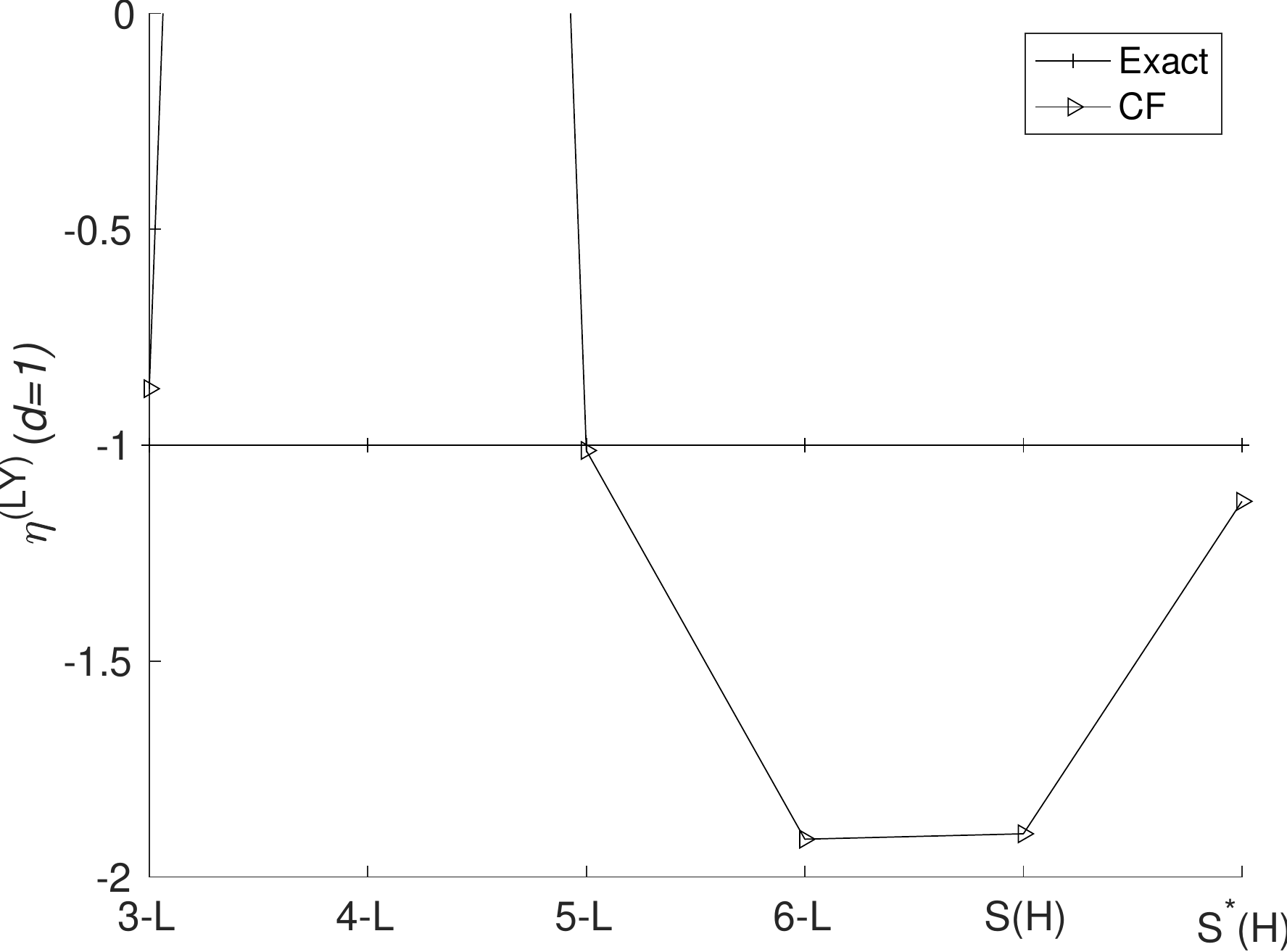} 
\caption{Estimates of one-dimensional $\eta^{(LY)}$ compared with exact result $\eta^{(LY)}=-1$ \cite{John1985}.}

\end{subfigure}
\begin{subfigure}{0.495\textwidth}
\includegraphics[width=1\linewidth, height=6cm]{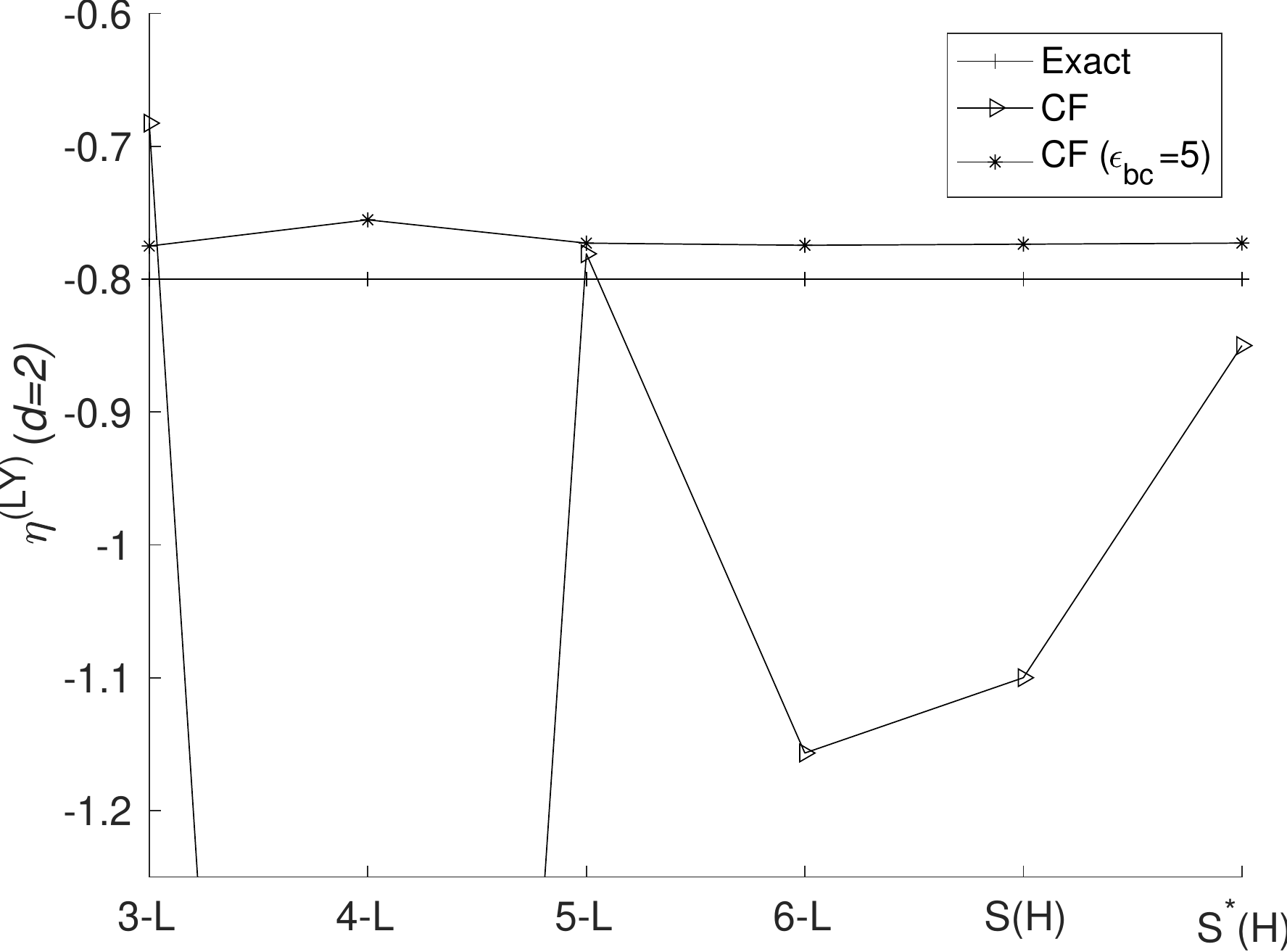}
\caption{Estimates of two-dimensional $\eta^{(LY)}$ compared with exact result $\eta^{(LY)}=-0.8$ \cite{John1985}.}

\end{subfigure}

\caption{Estimates of $\eta^{(LY)}$ at successive orders compared with exact results.}
\end{figure} 
 \begin{figure}[ht]
\centering
\begin{subfigure}{0.495\textwidth}
\includegraphics[width=1\linewidth, height=6cm]{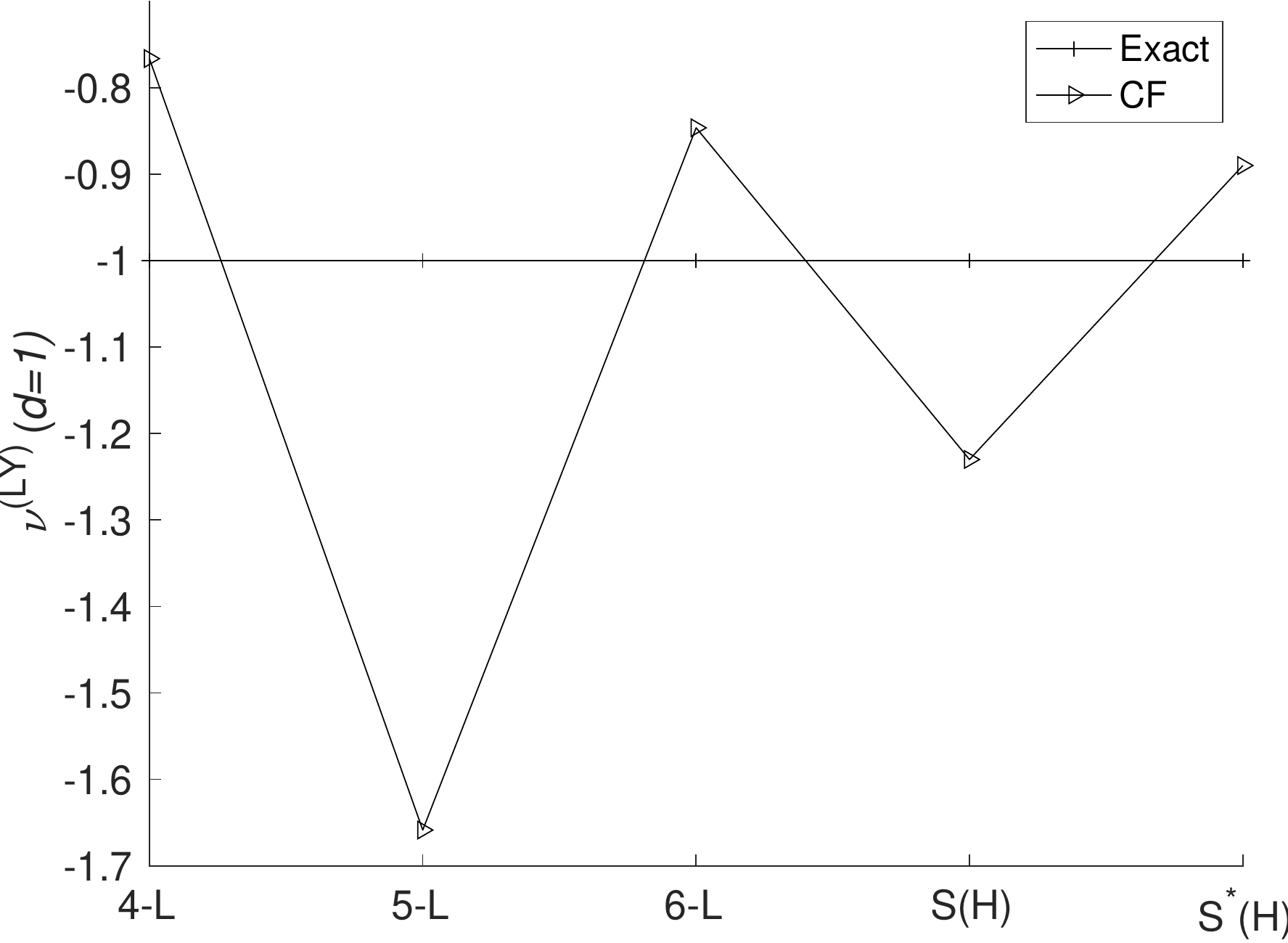} 
\caption{Estimates of one-dimensional $\nu^{(LY)}$ compared with exact result $\nu^{(LY)}=-1$ \cite{John1985}.}

\end{subfigure}
\begin{subfigure}{0.495\textwidth}
\includegraphics[width=1\linewidth, height=6cm]{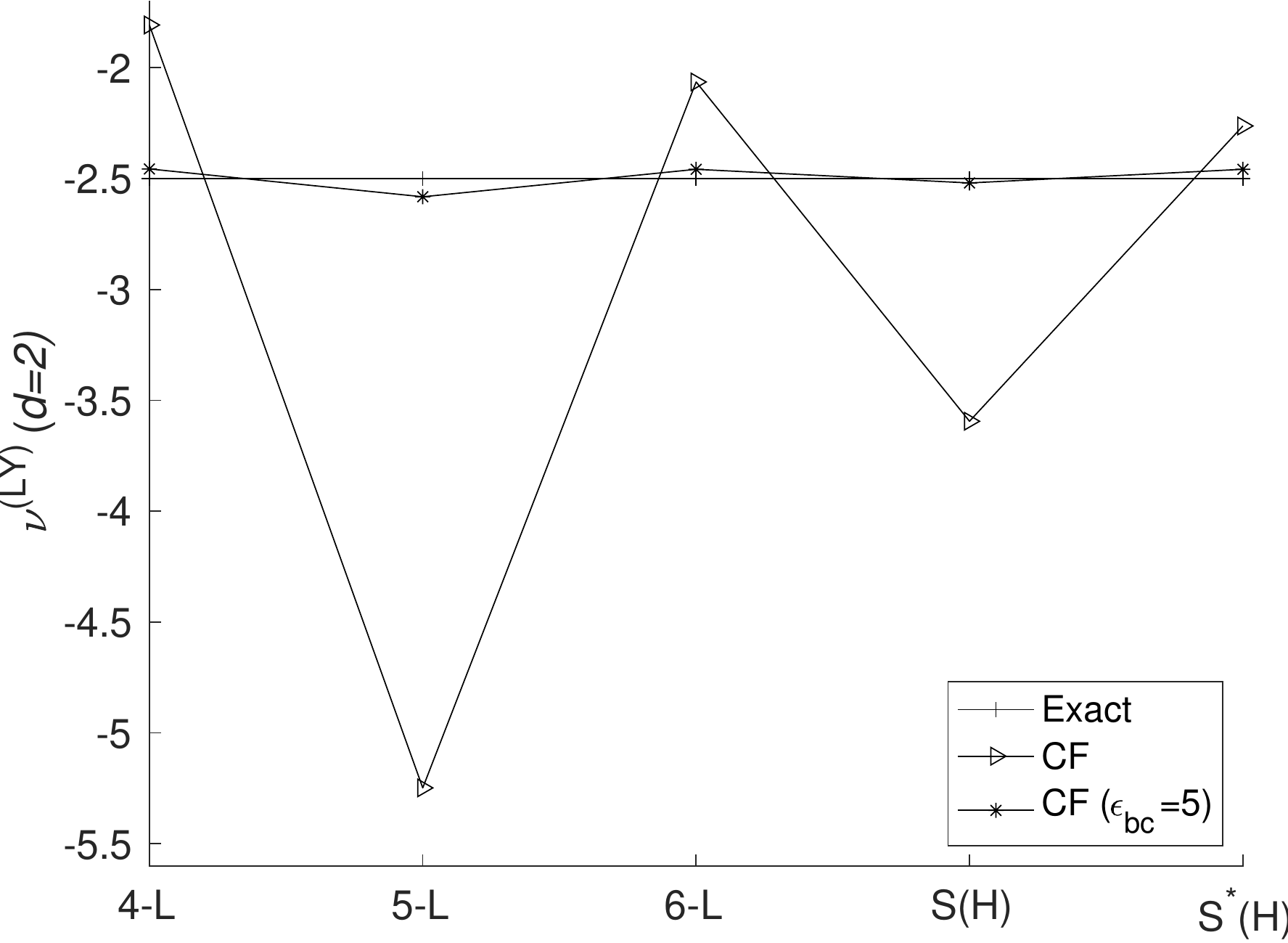}
\caption{Estimates of two-dimensional $\nu^{(LY)}$ compared with exact result $\nu^{(LY)}=-2.5$ \cite{John1985}.}

\end{subfigure}

\caption{Estimates of $\nu^{(LY)}$ at successive orders compared with exact results.}
\end{figure} 
 \begin{figure}[ht]
\centering
\begin{subfigure}{0.495\textwidth}
\includegraphics[width=1\linewidth, height=6cm]{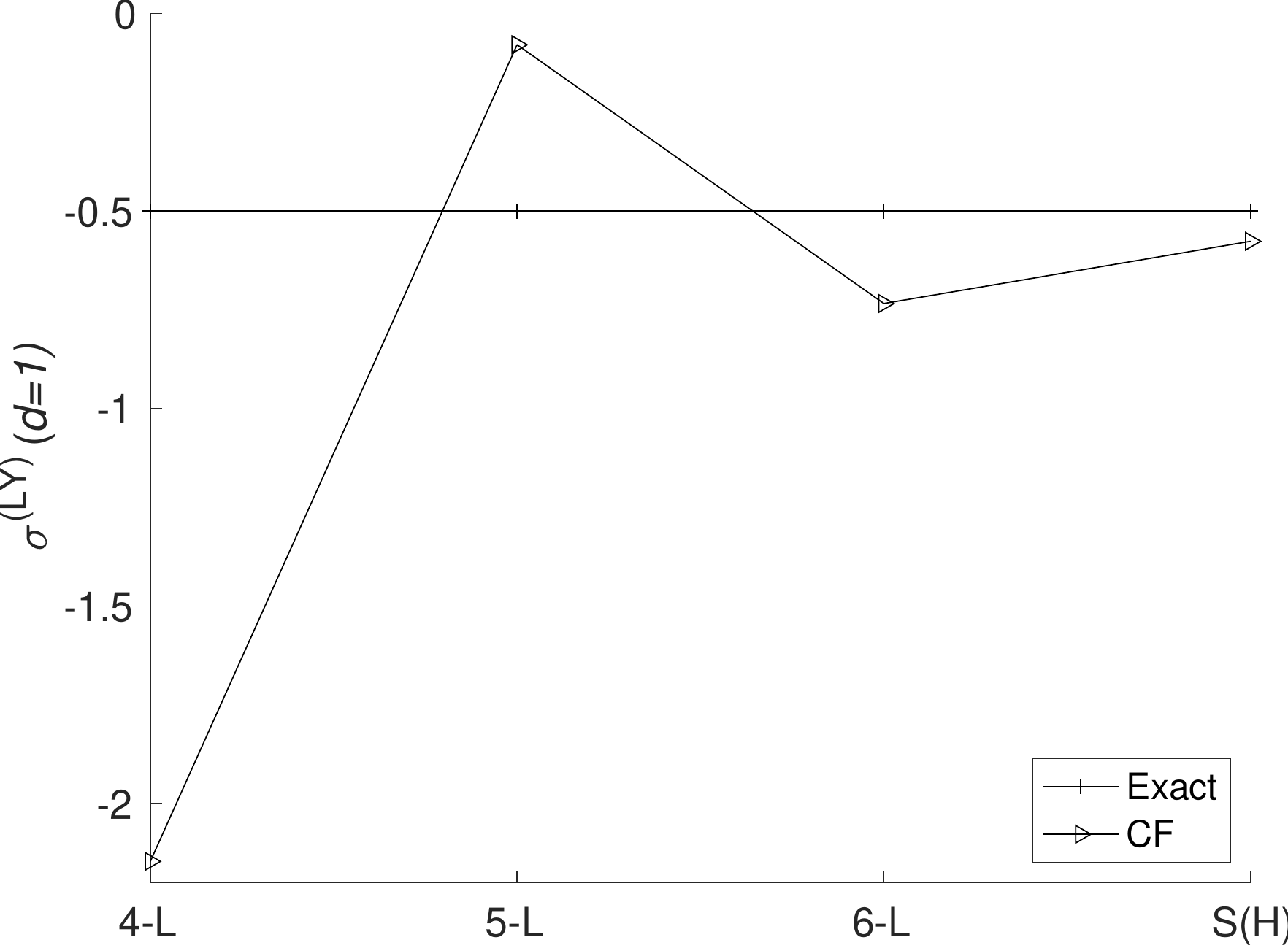} 
\caption{Estimates of one-dimensional $\sigma^{(LY)}$ compared with exact result $\sigma^{(LY)}=-0.5$ \cite{John1985}.}

\end{subfigure}
\begin{subfigure}{0.495\textwidth}
\includegraphics[width=1\linewidth, height=6cm]{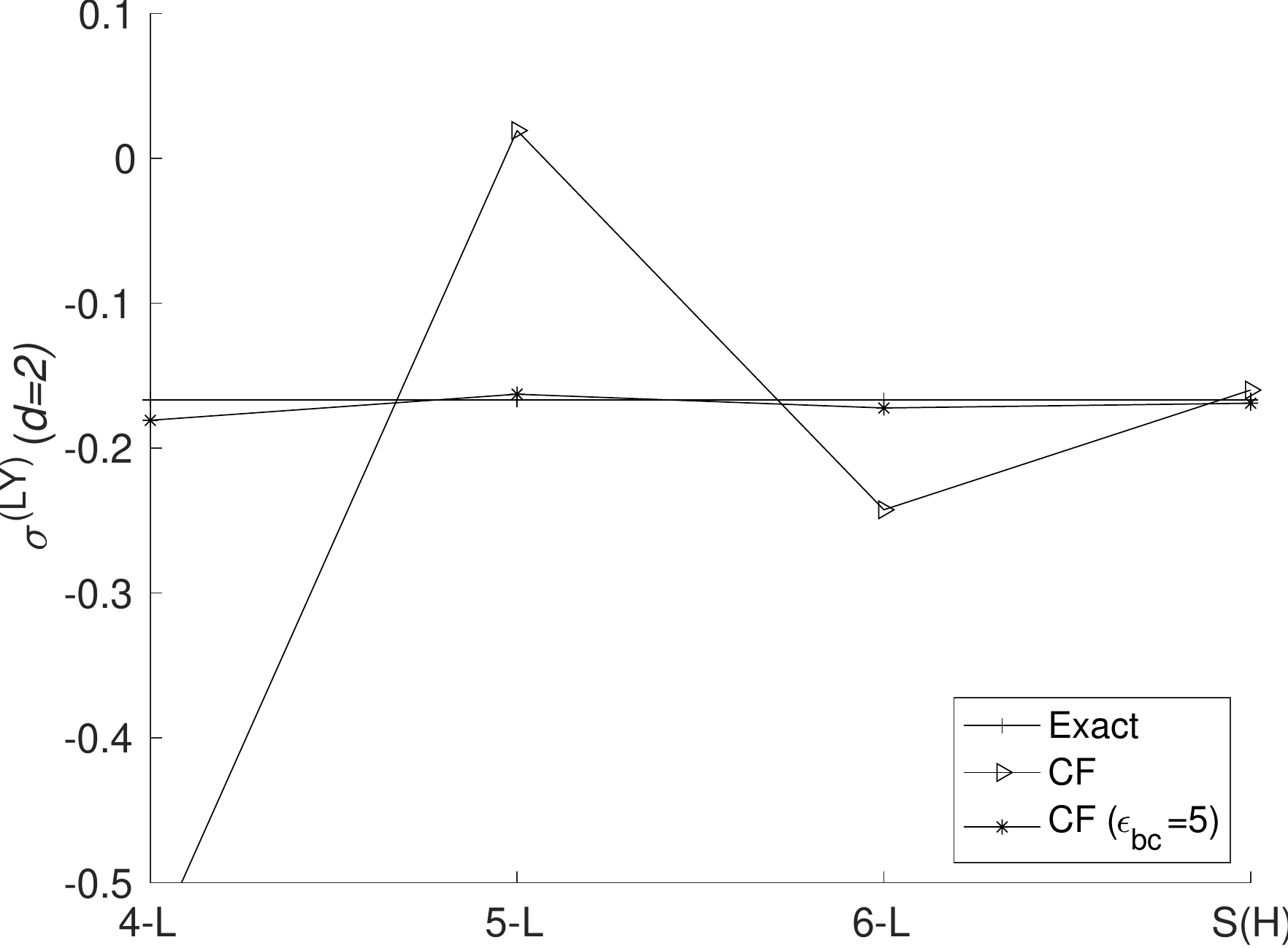}
\caption{Estimates of two-dimensional $\sigma^{(LY)}$ compared with exact result $\sigma^{(LY)}=-0.1667$ \cite{John1985}.}

\end{subfigure}

\caption{Estimates of $\sigma^{(LY)}$ at successive orders compared with exact results.}
\end{figure}

The series concerning the anomalous dimension which completely determines the scaling properties of Lee-Yang edge, $\eta^{(LY)}$ and $\nu^{(LY)}$, $\sigma^{(LY)}$ show slow convergence behaviour towards the exact values with large deviations due to larger values of $\epsilon=5,4$. Similar to previous section we observe one-sided convergence ($S^*$ from Eq. (52)) in case of $\eta^{(LY)}$ and $\nu^{(LY)}$ approach faster towards the exact value. However, the constrained series ($\epsilon_{bc}=5$) show remarkable convergence towards the exact values for two-dimensional estimates. In the same manner for three-dimensional ($\epsilon=3$) case the oscillating convergent sequence of $\eta^{(LY)}$, $\nu^{(LY)}$, $\sigma^{(LY)}$ and $\omega^{(LY)}$ are illustrated in Figs. (18), (19). The direct estimates seem to slightly undershoot or overshoot when compared to previous predictions from functional renormalization group (FRG) approach \cite{11frg}. However, the constrained series ($\epsilon_{bc}=5, 4$) estimates seem to be compatible with the existing results. 
\begin{figure}[ht]
\centering
\begin{subfigure}{0.495\textwidth}
\includegraphics[width=1\linewidth, height=6cm]{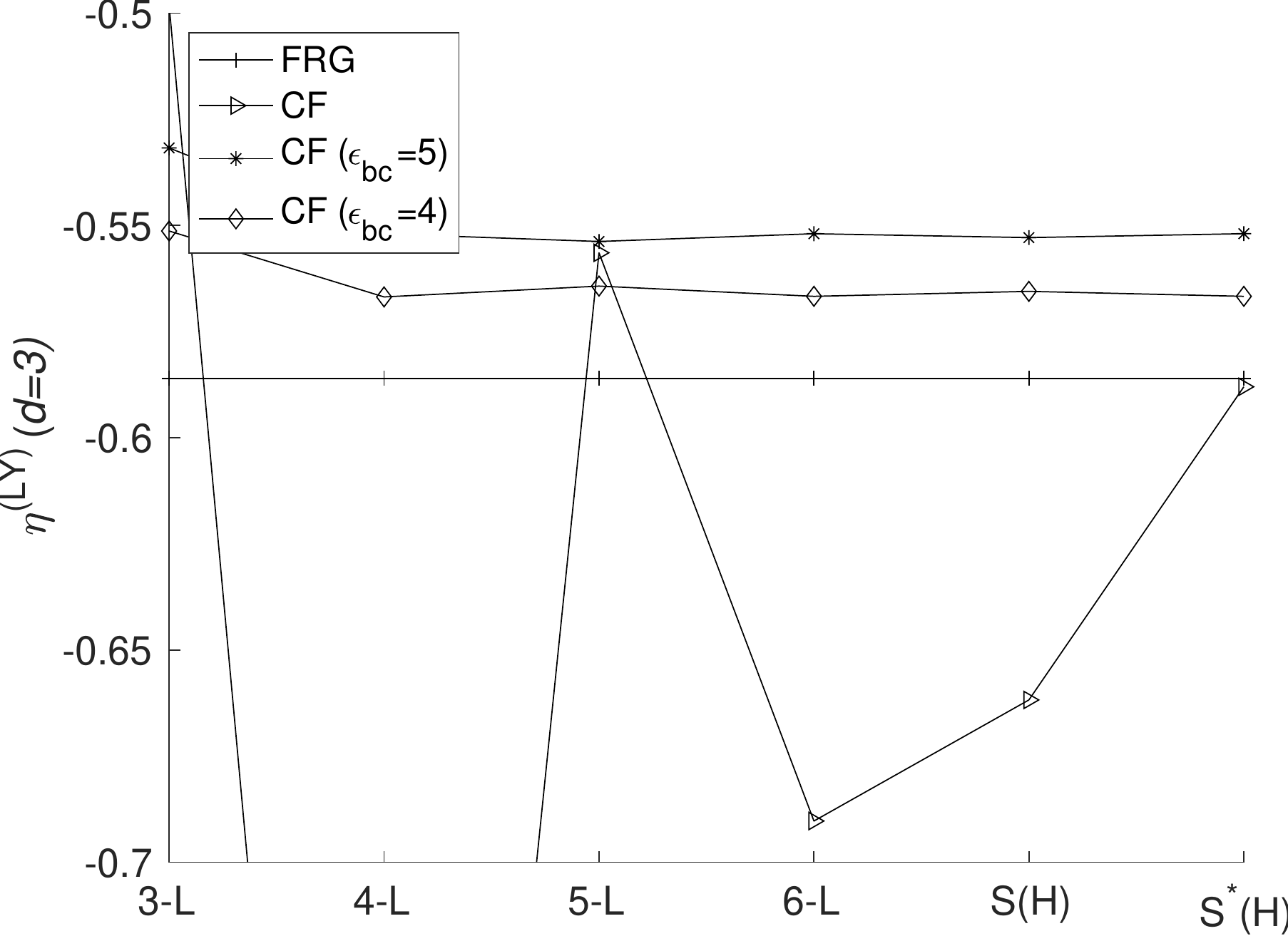} 
\caption{Estimates of three-dimensional $\eta^{(LY)}$ compared with FRG value $\eta^{(LY)}=-0.586$ \cite{11frg}.}

\end{subfigure}
\begin{subfigure}{0.495\textwidth}
\includegraphics[width=1\linewidth, height=6cm]{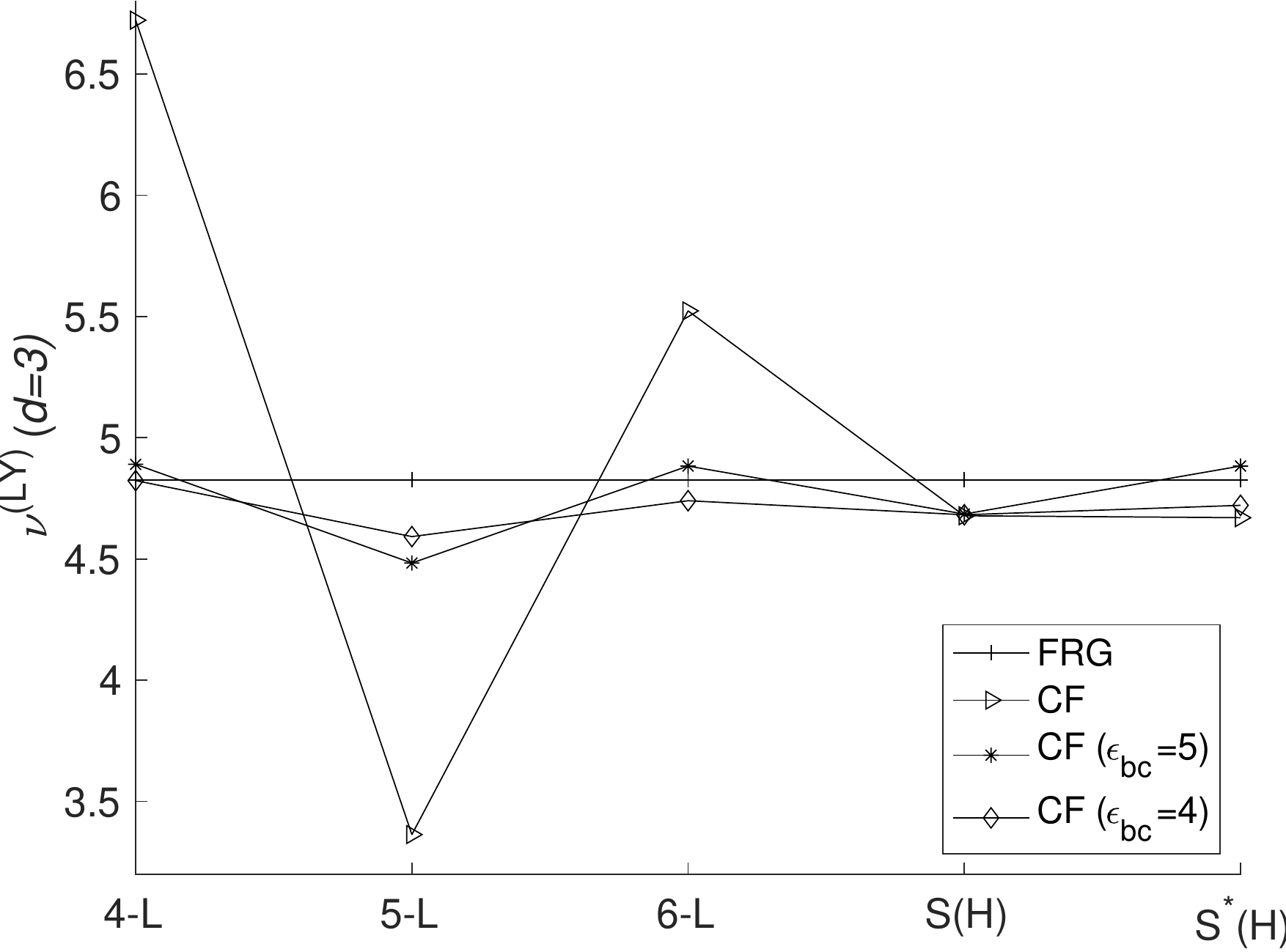}
\caption{Estimates of three-dimensional $\nu^{(LY)}$ compared with FRG value $\nu^{(LY)}=4.826146$ \cite{11frg}.}

\end{subfigure}

\caption{Estimates of $\eta^{(LY)}$ and $\nu^{(LY)}$ at successive orders compared with FRG values.}
\end{figure}
\begin{figure}[ht]
\centering
\begin{subfigure}{0.495\textwidth}
\includegraphics[width=1\linewidth, height=6cm]{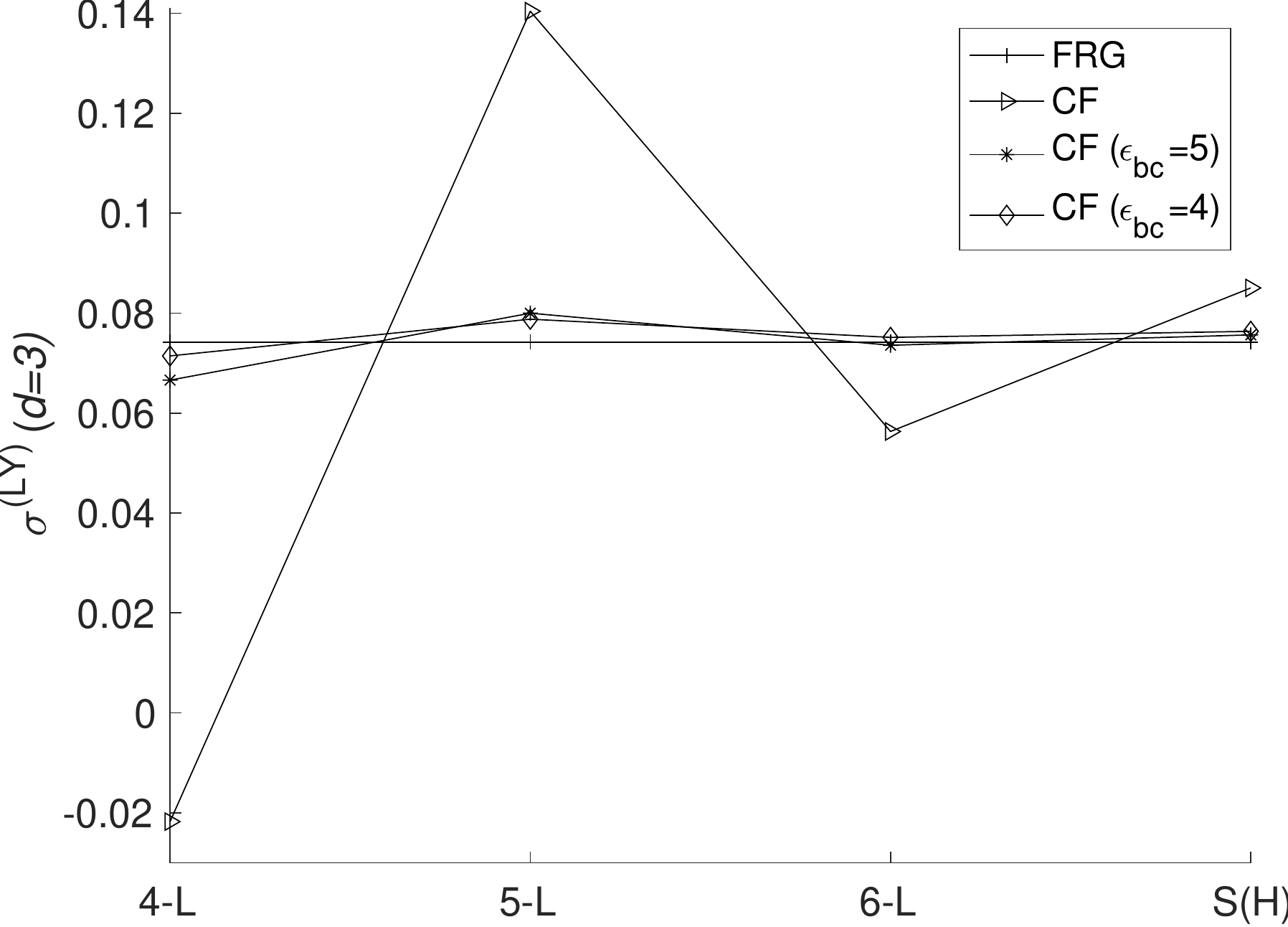} 
\caption{Estimates of three-dimensional $\sigma^{(LY)}$ compared with FRG value $\sigma^{(LY)}=0.0742$ \cite{11frg}.}

\end{subfigure}
\begin{subfigure}{0.495\textwidth}
\includegraphics[width=1\linewidth, height=6cm]{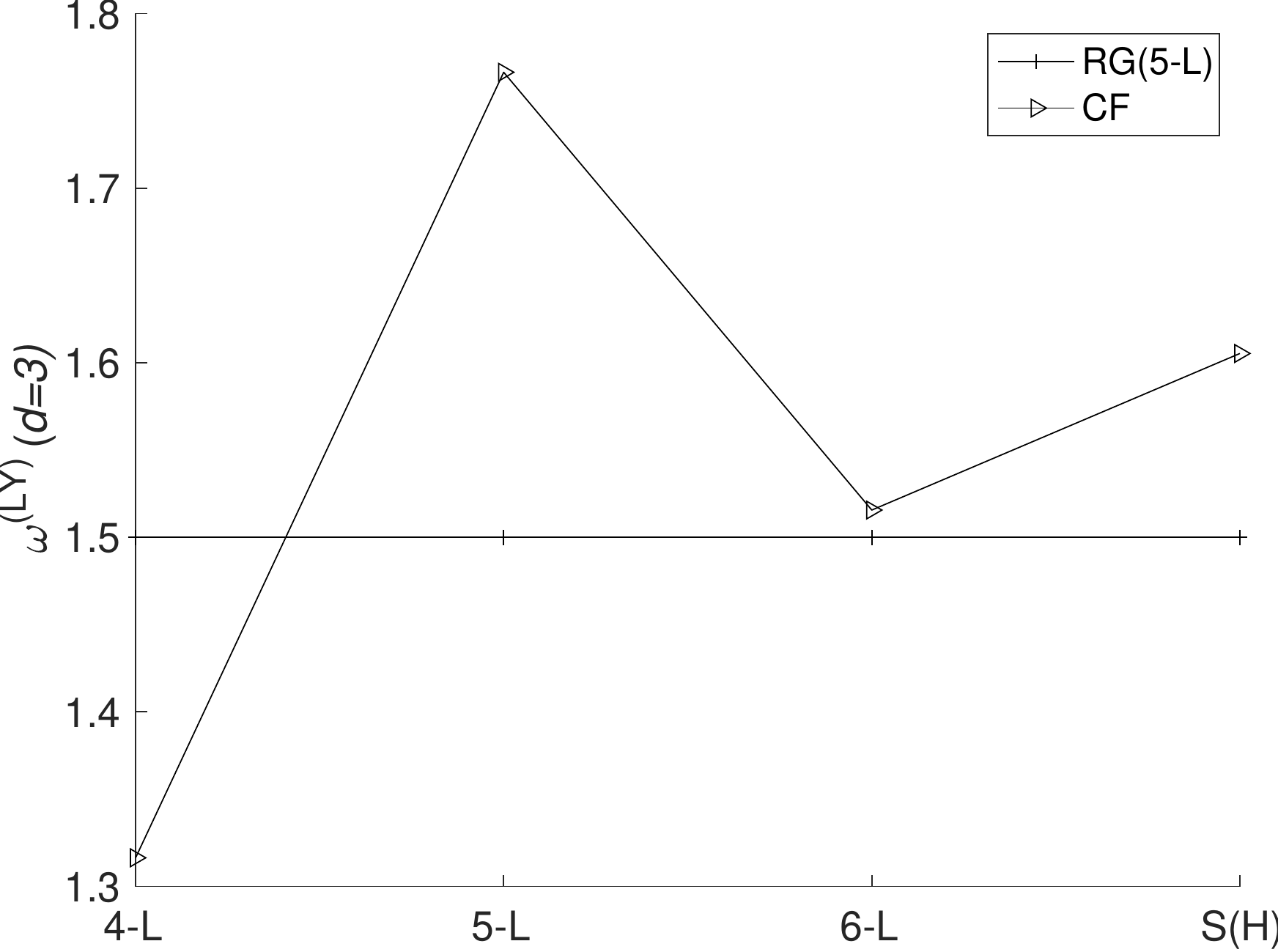}
\caption{Estimates of three-dimensional $\omega^{(LY)}$ compared with five-loop RG value $\omega^{(LY)}=1.5$ \cite{phi^3}.}

\end{subfigure}

\caption{Estimates of $\sigma^{(LY)}$ and $\omega^{(LY)}$ at successive orders compared with FRG and RG values.}
\end{figure}

 Using this way of analysis we predict estimates for exponents characterising the Lee-Yang edge and tabulate them in Table 8, for dimensions ranging from one to five. The estimates derived from CF show improvement and are compatible with existing predictions from five-loop resummation \cite{phi^3}, functional renormalization group approach \cite{11frg,12frg} and series expansions of lattice spin models \cite{PRE2012}.  From basic two exponents $\nu$ and $\eta$ all other exponents can be estimated from hyperscaling relations \cite{phi^3}. 
    \begingroup
\setlength{\tabcolsep}{0pt} 
\renewcommand{\arraystretch}{1} 
\begin{table}[htbp]
 \scriptsize
\begin{center}
\caption{CF estimates for Lee-Yang critical exponents  in one to five dimensions compared with existing predictions and hypergeometric prediction for their six-loop approximation.} 

 \begin{tabular}{||c c c c c c c||}
 
 \hline
Exponent &  \begin{tabular}{c c}
     &  Six-loop\\
     & approximation
\end{tabular} & $\epsilon=5$  & $\epsilon=4$   & $\epsilon=3$  & $\epsilon=2$  & $\epsilon=1$  \\ [0.5ex] 
 \hline\hline
 $\eta^{(LY)}$
   & $-0.68971 \epsilon^6$
   &   \begin{tabular}{c c} & \\
   & \\
   & \\
   & \\
     &  -1.9(5) (CF,$S$) \\
     & -1.1(1) (CF,$S^*$)\\
      & -1.2(3)\cite{phi^3} \\
     & -1 \cite{John1985} (exact)
\end{tabular}
   &  \begin{tabular}{c c} & \\
   & \\
     &  -0.774(1) ($\epsilon_{bc}=5$,$S$) \\
     &  -0.773 ($\epsilon_{bc}=5$,$S^*$) \\
     &  -1.1(2) (CF,$S$) \\
     & -0.85(8) (CF,$S^*$)\\
      & -0.78(3)\cite{phi^3} \\
     & -0.8 \cite{John1985} (exact)
\end{tabular}
   & \begin{tabular}{c c c}
   &  -0.566(2) ($\epsilon_{bc}=4$,$S$) \\
    &  -0.567 ($\epsilon_{bc}=4$,$S^*$) \\
   &  -0.553(1) ($\epsilon_{bc}=5$,$S$) \\
    &  -0.552 ($\epsilon_{bc}=5$,$S^*$) \\
     & -0.66(8) (CF,$S$) \\
     & -0.59(5) (CF,$S^*$)\\
      & -0.57(2) \cite{phi^3}\\
     & -0.586(29)\cite{11frg}
\end{tabular}
   & \begin{tabular}{c c c}
    &  -0.346(1) ($\epsilon_{bc}=4$,$S$) \\
    &  -0.347 ($\epsilon_{bc}=4$,$S^*$) \\
   &  -0.3419(8) ($\epsilon_{bc}=5$,$S$) \\
    &  -0.3413 ($\epsilon_{bc}=5$,$S^*$) \\
     &  -0.37(2) (CF,$S$) \\
     & -0.35(2) (CF,$S^*$)\\
      & -0.347(12)\cite{phi^3} \\
     & -0.316(16)\cite{11frg}
\end{tabular}
& \begin{tabular}{c c c}
 &  -0.150 ($\epsilon_{bc}=4$,$S$) \\
    &  -0.150 ($\epsilon_{bc}=4$,$S^*$) \\
   &  -0.150 ($\epsilon_{bc}=5$,$S$) \\
    &  -0.150 ($\epsilon_{bc}=5$,$S^*$) \\
     &  -0.152(2) (CF,$S$) \\
     & -0.151(3) (CF,$S^*$)\\
          & -0.151(2)\cite{phi^3} \\
     & -0.126(6) \cite{11frg}
\end{tabular}\\ 
 \hline
    $\nu^{(LY)}$
   & $-0.34409 \epsilon^6$
   &   \begin{tabular}{c c} & \\
   & \\
   & \\
   & \\
     & -1.2(6) (CF,$S$) \\
     & -0.9(1) (CF,$S^*$)\\
     & -1.0(3) \cite{phi^3}\\
     & -1 \cite{John1985} (exact)
\end{tabular}
   &  \begin{tabular}{c c} & \\
   & \\
     &  -2.52(9) ($\epsilon_{bc}=5$,$S$) \\
     &  -2.458(1) ($\epsilon_{bc}=5$,$S^*$) \\
     & -4(2) (CF,$S$) \\
     & -2.3(2) (CF,$S^*$)\\
     & -2.6(2) \cite{phi^3}\\
     & -2.5 \cite{John1985} (exact)
\end{tabular}
   & \begin{tabular}{c c c}
      &  4.7(1) ($\epsilon_{bc}=4$,$S$) \\
    &  4.72(5) ($\epsilon_{bc}=4$,$S^*$) \\
   &  4.7(3) ($\epsilon_{bc}=5$,$S$) \\
    &  4.883(4) ($\epsilon_{bc}=5$,$S^*$) \\
     &  4(1) (CF,$S$) \\
     & 4.67(4) (CF,$S^*$)\\
      & 4.6(2) \cite{phi^3} \\
     & 4.826146 \cite{11frg}
\end{tabular}
   & \begin{tabular}{c c c}
     &  1.212(4) ($\epsilon_{bc}=4$,$S$) \\
    &  1.214(2) ($\epsilon_{bc}=4$,$S^*$) \\
   &  1.21(1) ($\epsilon_{bc}=5$,$S$) \\
    &  1.219 ($\epsilon_{bc}=5$,$S^*$) \\
     &  1.21(3) (CF,$S$) \\
     & 1.22(1) (CF,$S^*$)\\
      & 1.211(8)\cite{phi^3} \\
     & 1.187477 \cite{11frg}
\end{tabular}
& \begin{tabular}{c c c}
    &  0.7020(2) ($\epsilon_{bc}=4$,$S$) \\
    &  0.7020(2) ($\epsilon_{bc}=4$,$S^*$) \\
   &  0.7021(4) ($\epsilon_{bc}=5$,$S$) \\
    &  0.7024 ($\epsilon_{bc}=5$,$S^*$) \\
     &  0.7019(7) (CF,$S$) \\
     & 0.7022(4) (CF,$S^*$)\\
      & 0.7020(4)\cite{phi^3} \\
     & 0.696008 \cite{11frg}
\end{tabular}\\ 
 \hline
    $\sigma^{(LY)}$
   & $-0.11594 \epsilon^6$
   &   \begin{tabular}{c c} & \\
  
     &  -0.5(4) (CF)\\
      & -0.2(2)\cite{phi^3} \\
     & -0.5 \cite{John1985} (exact) \\
     & \\
     &
\end{tabular}
   &  \begin{tabular}{c c c c} & \\
     & -0.169(6) ($\epsilon_{bc}=5$)\\
    &  -0.2(2) (CF)\\
      & -0.17(5)\cite{phi^3} \\
     & -0.1667 \cite{John1985} (exact) \\
     &  -0.193 \cite{12frg} \\
     & -0.1662(5) \cite{PRE2012} (sc) \\
     & -0.1662(5) \cite{PRE2012} (bcc) 
\end{tabular}
   & \begin{tabular}{c c c}
   & 0.076(2) ($\epsilon_{bc}=4$)\\
   & 0.0776(2) ($\epsilon_{bc}=5$)\\
     &  0.08(6) (CF)\\
      & 0.075(11) \cite{phi^3}\\
     & 0.0588 \cite{12frg}\\
     & 0.0742(56) \cite{11frg} \\
     & 0.077(2) \cite{PRE2012} (sc) \\
     & 0.076(2) \cite{PRE2012} (bcc) 
\end{tabular}
   & \begin{tabular}{c c c}
   & 0.260(1) ($\epsilon_{bc}=4$)\\
   & 0.260(1) ($\epsilon_{bc}=5$)\\
     &  0.26(1) (CF)\\
      & 0.259(5) \cite{phi^3}\\
     & 0.2648(6) \cite{12frg}\\
     & 0.2667(32) \cite{11frg} \\
     & 0.258(5) \cite{PRE2012} (sc) \\
     & 0.261(4) \cite{PRE2012} (bcc) 
\end{tabular}
& \begin{tabular}{c c c}
 & 0.3984(1) ($\epsilon_{bc}=4$)\\
 & 0.3984(1) ($\epsilon_{bc}=5$)\\
     &  0.3986(6) (CF)\\
      & 0.3986(13) \cite{phi^3}\\
     & 0.40166(2) \cite{12frg}\\
     & 0.4033(12)) \cite{11frg} \\
     & 0.402(2) \cite{PRE2012} (sc) \\
     & 0.402(5) \cite{PRE2012} (bcc) 
\end{tabular}
   \\
 \hline
   $\omega^{(LY)}$
   & $-58.443 \epsilon^6$
   &   \begin{tabular}{c c}
     &  2.3(5) (CF) \\
     &  1.9(6) \cite{phi^3}
\end{tabular}
   &  \begin{tabular}{c c}
     &  1.9(3) (CF)\\
     & 1.8(4) \cite{phi^3}
\end{tabular}
   & \begin{tabular}{c c c}
     &  1.6(2) (CF)\\
      & 1.5(2) \cite{phi^3}
\end{tabular}
   & \begin{tabular}{c c c}
     &  1.19(6) (CF)\\
      & 1.15(7)\cite{phi^3}
\end{tabular}
& \begin{tabular}{c c c}
     &  0.697(8) (CF)\\
      & 0.691(8)\cite{phi^3}
\end{tabular}\\ 
 \hline
\end{tabular}
\end{center}
\end{table}
\begin{figure}[ht]
\centering
\begin{subfigure}{0.495\textwidth}
\includegraphics[width=1\linewidth, height=6cm]{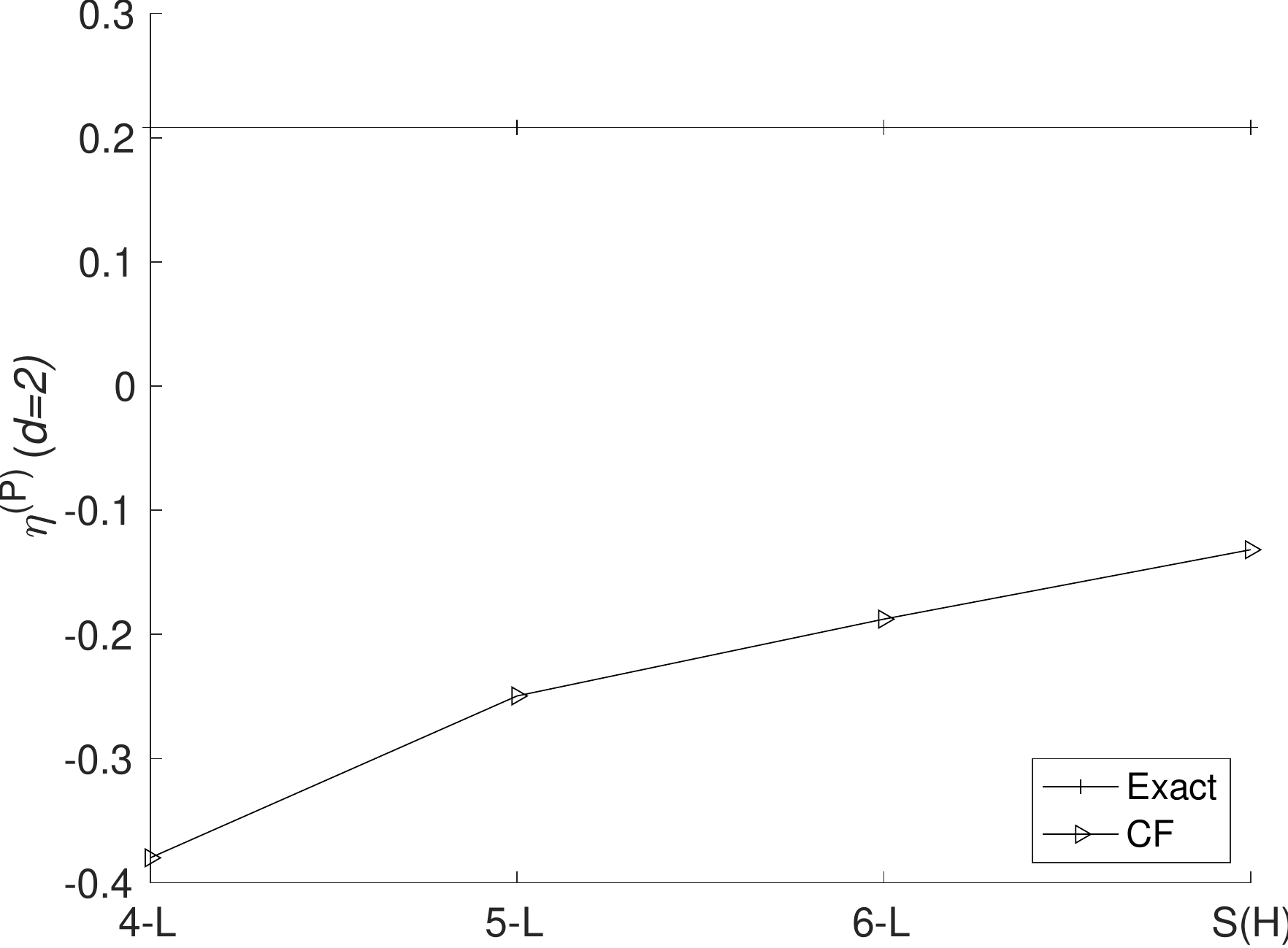} 
\caption{Estimates of two-dimensional $\eta^{(P)}$ compared with exact result $\eta^{(P)}=0.2083$ \cite{PhysRevB1980,PhysRevLett1984}.}

\end{subfigure}
\begin{subfigure}{0.495\textwidth}
\includegraphics[width=1\linewidth, height=6cm]{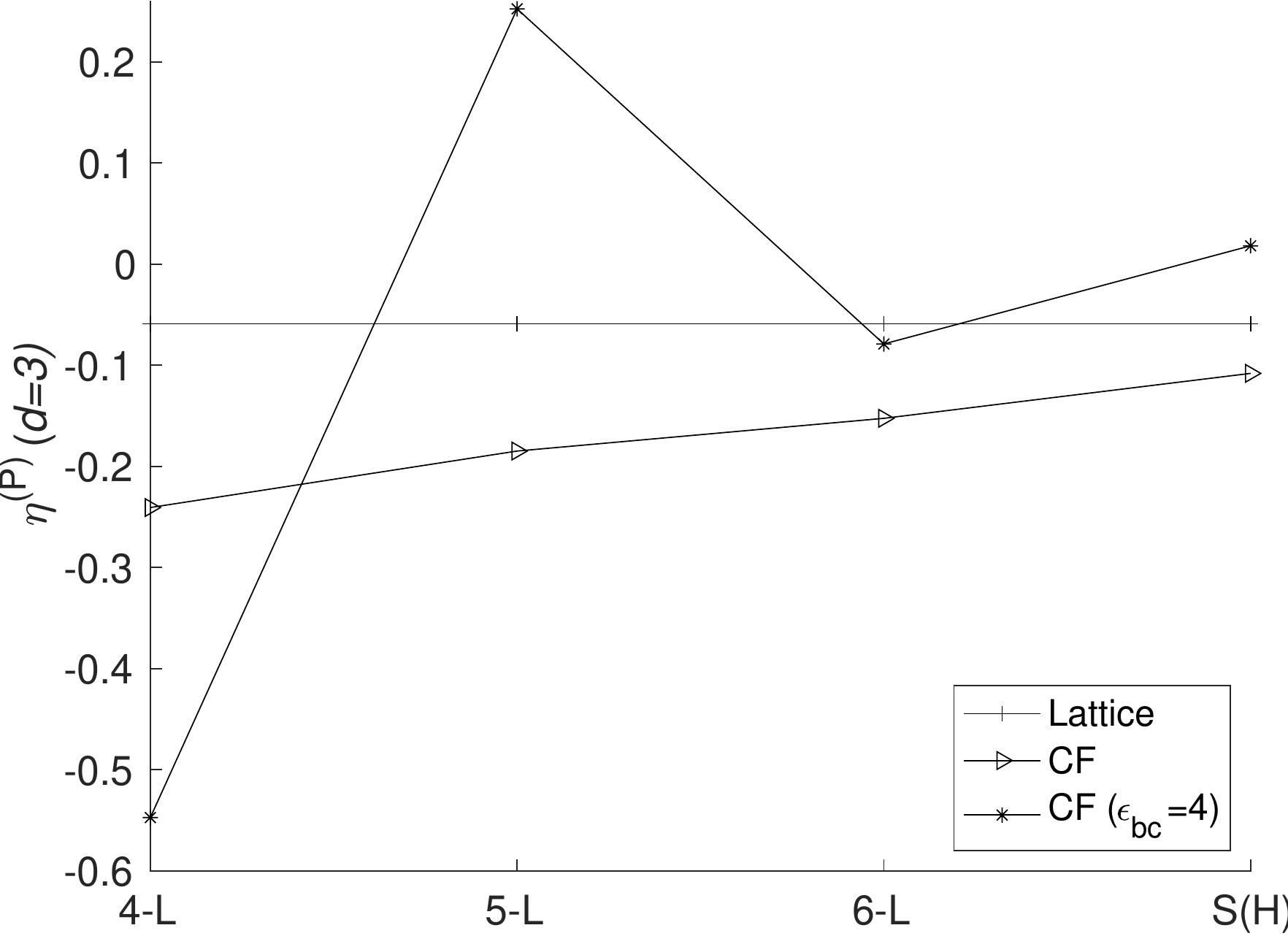}
\caption{Estimates of three-dimensional $\eta^{(P)}$ compared with lattice result $\eta^{(P)}=-0.059$ \cite{90-1998}.}

\end{subfigure}

\caption{Estimates of $\eta^{(LY)}$ at successive orders compared with existing results.}
\end{figure} 
 \begin{figure}[ht]
\centering
\begin{subfigure}{0.495\textwidth}
\includegraphics[width=1\linewidth, height=6cm]{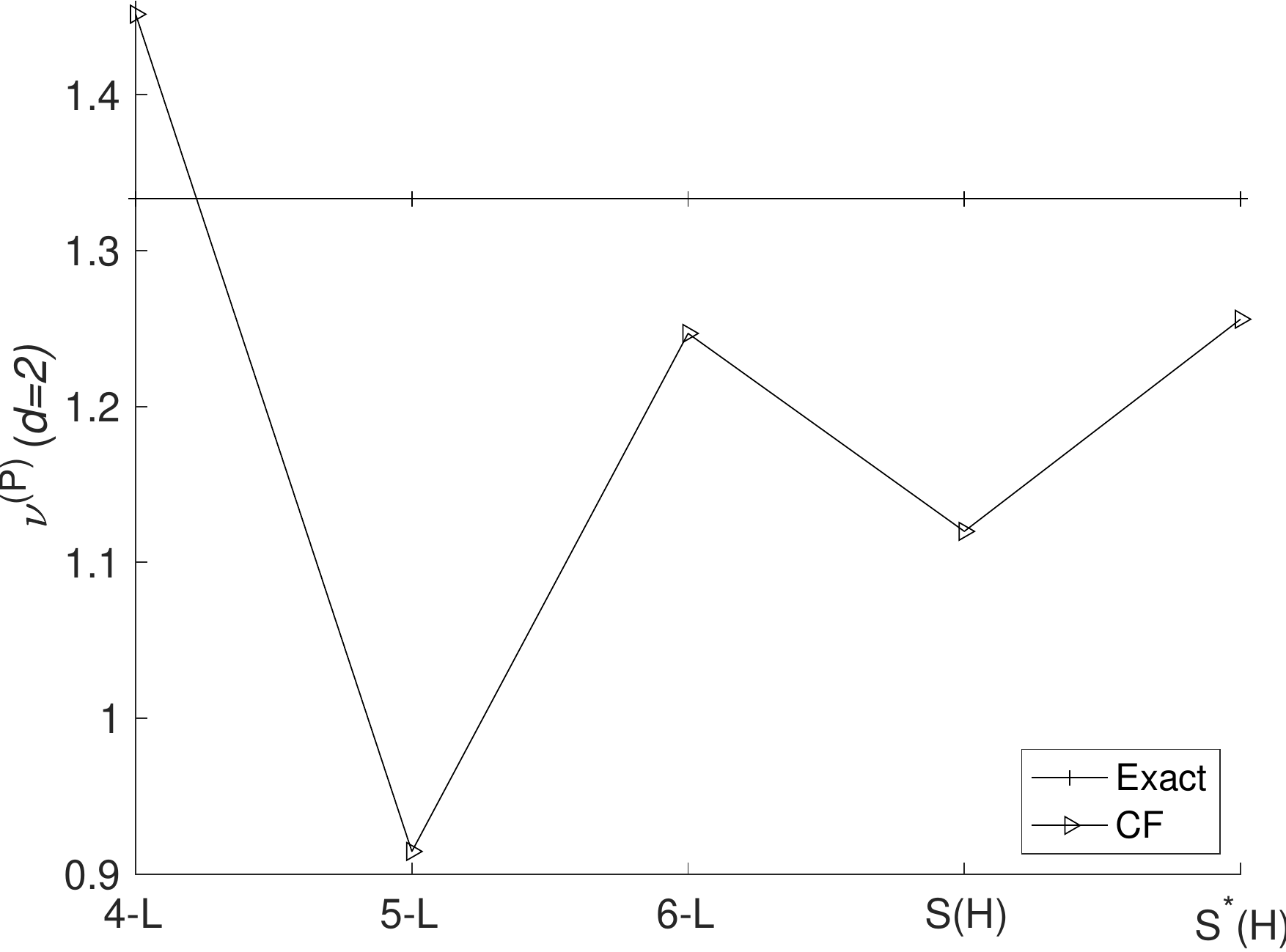} 
\caption{Estimates of two-dimensional $\nu^{(P)}$ compared with exact result $\nu^{(P)}=1.333$ \cite{PhysRevB1980,PhysRevLett1984}.}

\end{subfigure}
\begin{subfigure}{0.495\textwidth}
\includegraphics[width=1\linewidth, height=6cm]{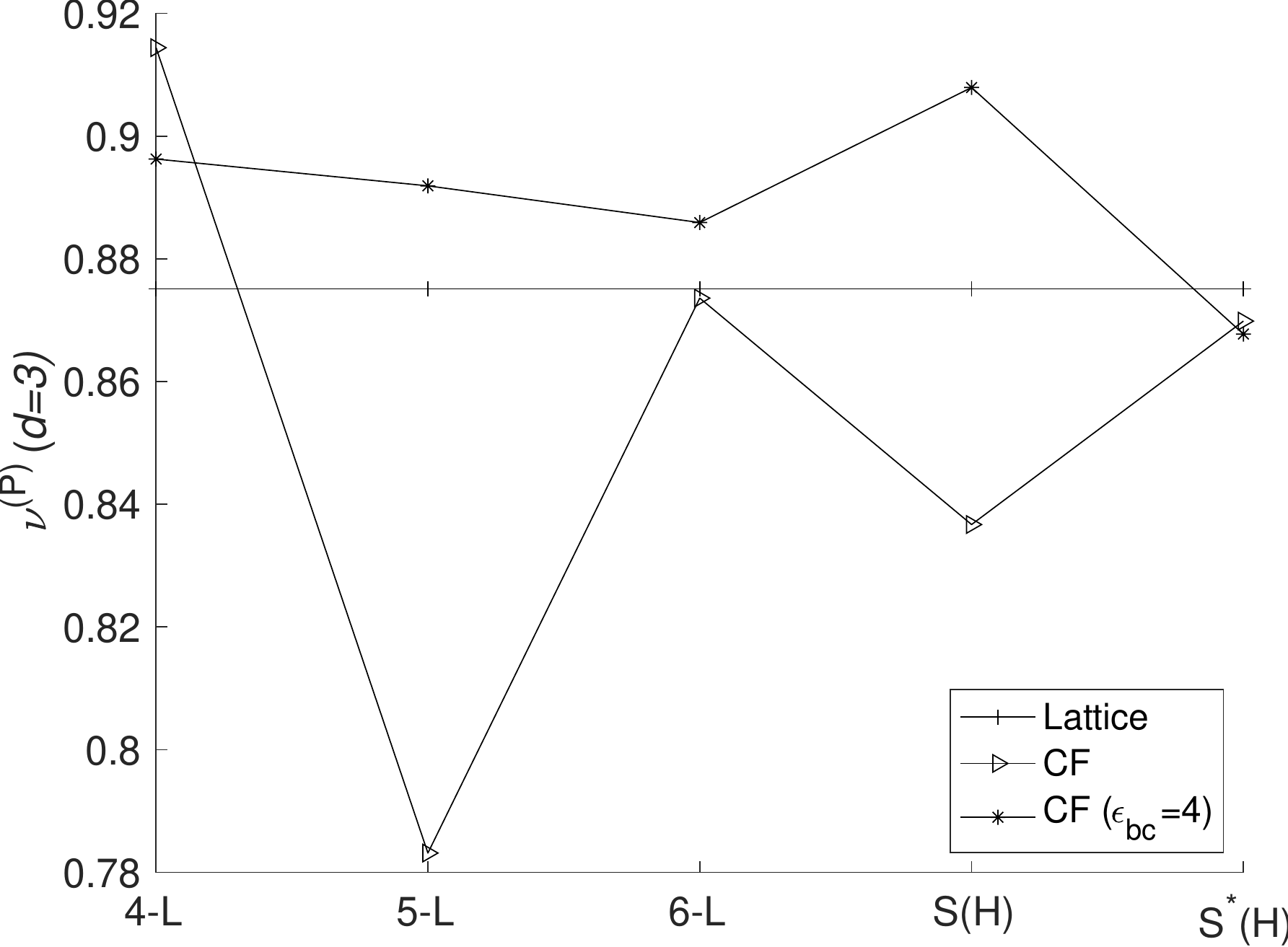}
\caption{Estimates of three-dimensional $\nu^{(P)}$ compared with lattice result $\nu^{(P)}=0.8751$ \cite{PhysRevE95}.}

\end{subfigure}

\caption{Estimates of $\nu^{(LY)}$ at successive orders compared with existing results.}
\end{figure} 
 \begin{figure}[ht]
\centering
\begin{subfigure}{0.495\textwidth}
\includegraphics[width=1\linewidth, height=6cm]{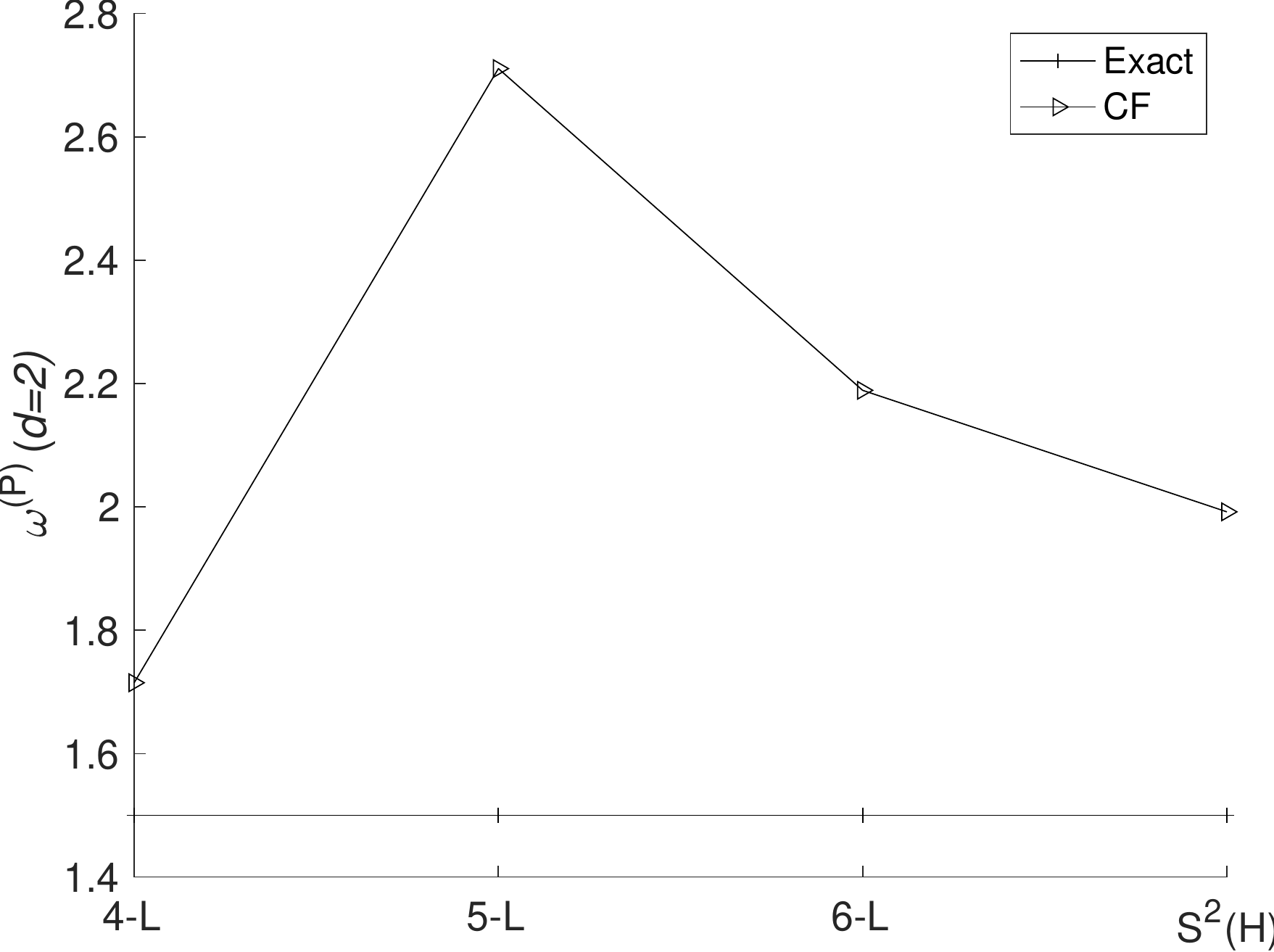} 
\caption{Estimates of two-dimensional $\omega^{(P)}$ compared with exact result $\omega^{(P)}=1.5$ \cite{PhysRevB1980,PhysRevLett1984}.}

\end{subfigure}
\begin{subfigure}{0.495\textwidth}
\includegraphics[width=1\linewidth, height=6cm]{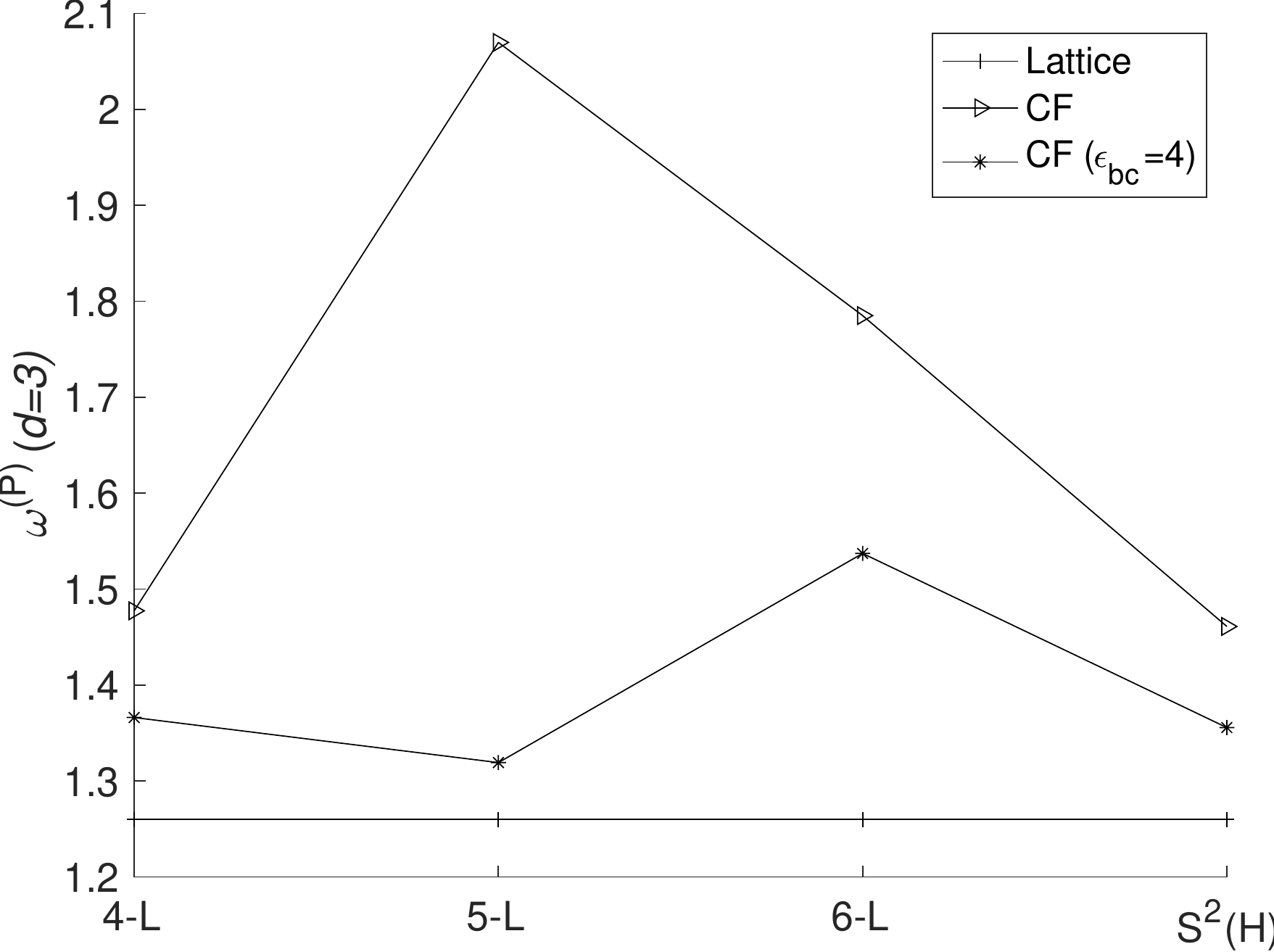}
\caption{Estimates of three-dimensional $\omega^{(P)}$ compared with lattice result $\omega^{(P)}=1.26$ \cite{86-1990}.}

\end{subfigure}

\caption{Estimates of $\sigma^{(LY)}$ at successive orders compared with existing results.}
\end{figure}
\\Similarly, using CF we obtain percolation exponents for dimensions ranging from two to five in Table 9. For percolation exponent $\nu^{(P)}$ direct estimates and constrained series estimates show improvement and are compatible with five-loop \cite{phi^3} and four-loop \cite{four-loop-phi^3} resummation results. Also these estimates seem to be comparable with predictions from discrete lattice models \cite{Koza2016,PhysRevB93,PhysRevE94,PhysRevE95,86-1990}, especially in physically relevant three-dimensions. We illustrate the convergence nature of these estimates in Figs. (20), (21), (22) and compare with their corresponding exact results for two-dimensional case and compare with lattice results for three-dimensional case. In case of $\eta^{(P)}$ the estimates seem to undershoot or overshoot similar to five-loop resummation results \cite{phi^3}, which is quite evident in direct estimate of two-dimensional case where $\eta^{(P)}=-0.1(1)$ from CF in Fig. 20(a) and $\eta^{(P)}=-0.28(15)$ \cite{phi^3} do not match with exact numerical result $\eta^{(P)}=.2083$ \cite{PhysRevB1980,PhysRevLett1984}. Using this value, constrained series estimate seems to overshoot as seen in Fig. 20(b) for three-dimensional case when compared with existing predictions from lattice models \cite{86-1990,90-1998}. Constrained series estimates of $\omega^{(P)}$ show improvement when compared to five-loop resummation results \cite{phi^3}. 

\begingroup
\setlength{\tabcolsep}{0.6pt} 
\renewcommand{\arraystretch}{1} 
\begin{table}[htbp]
 \scriptsize
\begin{center}
\caption{CF estimates for percolation critical exponents  in two to five dimensions compared with existing predictions and hypergeometric prediction for their six-loop approximation.} 

 \begin{tabular}{||c c c c c c||}
 
 \hline
Exponent &  \begin{tabular}{c c}
     &  Six-loop\\
     & approximation
\end{tabular} & $\epsilon=4$   & $\epsilon=3$  & $\epsilon=2$  & $\epsilon=1$  \\ [0.5ex] 
 \hline\hline
   $\eta^{(P)}$
   & $-0.76835 \epsilon^6$
   &   \begin{tabular}{c c}
     &  -0.1(1) (CF)\\
     & -0.28(15) \cite{phi^3}\\
     & 0.2083 \cite{PhysRevB1980,PhysRevLett1984} (exact)
\end{tabular}
   &  \begin{tabular}{c c}
   &  0.02(21) ($\epsilon_{bc}=4$) \\ 
     &  -0.10(4) (CF)\\
      & -0.06(10)\cite{phi^3} \\
     & -0.059(9) \cite{90-1998} \\
     & -0.07(5) \cite{86-1990}
\end{tabular}
   &  \begin{tabular}{c c}
   &  -0.09(3) ($\epsilon_{bc}=4$) \\ 
     &  -0.09(2) (CF)\\
      & -0.10(3)\cite{phi^3} \\
     & -0.12(4) \cite{86-1990}  \\
     &
\end{tabular}
  &  \begin{tabular}{c c}
   &  -0.0574(3) ($\epsilon_{bc}=4$) \\  &  -0.0571(6) (CF)\\
      & -0.056(3)\cite{phi^3} \\
     & -0.075(20) \cite{86-1990}  \\
     &
\end{tabular}

   \\
   \hline
    $\nu^{(P)}$
   & $-1.7704 \epsilon^6$
   &   \begin{tabular}{c c}
     & 1.1(2) (CF,$S$) \\
     & 1.3(1) (CF,$S^*$)\\
     & 1.1(2) \cite{phi^3}\\
     & 1.3333 \cite{PhysRevB1980,PhysRevLett1984} (exact) \\
     & \\
     & \\
     & \\
\end{tabular}
   &  \begin{tabular}{c c}
    &  0.91(1) ($\epsilon_{bc}=4$,$S$) \\
    &  0.867(14) ($\epsilon_{bc}=4$,$S^*$) \\
     & 0.84(6) (CF,$S$) \\
     & 0.87(2) (CF,$S^*$)\\
     & 0.89(3) \cite{phi^3}\\
     & 0.8960 \cite{four-loop-phi^3} \\
     & 0.8774(13) \cite{Koza2016} \\
     & 0.8765(18) \cite{PhysRevB93} \\
     & 0.8764(12) \cite{PhysRevE94} \\
     & 0.8751(11) \cite{PhysRevE95} \\
     & 0.872(70) \cite{86-1990}
\end{tabular}
   & \begin{tabular}{c c c}
   &  0.704(9) ($\epsilon_{bc}=4$,$S$) \\
    &  0.685(3) ($\epsilon_{bc}=4$,$S^*$) \\
     & 0.68(1) (CF,$S$) \\
     & 0.685(5) (CF,$S^*$)\\
      & 0.689(5) \cite{phi^3} \\
     &  0.6920 \cite{four-loop-phi^3} \\
     & 0.6852(28) \cite{Koza2016} \\
     & 0.678(50) \cite{86-1990}
     & \\
     & \\
     & \\
     &
\end{tabular}
    & \begin{tabular}{c c c}
   &  0.572(1) ($\epsilon_{bc}=4$,$S$) \\
    &  0.5738(4) ($\epsilon_{bc}=4$,$S^*$) \\
     & 0.5735(9) (CF,$S$) \\
     & 0.5739(4) (CF,$S^*$)\\
      & 0.5741(6) \cite{phi^3} \\
     &  0.5746 \cite{four-loop-phi^3} \\
     & 0.5723(18) \cite{Koza2016} \\
     & 0.571(3) \cite{86-1990}
     & \\
     & \\
     & \\
     &
\end{tabular}\\ 
 
 \hline
   $\omega^{(P)}$
   & $-150.01 \epsilon^6$
   &   \begin{tabular}{c c}
     &  1.9(2) (CF) \\
     &  2.4(2) \cite{phi^3} \\
     & 1.5 \cite{PhysRevB1980,PhysRevLett1984} (exact)
\end{tabular}
   &  \begin{tabular}{c c}
   &  1.356(8) ($\epsilon_{bc}=4$) \\ 
     &  1.46(24) (CF) \\
     & 1.7(3) \cite{phi^3} \\
     & 1.6334 \cite{four-loop-phi^3} \\
     & 1.0(2) \cite{KOZLOV2010} \\
     & 1.26(23) \cite{86-1990}
\end{tabular}
    &  \begin{tabular}{c c}
   &  1.104(8) ($\epsilon_{bc}=4$) \\ 
     &  1.41(5) (CF) \\
     & 1.31(9) \cite{phi^3} \\
     & 1.2198 \cite{four-loop-phi^3} \\
     & 0.94(15) \cite{86-1990} \\ &
\end{tabular}
   &  \begin{tabular}{c c}
   &  0.694(2) ($\epsilon_{bc}=4$) \\ 
     &  0.76(1) (CF) \\
     & 0.746(11) \cite{phi^3} \\
     & 0.7178 \cite{four-loop-phi^3} \\
     & 0.96(26) \cite{86-1990}
\\ & \end{tabular} \\
 \hline
\end{tabular}
\end{center}
\end{table}
\section{Conclusion}
 The asymptotic nature of different perturbative expansions is studied using properties of ${}_{j+2}F_j$ hypergeometric functions and, by using continued functions such as continued exponential, continued fraction and continued exponential fraction. The discussed methods can be used to extrapolate strong coupling information, large order information and use them to approximate meaningful estimates from perturbation series of field theories. The effective approach of extrapolation was examined in detail by using the method in different instances of field theories. Existing perturbative renormalization group information was implemented in hypergeometric functions to predict the six-loop approximation results in $\epsilon$-expansions of $\phi^3$ models and eight-loop approximation results in $\epsilon$-expansions of $\phi^4$ models. Further to check the validity of these predictions, they were implemented in appropriate resummation procedure using continued functions to extract critical exponents relevant to physical systems. The region of strong coupling limit was discussed in Gell-Mann-Low functions of $\phi^4$ field theory. While there exist different sophisticated ways to approximate information through resummation techniques, extrapolation of relevant unknown large parameter information regarding the sought quantities with only minimal available small parameter information is an interesting new application \cite{Shalaby2022,physrevdself2}.
 
Perturbative field-theoretic results are not exact, and usually, rigorous resummation methods are used to check their validity. The methods discussed using hypergeometric functions and continued functions are quite general once a sequence of equations are constructed for arbitrary order of perturbative information. In such field-theoretic works, hypergeometric functions can be first used to check the direct results numerically, and the simple, straightforward continued functions can be used to check the resummed results. However, further, generalized convergence analysis of different continued functions based on the perturbation series's numerical structure and their convergence limits would be helpful. Also, one can further study efficiently implementing ${}_{4}F_2$ or higher hypergeometric functions \cite{Shalaby2022arxiv}, which requires more than normally used computing to solve the system of non-linear equations for hypergeometric function coefficients. 
\\ \\
Data sharing not applicable to this article as no datasets were generated or analysed during the current study.

\bibliographystyle{ieeetr}
\bibliography{sample.bib}
\end{document}